\documentclass[%
 reprint,
superscriptaddress,
nofootinbib,
 amsmath,amssymb,
 aps,
prd,
]{revtex4-2}

\usepackage{graphicx}
\usepackage{dcolumn}
\usepackage{bm}
\usepackage{hyperref}

\hypersetup{
	colorlinks,
	linkcolor={blue!70!black},
	citecolor={blue!70!black},
	urlcolor={blue!70!black}
}

\usepackage{pifont}
\usepackage{orcidlink}
\usepackage{ragged2e}
\usepackage{multirow}
\usepackage{booktabs}

\definecolor{MidnightBlue}{HTML}{104e8b}

\newcommand{\ion}[2]{#1\,\textsc{#2}}
\newcommand{\GRASP}{\texttt{GRASP}}
\newcommand{\FAC}{\texttt{FAC}}
\newcommand{\HFR}{\texttt{HFR}}
\newcommand{\HULLAC}{\texttt{HULLAC}}
\newcommand{\conf}[3]{$#1$\textit{#2}$^{\,#3}$}
\newcommand{\conb}[3]{\textcolor{MidnightBlue}{$#1$\textit{#2}$^{\,#3}$}}

\newcommand\aap{Astron. \& Astrophys.}                      
\newcommand\aj{AJ}                                          
\newcommand\aplett{Astrophys.~Lett.}                        
\newcommand\apjl{Astrophys, J.}                             
\newcommand\apjs{ApJS}                                      
\newcommand\jqsrt{J.~Quant. Spectrosc. Radiative Transfer}  
\newcommand\mnras{Mon. Not. R. Astron. Soc.}                
\newcommand\pasj{PASJ}                                      
\newcommand\physscr{Phys.~Scr.}                             
\renewcommand\nat{Nature}                                   

\begin{document}


\title{Calibrated Lanthanide Atomic Data for Kilonova Radiative Transfer. \\ I.~Atomic Structure and Opacities}

\author{Andreas~Fl\"ors\,\orcidlink{0000-0003-2024-2819}}
\email[]{a.floers@gsi.de}
\affiliation{GSI Helmholtzzentrum f{\"u}r Schwerionenforschung,
	Planckstra{\ss}e 1, D-64291 Darmstadt, Germany}
  
\author{Ricardo~Ferreira~da~Silva\,\orcidlink{0000-0003-3030-0496}}
\affiliation{Laboratório de Instrumentação e Física Experimental de Partículas (LIP), Av. Prof. Gama Pinto 2, 1649-003 Lisboa, Portugal}
\affiliation{Faculdade de Ciências da Universidade de Lisboa, Rua Ernesto de Vasconcelos, Edifício C8, 1749-016, Lisboa, Portugal}

\author{José~Pires~Marques\,\orcidlink{0000-0002-3797-3880}}
\affiliation{Laboratório de Instrumentação e Física Experimental de Partículas (LIP), Av. Prof. Gama Pinto 2, 1649-003 Lisboa, Portugal}
\affiliation{Faculdade de Ciências da Universidade de Lisboa, Rua Ernesto de Vasconcelos, Edifício C8, 1749-016, Lisboa, Portugal}

\author{Jorge~Miguel~Sampaio\,\orcidlink{0000-0003-4359-493X}}
\affiliation{Laboratório de Instrumentação e Física Experimental de Partículas (LIP), Av. Prof. Gama Pinto 2, 1649-003 Lisboa, Portugal}
\affiliation{Faculdade de Ciências da Universidade de Lisboa, Rua Ernesto de Vasconcelos, Edifício C8, 1749-016, Lisboa, Portugal}

\author{Gabriel~Mart{\'i}nez-Pinedo\,\orcidlink{0000-0002-3825-0131}}
\affiliation{GSI Helmholtzzentrum f{\"u}r Schwerionenforschung,
	Planckstra{\ss}e 1, D-64291 Darmstadt, Germany}
\affiliation{Institut für Kernphysik (Theoriezentrum),  Fachbereich Physik,
	Technische Universität Darmstadt, Schlossgartenstra{\ss}e 2, D-64289
Darmstadt, Germany}
\affiliation{Helmholtz Forschungsakademie Hessen f\"ur FAIR (HFHF),
	GSI Helmholtzzentrum f\"ur Schwerionenforschung, Planckstra{\ss}e~1,
D-64291 Darmstadt, Germany}

\date{\today}

\begin{abstract}
	The early spectra of the kilonova (KN) AT2017gfo following the binary neutron star merger GW170817 exhibit numerous features shaped by r-process nucleosynthesis products. Although a few species were tentatively detected, no third-peak elements were unambiguously identified, as the amount of atomic data required for radiative transfer modeling is immense. Although comprehensive atomic data, including atomic opacities, is now available for many elements, wavelength-calibrated data remains limited to a select few ions.
		
	To examine the atomic opacities of all singly and doubly ionized lanthanides, from La (Z = 57) to Yb (Z = 70), we perform atomic structure calculations using the \FAC\ code. Our calculations incorporate an innovative optimization of the local central potential and the number of configurations considered, alongside a calibration technique aimed at enhancing agreement between theoretical and experimental atomic energy levels. We assess the accuracy of the computed data, including energy levels and electric dipole (E1) transition strengths, as well as their impact on KN opacities. We find that strong transitions ($\log(gf)>-1$) are in good agreement with both experiments and semi-empirical calculations. For ions with substantial experimental data, the computed opacities exhibit good agreement with prior calculations. By calibrating low-lying energy levels with experimental data, we have identified 66\,591 transitions with experimentally calibrated wavelength information, rendering future lanthanide line identifications through radiative transfer modeling feasible. In total, our calculations encompass 28 ions, yielding 146\,856 energy levels below the ionization threshold and 28\,690\,443 transitions among these levels. 
\end{abstract}
\maketitle
\section{Introduction}
The discovery of the kilonova (KN) AT2017gfo \cite{2017Sci...358.1556C, 2017ApJ...848L..17C, 2017Natur.551...75S, 2017Sci...358.1570D, 2017PASJ...69..101U}, associated with the gravitational wave event GW170817 \cite{2017ApJ...848L..12A, 2017PhRvL.119p1101A}, marked a major milestone in astrophysics. This multi-messenger observation provided direct evidence that binary neutron star mergers are a site of rapid neutron-capture (r-process) nucleosynthesis \cite{1974ApJ...192L.145L, 1982ApL....22..143S, 1989Natur.340..126E, 1999ApJ...525L.121F, 2011ApJ...738L..32G, 2013PhRvL.111m1101B, 2014ApJ...789L..39W}, and that the electromagnetic emission from such events is powered by the radioactive decay of freshly synthesized heavy elements \cite{1998ApJ...507L..59L, 2005astro.ph.10256K, 2010MNRAS.406.2650M}.

The kilonova light curve undergoes rapid temporal evolution. For AT2017gfo, this manifested as a spectral transition from blue optical emission at early times to redder, near-infrared wavelengths at later epochs \citep{2013ApJ...774...25K, 2013ApJ...775..113T, 2014MNRAS.441.3444M, 2017Natur.551...80K}. This spectral evolution is driven by the opacity of the ejected material, which depends sensitively on the atomic structure of the r-process elements. In particular, lanthanides with their partially filled \conf{4}{f}{} electron shells strongly influence the opacity due to their complex spectra and large number of low-energy transitions \citep{2016ApJ...829..110B, 2017Natur.551...80K, 2018ApJ...852..109T, 2020MNRAS.496.1369T}. These transitions lead to strong line blanketing in the optical and NIR bands, reducing optical flux and shifting emission toward longer wavelengths \citep{2017Natur.551...80K, 2018ApJ...852..109T, 2020MNRAS.496.1369T}.

While the distinction between “blue” and “red” KNe is often linked to the presence or absence of lanthanides in the ejecta, recent multidimensional radiative transfer simulations have shown that this dichotomy may oversimplify the complex interplay between ejecta composition, geometry, and viewing angle \citep{2023ApJ...954L..41S, 2024MNRAS.529.1333C}. Nonetheless, lanthanide-rich ejecta remain a critical factor in shaping KN spectra and must be accurately treated in radiative transfer models \citep{2013ApJ...774...25K, 2017Natur.551...80K, 2021ApJ...918...44B}.

Even the simplest KN radiative transfer models, assuming local thermodynamic equilibrium (LTE), spherical symmetry, and steady-state conditions, require accurate and complete atomic data to model opacities \citep{2020MNRAS.496.1369T}. For LTE calculations, knowledge of energy levels and transition probabilities is sufficient, but the computation of these quantities for lanthanides is highly non-trivial. Their complex atomic structure, driven by a large number of electrons, strong electron-electron correlations, partially filled \conf{4}{f}{} orbitals, and relativistic effects, limits both the accuracy and completeness of theoretical predictions.

Prior to the observational breakthrough provided by AT2017gfo, the opacity of r-process ejecta had already been the subject of theoretical investigation through atomic structure calculations \citep{2013ApJ...774...25K, 2013ApJ...775..113T, 2013ApJ...775...18B, 2015MNRAS.450.1777K, 2015HEDP...16...53F}. Nonetheless, the detection of AT2017gfo significantly accelerated efforts to compute atomic data for lanthanides and other second or third r-process peak elements. These initiatives have spanned a range of ionization states: some groups \citep[e.g.][]{2020ApJ...901...29B, 2022MNRAS.509.6138C, 2022MNRAS.513.2302C, 2022ApJ...934..117B, 2023MNRAS.518..332C, 2024ApJ...968...64B, 2024A&A...685A..91C} have focused on highly charged ions (5$^{\rm th}$ to 10$^{\rm th}$ ionization stages) that dominate the early-time ($<$1 day) KN emission, while others have concentrated on low-charge species (neutral through 4$^{\rm th}$ ionization), which are most relevant for interpreting light curves and spectra at later epochs, such as those observed in AT2017gfo. These studies can be further categorized into those targeting individual elements \citep{2020ApJS..248...13G, 2021MNRAS.501.1440C, 2021MNRAS.506.3560G, 2022Atoms..10...18S, 2023MNRAS.524.3083F, 2024MNRAS.530.5220G} and those aiming to cover a broader set of elements or ionization stages \citep[e.g.][]{2017Natur.551...80K, 2020MNRAS.493.4143F, 2020MNRAS.496.1369T, 2020ApJS..248...17R, 2021ApJS..257...29R, 2023MNRAS.519.2862F, 2024MNRAS.535.2670K, 2025A&A...696A..32D}. A substantial portion of this work has concentrated on lanthanide ions, owing to their characteristically high line opacities, which play a dominant role in shaping KN spectra. Motivated by similar considerations, actinide ions have also been investigated, albeit to a lesser extent and in a smaller number of studies \citep{2020ApJ...899...24E, 2023MNRAS.519.2862F, 2023MNRAS.524.3083F}.

To support these efforts, a variety of atomic structure codes have been employed, including \texttt{AUTOSTRUCTURE} \citep{2016ascl.soft12014B}, \texttt{FAC} \citep{2008CaJPh..86..675G}, \texttt{GRASP} \citep{2019CoPhC.237..184F}, \texttt{HULLAC} \citep{2001JQSRT..71..169B}, \texttt{HFR} \citep{1981tass.book.....C}, \texttt{LASER} \citep{2015JPhB...48n4014F}, and \texttt{MCDFGME} \citep{1975CoPhC...9...31D}. These codes adopt differing approaches to strike a balance between computational efficiency and the level of physical accuracy required to model the complex electron configurations of heavy, open-shell ions. Some studies have targeted individual ions in detail to assess and improve the accuracy of computed atomic data, while others have aimed to produce extensive datasets covering many species, thereby enabling comprehensive radiative transfer simulations of KNe.

Despite these advances, KN models remain subject to considerable uncertainties, primarily due to the limitations of the available atomic data. Accurate modeling of heavy element spectra requires sophisticated configuration interaction (CI) methods that include relativistic effects. However, the validation of such theoretical predictions is constrained by the scarcity of experimental data, especially for low-ionization lanthanide species. Experimental benchmarks, such as measured energy levels or transition wavelengths, are limited, and given the vast number of relevant ions, this shortfall is unlikely to be fully addressed in the near future. As a result, theoretical results must be carefully validated through cross-comparison with available experimental data, spectral features observed in stellar atmospheres, and consistency checks against previous theoretical calculations.

One of the principal goals of KN modeling is the secure identification of spectral signatures corresponding to second- and third-peak r-process elements, which are synthesized in the aftermath of neutron star mergers. These identifications are essential for understanding the detailed nucleosynthetic pathways and for constraining the elemental yields produced in such extreme environments. For instance, tentative detections of lanthanide elements in KN spectra have already been reported \citep[see][]{2022ApJ...939....8D}, demonstrating the feasibility of using KN observations to probe heavy element formation. However, to move beyond tentative identification toward systematic abundance measurements, models require highly reliable atomic data, particularly accurate opacities that govern radiative transfer in the ejecta.

The calculation of these opacities hinges on both the completeness and accuracy of the atomic structure data, namely, energy levels, oscillator strengths, and transition wavelengths. While reasonably complete atomic datasets exist for many first- and second-peak r-process elements, these datasets often rely on theoretical structure calculations which, despite being state-of-the-art, can suffer from substantial wavelength inaccuracies. Recent work \citep[e.g.][]{2021PhRvA.103b2808L, 2022MNRAS.509.4723M, 2023MNRAS.524.3083F, 2023ApJ...954L..41S} has highlighted that even small errors in predicted wavelengths can significantly impair the ability to match synthetic spectral features with observed lines. In extreme cases, discrepancies exceeding 50\% between theoretical and experimental wavelengths have been found, rendering confident line identifications effectively impossible.

To mitigate this issue, theoretical energy levels can be corrected, or calibrated, using experimental spectroscopic data. Such calibrations serve to anchor theoretical predictions in empirically verified energy structures, thereby improving the fidelity of predicted wavelengths. In previous work, we developed a calibration scheme that grouped levels by their parity and total angular momentum, applying a uniform shift within each group to align the theoretical level structure with available experimental energies \citep[][]{2023MNRAS.524.3083F}. This method enhanced the agreement between calculated and experimental transition wavelengths, but the resulting accuracy -- while improved -- was still insufficient to enable unambiguous line identifications in most cases.

A fundamental challenge lies in the absence of a general, automated method for associating theoretical energy levels with experimental ones. The quantum mechanical labeling of levels (e.g., in terms of configuration, term symbols, and mixing coefficients) is often ambiguous or incomplete, particularly in complex open-shell systems typical of lanthanides and actinides. As a result, identifications between calculated and measured levels must be done manually, requiring careful inspection of level energies, their LS term contributions, and quantum numbers. Reliable wavelength predictions are central to the interpretation of astrophysical observations and to our understanding of the cosmic origin of the heavy elements.

The objective of the present study is to generalize and extend our previous atomic structure calculations for \ion{Nd}{ii} and \ion{Nd}{iii} to encompass all 28 singly and doubly ionized lanthanide species. For each ion for which experimental spectroscopic data are available, we apply a level calibration procedure to align the computed energy levels with measured values. In Section~\ref{sec:methods}, the general theory of atomic structure, the \FAC\ code and our calibration procedure are reviewed. In Section~\ref{sec:calculations}, we present the results of our atomic structure calculations for all singly and doubly ionized lanthanides. For each ion, we also discuss the accuracy of the calculated energy levels compared to available experimental data. In Section~\ref{sec:opacity}, we focus on the permitted E1 transitions between our calculated and calibrated energy levels. We compare oscillator strengths to both experimental data used in stellar spectroscopy as well as other theoretical works. From the oscillator strengths, we derive opacities for LTE conditions, which will be compared to the literature. Finally, in Section~\ref{sec:summary}, we summarize our findings and conclude with an outlook.

\section{Methods}
\label{sec:methods}
\subsection{Atomic structure calculations}
\label{sec:fac}
For our computations, we employ the \texttt{Flexible Atomic Code} \citep[\texttt{FAC, }][]{2008CaJPh..86..675G}, an open-source package for relativistic atomic structure. \FAC\ operates by diagonalizing the Dirac-Coulomb Hamiltonian
\begin{equation}
	\label{eq:hdc}
	H_{D C}=\sum_{i=1}^{N}\left(c \bm{\alpha}_{i} \cdot \bm{p}_{i}+\left(\beta_{i}-1\right) c^{2}+V_{i}\right)+\sum_{i<j}^{N} \frac{1}{r_{i j}} \, , 
\end{equation}
where $\bm{\alpha}_{i}$ and $\beta_i$ represent the $ 4 \, \times \, 4$ Dirac matrices, and $V_i$ is the potential from the nuclear charge. The screened hydrogenic approximation accounts for vacuum polarization and self-energy corrections. The Breit interaction, included in the zero-energy approximation for the exchanged photon, accounts for recoil and retardation effects.

The \FAC{} code performs relativistic configuration interaction (CI) calculations. In this approach, an Atomic State Function (ASF), $\Psi$, is expressed as a superposition of $N_{\texttt{CSF}}$ Configuration State Functions (CSFs), $\Phi_i$, that share the same total angular momentum $J$ and parity $P$:
\begin{equation}
	\Psi (n J M_J P)=\sum_{i=1}^{N_{\texttt{CSF}}}c_i^n \, \Phi_i (\gamma_i J M_J P),
\end{equation}
where $c_i$ are the expansion coefficients and $\gamma_i$ refers to the additional quantum numbers that uniquely characterise each CSF.

The entire calculation is performed within a $jj$-coupling scheme, and the output is presented accordingly. 

The ASFs and the corresponding eigenvalues are obtained by diagonalisation of the Hamiltonian in Equation~\eqref{eq:hdc} in the basis of N$_{\text{CSF}}$ configuration state functions.
For a specific collection of configurations, these eigenvalues provide the optimal energy estimations. Typically, including more configurations enhances the precision of the wave functions, albeit with a significant increase in both computational time and memory demand. In this study, the quantity of configurations is chosen based on the ion's complexity. For ions with a half-filled $f$-shell, we include fewer than 10 configurations, while ions at the extremes of the lanthanide series may involve hundreds. The influence of additional configurations on the precision of the computed energies was examined by \cite{2023MNRAS.524.3083F} for neodymium ions.

A significant benefit over the multiconfiguration Dirac-Fock method as implemented in codes such as \GRASP\ \citep{2019CoPhC.237..184F} or \texttt{MCDFGME}\,\citep{1975CoPhC...9...31D} is the inherent orthogonality of all electron orbitals, as they are all subjected to the same potential. Consequently, the eigenvalues and mixing coefficients are computed only once, which dramatically lowers the numerical complexity of the task. 

The computation of the FMC is achieved through the self-consistent Dirac-Fock-Slater iteration process. One-electron radial orbitals can be computed from the Dirac equation, although the potential is not optimized for any specific configuration. This is very computationally efficient, as orthogonality is achieved directly; however, it can compromise on accuracy.

Our approach constructs FMCs to account for the number of active electrons, distributed across various configurations. By altering the fractional occupation numbers, the weight of each configuration in the mean configuration construction can be adjusted. Consistent with earlier work on neodymium, the number of configurations is chosen such that further additions produce only minimal changes to the wave functions and, consequently, to the resulting level energies \cite{2023MNRAS.524.3083F}. Due to computational constraints, we could not extend the number of configurations beyond 4\,--\,10 for ions with a half-filled f-subshell. We show all included configurations in table~\ref{tab:FAC_configs1}. For further details on the \FAC\ calculations and the local central potential employed by \FAC\ see refs.~\cite{2023MNRAS.524.3083F, 2025arXiv250213250F}.
\subsection{Determination of FMC weights}
\label{sec:optimisation}

To enhance the accuracy of our calculations, we implemented a systematic optimization of the FMC used to determine the optimal potential \citep[for details see][]{2025arXiv250213250F}.  The optimization procedure employs a Sequential Model-Based Optimization (SMBO) algorithm, which is particularly effective for high-dimensional optimization problems. The algorithm constructs a Gaussian surrogate model to approximate the objective function, balancing exploration of the parameter space with exploitation of known low-error regions.

The optimization process begins with an initial dataset derived from energetically ordered configurations. Subsequent points are determined through Latin hypercube sampling. The algorithm utilizes multiple acquisition functions simultaneously -- Lower Confidence Bounds, Expected Improvement, and Probability of Improvement -- combined through a GP-Hedge strategy \citep[]{10.5555/3020548.3020587}. This approach probabilistically selects the best candidate for the next iteration using a softmax function applied to the computed acquisition functions.

To evaluate the performance of each iteration, we employ a weighted Root Mean Square Deviation (RMSD) between calculated values and reference data. The loss function ($\mathcal{L}$) incorporates a Boltzmann factor accounting for the excitation energy of reference levels, promoting convergence to physically meaningful values:

\begin{equation}
	\mathcal{L} = \sqrt{\frac{1}{N}\sum_{i=1}^{N} e^{-E^{\text{ref}}_i/kT} (\Delta E_i)^2}
\end{equation}

where $\Delta E_i$ represents the difference between calculated and reference energy levels $\Delta E_i^{\mathrm{ref}}$, $N$ is the number of levels considered, and $T$ is the excitation temperature. The squared differences are weighted by a Boltzmann factor, thereby assigning greater importance to low-lying energy levels, which are more significantly populated under thermodynamic conditions relevant to the plasma.

To avoid unphysical solutions, constraints are applied to the parameter space. In particular, the $f$-subshell is restricted between $n_e-2$ and $n_e+2$, (where $n_e$ is the number of electrons in the $n = 4$ and $n = 5$ $f$-subshells in the ground configuration. Similarly, the $n = 6$ and $n = 7$ $p$-subshells are limited between 0 and 1. These constraints ensure that the optimization explores physically meaningful configurations while allowing sufficient flexibility to improve the accuracy of the calculations. These constraints ensure that the optimization explores physically meaningful configurations while allowing sufficient flexibility to improve the accuracy of the calculations.

\subsection{Calibration to experimental data}
\label{sec:calibration}
\begin{figure*}
	\includegraphics[width=0.97\textwidth]{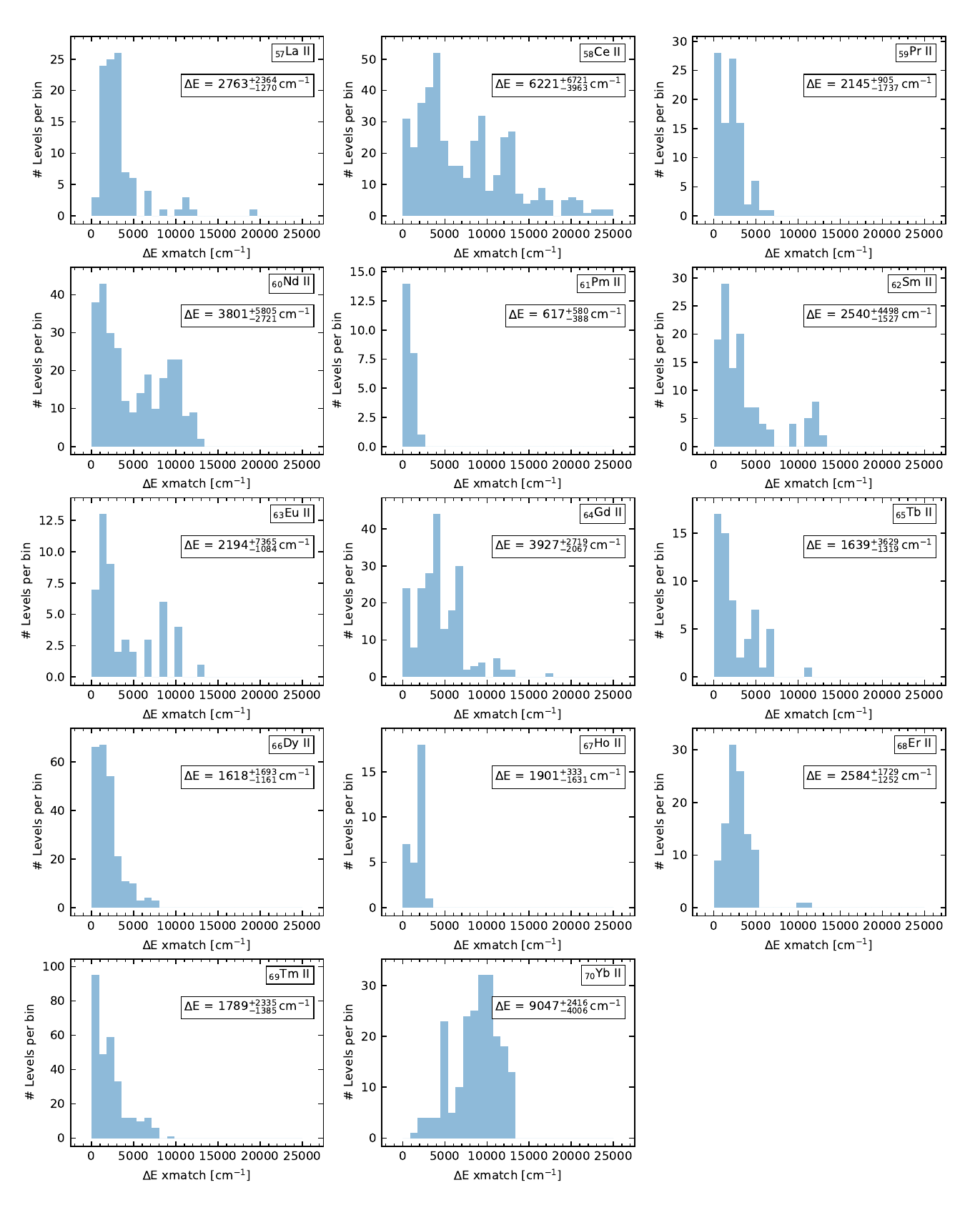}
	\caption{\justifying Deviations between theoretical and experimental energy levels for doubly ionized lanthanide ions. The figure includes only those levels for which a reliable cross-match could be established. Shown are the individual deviations, along with the median correction and the 68\% confidence interval of the applied calibration. 
	}
	\label{fig:xmatch_accuracy_singly}
\end{figure*}

\begin{figure*}
	\includegraphics[width=0.97\textwidth]{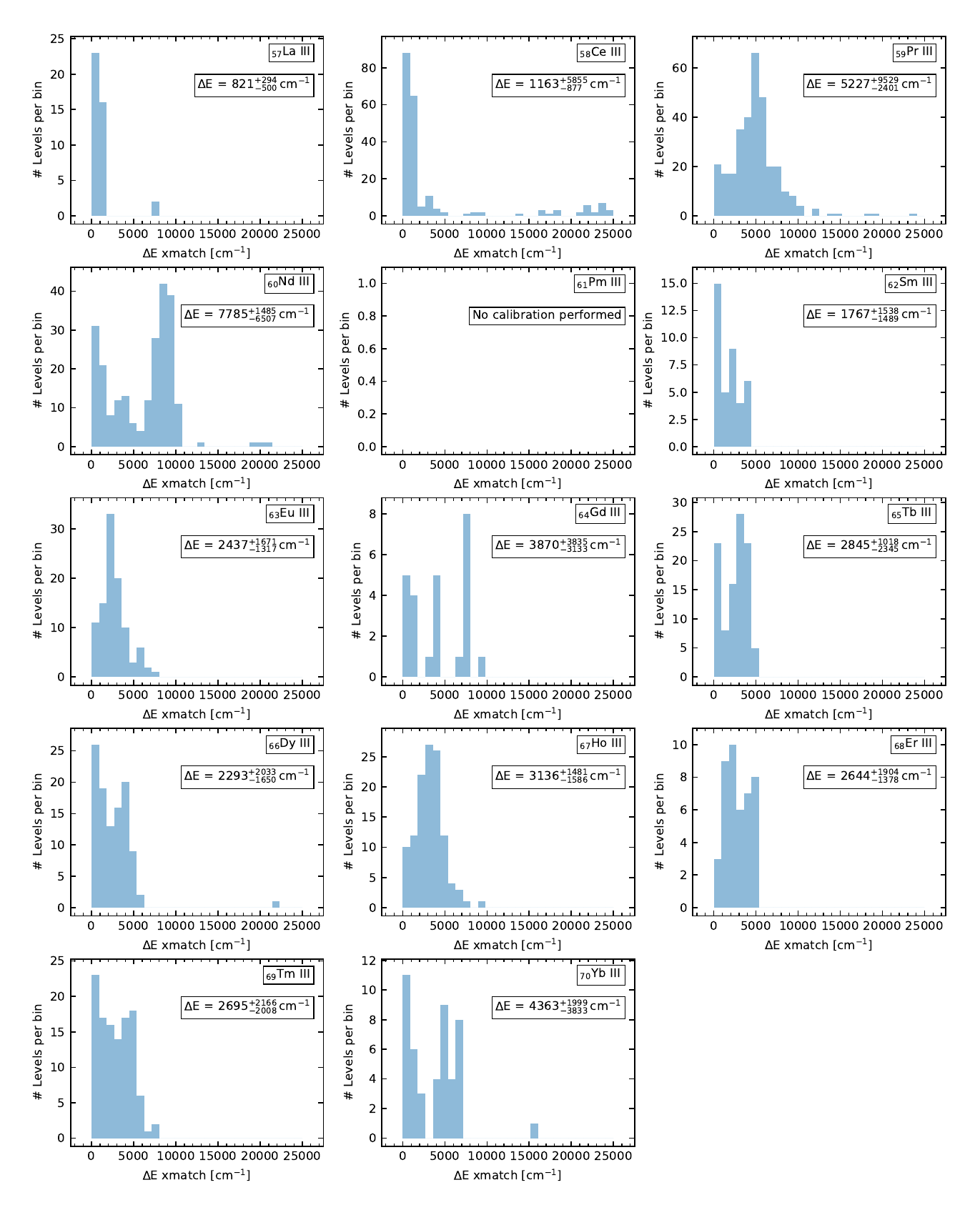}
	\caption{\justifying Deviations between theoretical and experimental energy levels for doubly ionized lanthanide ions. The figure includes only those levels for which a reliable cross-match could be established. Shown are the individual deviations, along with the median correction and the 68\% confidence interval of the applied calibration. No calibration was performed for $_{61}$Pm due to the absence of experimental data. For $_{60}$Nd, calibration results are presented using experimental values from \citet[][]{2024AA...684A.149D}.
	}
	\label{fig:xmatch_accuracy_doubly}
\end{figure*}

The use of mean configurations in atomic structure calculations can inadvertently introduce significant discrepancies between theoretical and experimental energy levels, as demonstrated in several recent studies \citep{2021PhRvA.103b2808L, 2022MNRAS.509.4723M, 2023MNRAS.524.3083F}. \citet{2023ApJ...954L..41S} emphasized that transition wavelengths, determined by energy differences between levels connected by selection rules, can differ from experimental values by factors of several orders of magnitude, even for relatively simple elements such as strontium, which is generally considered more tractable than the lanthanides or actinides. Their results also demonstrated that synthetic spectra generated using uncalibrated ab initio atomic data for Sr, Y, and Zr differ markedly from those based on experimental data. Owing to the complex interplay between atomic transitions, plasma conditions, and the radiation field in radiative transfer simulations, these discrepancies propagated beyond the specific ions in question, influencing the emergent spectrum produced by all species included in the model. As a result, the reliable identification of r-process material in observed spectra depends critically on the precision of transition wavelength accuracy that is difficult to attain through \textit{ab-initio} calculations alone.

To address this, we apply a calibration to final energy levels and corresponding transition wavelengths, based on available low-lying experimental energy levels. This calibration procedure leaves the underlying orbitals and atomic state function (ASF) compositions unchanged and instead modifies the transition energies and derived quantities, such as partition functions, through corrected excitation energies.

The calibration is performed by aligning computed levels with experimental energy levels from the NIST Atomic Spectra Database (ASD) \citep{NIST_ASD} and, where applicable, theoretical data from DREAM \citep[Database on Rare Earths at Mons University,][]{2020Atoms...8...18Q}. This procedure is not applicable to ions for which experimental data beyond the ground state are unavailable. In atomic spectroscopy, LS coupling is typically used to classify energy levels, whereas theoretical calculations may employ $jj$- or LS-coupling schemes. The \FAC\ code adopts $jj$-coupling by default. To facilitate consistent labeling, we utilize the \texttt{JJ2LSJ} code \citep{2017Atoms...5....6G}, which transforms the $jj$-coupled configuration state functions (CSFs) produced by \FAC\ into the LS-coupling scheme and assigns LS term labels to all calculated energy levels. This transformation is not bijective; multiple ASFs may share the same dominant LS label, particularly at higher excitation energies, where level mixing increases. We consider an LS label to be reliable when the squared leading LS composition exceeds 0.7.

The matching of theoretical and experimental levels is carried out within groups defined by parity ($\pi$) and total angular momentum ($J$), prioritizing those levels with leading LS-squared compositions above 0.7. For levels with leading LS contributions between 0.5 and 0.7, a scaled correction is applied. Levels that cannot be directly matched to experiment, due to the absence of data, are adjusted using average corrections derived from levels within the same $\pi$/J symmetry group and leading configuration. For states with LS contributions below 0.5, only the average $\pi$/J correction is used; the configuration-dependent correction increases linearly for leading contributions approaching 0.7. For LS-squared values at the threshold (0.7), the applied correction is computed as

\begin{equation}
	\Delta E = 0.5 \times \left( \overline{\Delta E_{\pi/J}} + \overline{\Delta E_{\mathrm{conf.},i}} \right),
\end{equation}
where $\overline{\Delta E_{\pi/J}}$ is the average correction for matched levels of the same parity and total angular momentum, and $\overline{\Delta E_{\mathrm{conf.},i}}$ is the average correction for levels with leading configuration $i$.

All levels are corrected individually, either by direct identification with experimental values or based on their symmetry and leading configuration. At present, no automated calibration procedure exists that simultaneously accounts for level spacing, dominant electron configurations, and quantum numbers with sufficient reliability. As a result, all calibrations were performed manually. Once energies are corrected, transition probabilities are recalculated accordingly. The published atomic data include metadata describing the energy calibration status for each transition, indicating whether both, one, or neither of the upper and lower levels were matched to experimental values. In cases where direct identification was not possible, calibrations were applied based on the symmetry properties of the levels. This information provides a quantitative basis for assessing the reliability of specific transitions in radiative transfer simulations and for evaluating the robustness of associated spectral features in model predictions.

The details of the correction methodology are retained to inform radiative transfer modeling and to facilitate uncertainty estimates on spectral features, depending on whether both levels are matched, only one is matched, or both are adjusted based on their symmetry characteristics.

An overview of the magnitude of energy corrections applied to singly and doubly ionized lanthanide species is provided in figures~\ref{fig:xmatch_accuracy_singly} and \ref{fig:xmatch_accuracy_doubly}, respectively. Each figure displays the median correction along with the corresponding 34\,\% intervals above and below the median. It is important to note that the majority of the energy correction distributions exhibit significant departures from Gaussianity, usually in the form of extended tails to higher energies.

\section{Atomic Structure Calculations}
\label{sec:calculations}

\begin{table*}
	\caption{\justifying Configurations included in the \FAC\  calculations for lanthanide ions. The presented configurations are sorted by energy. The FMC weights used for the radial optimization of the potential are given for each ion. Configurations are displayed in alternating black and blue to improve visual clarity and facilitate distinction between successive entries.}
	\label{tab:FAC_configs1}
    \begin{ruledtabular}
	\begin{tabular}{@{}cccl@{}}
		Ion & FMC weights & \#\footnote{Number of included configurations} & Configurations \\ 
		\midrule
\multirow{1}{*}{\ion{La}{ii}} &
\multirow{1}{*}{4\textit{f}$^{1.62}$ 5\textit{d}$^{\,0.38}$} &
111  &
\parbox[t]{0.70\textwidth}{\justifying\noindent
\conf{5}{d}{2}, \conb{5}{d}{1}\,\conb{6}{s}{1}, \conf{6}{s}{2}, \conb{4}{f}{1}\,\conb{6}{s}{1}, \conf{4}{f}{1}\,\conf{5}{d}{1}, \conb{5}{d}{1}\,\conb{6}{p}{1}, \conf{6}{s}{1}\,\conf{6}{p}{1}, \conb{4}{f}{1}\,\conb{6}{p}{1}, \conf{5}{d}{1}\,\conf{7}{s}{1}, \conb{5}{d}{1}\,\conb{6}{d}{1}, \conf{4}{f}{1}\,\conf{7}{s}{1}, \conb{5}{d}{1}\,\conb{7}{p}{1}, \conf{4}{f}{1}\,\conf{6}{d}{1}, \conb{6}{p}{2}, \conf{5}{d}{1}\,\conf{5}{f}{1}, \conb{6}{s}{1}\,\conb{7}{s}{1}, \conf{4}{f}{1}\,\conf{7}{p}{1}, \conb{4}{f}{2}, \conf{4}{f}{1}\,\conf{5}{f}{1}, \conb{6}{s}{1}\,\conb{6}{d}{1}, \conf{5}{d}{1}\,\conf{8}{s}{1}, \conb{5}{d}{1}\,\conb{7}{d}{1}, \conf{6}{s}{1}\,\conf{7}{p}{1}, \conb{5}{d}{1}\,\conb{8}{p}{1}, \conf{5}{d}{1}\,\conf{6}{f}{1}, \conb{5}{d}{1}\,\conb{5}{g}{1}, \conf{4}{f}{1}\,\conf{8}{s}{1}, \conb{4}{f}{1}\,\conb{7}{d}{1}, \conf{5}{f}{1}\,\conf{6}{s}{1}, \conb{4}{f}{1}\,\conb{8}{p}{1}, \conf{4}{f}{1}\,\conf{6}{f}{1}, \conb{5}{d}{1}\,\conb{8}{d}{1}, \conf{4}{f}{1}\,\conf{5}{g}{1}, \conb{6}{s}{1}\,\conb{8}{s}{1}, \conf{4}{f}{1}\,\conf{8}{d}{1}, \conb{5}{d}{1}\,\conb{9}{s}{1}, \conf{6}{s}{1}\,\conf{7}{d}{1}, \conb{5}{d}{1}\,\conb{9}{p}{1}, \conf{6}{s}{1}\,\conf{8}{p}{1}, \conb{6}{s}{1}\,\conb{6}{f}{1}, \conf{5}{g}{1}\,\conf{6}{s}{1}, \conb{5}{d}{1}\,\conb{9}{d}{1}, \conf{6}{s}{1}\,\conf{8}{d}{1}, \conb{6}{p}{1}\,\conb{7}{s}{1}, \conf{6}{s}{1}\,\conf{9}{s}{1}, \conb{6}{s}{1}\,\conb{9}{p}{1}, \conf{6}{p}{1}\,\conf{6}{d}{1}, \conb{6}{s}{1}\,\conb{9}{d}{1}, \conf{6}{p}{1}\,\conf{7}{p}{1}, \conb{5}{f}{1}\,\conb{6}{p}{1}, \conf{6}{p}{1}\,\conf{8}{s}{1}, \conb{6}{p}{1}\,\conb{7}{d}{1}, \conf{6}{p}{1}\,\conf{8}{p}{1}, \conb{6}{p}{1}\,\conb{6}{f}{1}, \conf{5}{g}{1}\,\conf{6}{p}{1}, \conb{6}{p}{1}\,\conb{8}{d}{1}, \conf{7}{s}{2}, \conb{6}{d}{1}\,\conb{7}{s}{1}, \conf{6}{d}{2}, \conb{7}{s}{1}\,\conb{7}{p}{1}, \conf{6}{d}{1}\,\conf{7}{p}{1}, \conb{5}{f}{1}\,\conb{7}{s}{1}, \conf{5}{f}{1}\,\conf{6}{d}{1}, \conb{7}{p}{2}, \conf{5}{f}{1}\,\conf{7}{p}{1}, \conb{7}{s}{1}\,\conb{8}{s}{1}, \conf{6}{d}{1}\,\conf{8}{s}{1}, \conb{7}{s}{1}\,\conb{7}{d}{1}, \conf{5}{f}{2}, \conb{6}{d}{1}\,\conb{7}{d}{1}, \conf{7}{s}{1}\,\conf{8}{p}{1}, \conb{6}{d}{1}\,\conb{8}{p}{1}, \conf{6}{f}{1}\,\conf{7}{s}{1}, \conb{5}{g}{1}\,\conb{7}{s}{1}, \conf{6}{d}{1}\,\conf{6}{f}{1}, \conb{5}{g}{1}\,\conb{6}{d}{1}, \conf{7}{s}{1}\,\conf{8}{d}{1}, \conb{7}{p}{1}\,\conb{8}{s}{1}, \conf{6}{d}{1}\,\conf{8}{d}{1}, \conb{7}{p}{1}\,\conb{7}{d}{1}, \conf{5}{f}{1}\,\conf{8}{s}{1}, \conb{5}{f}{1}\,\conb{7}{d}{1}, \conf{7}{p}{1}\,\conf{8}{p}{1}, \conb{5}{f}{1}\,\conb{8}{p}{1}, \conf{5}{g}{1}\,\conf{7}{p}{1}, \conb{5}{f}{1}\,\conb{5}{g}{1}, \conf{6}{f}{1}\,\conf{7}{p}{1}, \conb{5}{f}{1}\,\conb{6}{f}{1}, \conf{7}{p}{1}\,\conf{8}{d}{1}, \conb{5}{f}{1}\,\conb{8}{d}{1}, \conf{8}{s}{2}, \conb{7}{d}{1}\,\conb{8}{s}{1}, \conf{7}{d}{2}, \conb{8}{s}{1}\,\conb{8}{p}{1}, \conf{7}{d}{1}\,\conf{8}{p}{1}, \conb{5}{g}{1}\,\conb{8}{s}{1}, \conf{6}{f}{1}\,\conf{7}{d}{1}, \conb{5}{g}{1}\,\conb{7}{d}{1}, \conf{6}{f}{1}\,\conf{8}{s}{1}, \conb{5}{g}{1}\,\conb{8}{p}{1}, \conf{5}{g}{1}\,\conf{6}{f}{1}, \conb{8}{p}{2}, \conf{5}{g}{2}, \conb{6}{f}{1}\,\conb{8}{p}{1}, \conf{6}{f}{2}, \conb{8}{s}{1}\,\conb{8}{d}{1}, \conf{7}{d}{1}\,\conf{8}{d}{1}, \conb{5}{g}{1}\,\conb{8}{d}{1}, \conf{8}{p}{1}\,\conf{8}{d}{1}, \conb{6}{f}{1}\,\conb{8}{d}{1}, \conf{8}{d}{2}
}\\
\midrule

\multirow{1}{*}{\ion{La}{iii}} &
\multirow{1}{*}{4\textit{f}$^{1.00}$} &
23  &
\parbox[t]{0.70\textwidth}{\justifying\noindent
\conf{5}{d}{1}, \conb{6}{s}{1}, \conf{4}{f}{1}, \conb{6}{p}{1}, \conf{7}{s}{1}, \conb{6}{d}{1}, \conf{7}{p}{1}, \conb{5}{f}{1}, \conf{8}{s}{1}, \conb{7}{d}{1}, \conf{5}{g}{1}, \conb{6}{f}{1}, \conf{8}{p}{1}, \conb{9}{s}{1}, \conf{8}{d}{1}, \conb{6}{g}{1}, \conf{7}{f}{1}, \conb{9}{p}{1}, \conf{10}{s}{1}, \conb{9}{d}{1}, \conf{7}{g}{1}, \conb{8}{f}{1}, \conf{8}{g}{1}
}\\
\midrule

\multirow{1}{*}{\ion{Ce}{ii}} &
\multirow{1}{*}{4\textit{f}$^{1.86}$ 5\textit{d}$^{\,0.96}$  6\textit{s}$^{0.10}$  6\textit{p}$^{0.08}$} &
277  &
\parbox[t]{0.70\textwidth}{\justifying\noindent
\conf{4}{f}{1}\,\conf{5}{d}{2}, \conb{4}{f}{1}\,\conb{5}{d}{1}\,\conb{6}{s}{1}, \conf{4}{f}{2}\,\conf{6}{s}{1}, \conb{4}{f}{1}, \conf{4}{f}{2}\,\conf{5}{d}{1}, \conb{4}{f}{1}\,\conb{5}{d}{1}\,\conb{6}{p}{1}, \conf{4}{f}{2}\,\conf{6}{p}{1}, \conb{4}{f}{1}\,\conb{6}{s}{1}\,\conb{6}{p}{1}, \conf{4}{f}{1}\,\conf{5}{d}{1}\,\conf{7}{s}{1}, \conb{5}{d}{3}, \conf{4}{f}{1}\,\conf{5}{d}{1}\,\conf{6}{d}{1}, \conb{4}{f}{2}\,\conb{7}{s}{1}, \conf{5}{d}{2}\,\conf{6}{s}{1}, \conb{4}{f}{2}\,\conb{6}{d}{1}, \conf{4}{f}{1}\,\conf{5}{d}{1}\,\conf{7}{p}{1}, \conb{4}{f}{2}\,\conb{7}{p}{1}, \conf{4}{f}{2}\,\conf{5}{f}{1}, \conb{4}{f}{1}\,\conb{5}{d}{1}\,\conb{5}{f}{1}, \conf{4}{f}{1}\,\conf{6}{s}{1}\,\conf{7}{s}{1}, \conb{4}{f}{1}\,\conb{6}{p}{2}, \conf{4}{f}{1}\,\conf{5}{d}{1}\,\conf{8}{s}{1}, \conb{4}{f}{1}\,\conb{5}{d}{1}\,\conb{7}{d}{1}, \conf{4}{f}{2}\,\conf{8}{s}{1}, \conb{4}{f}{1}\,\conb{6}{s}{1}\,\conb{6}{d}{1}, \conf{4}{f}{1}\,\conf{5}{d}{1}\,\conf{8}{p}{1}, \conb{4}{f}{2}\,\conb{7}{d}{1}, \conf{4}{f}{2}\,\conf{8}{p}{1}, \conb{4}{f}{1}\,\conb{5}{d}{1}\,\conb{6}{f}{1}, \conf{4}{f}{2}\,\conf{6}{f}{1}, \conb{4}{f}{1}\,\conb{5}{d}{1}\,\conb{5}{g}{1}, \conf{4}{f}{2}\,\conf{5}{g}{1}, \conb{4}{f}{1}\,\conb{6}{s}{1}\,\conb{7}{p}{1}, \conf{4}{f}{1}\,\conf{5}{d}{1}\,\conf{8}{d}{1}, \conb{4}{f}{2}\,\conb{8}{d}{1}, \conf{4}{f}{1}\,\conf{5}{f}{1}\,\conf{6}{s}{1}, \conb{4}{f}{1}\,\conb{5}{d}{1}\,\conb{9}{s}{1}, \conf{5}{d}{2}\,\conf{6}{p}{1}, \conb{4}{f}{1}\,\conb{5}{d}{1}\,\conb{9}{p}{1}, \conf{4}{f}{1}\,\conf{6}{s}{1}\,\conf{8}{s}{1}, \conb{4}{f}{1}\,\conb{6}{s}{1}\,\conb{7}{d}{1}, \conf{4}{f}{1}\,\conf{5}{d}{1}\,\conf{9}{d}{1}, \conb{4}{f}{1}\,\conb{6}{s}{1}\,\conb{8}{p}{1}, \conf{5}{d}{1}, \conb{4}{f}{1}\,\conb{6}{s}{1}\,\conb{6}{f}{1}, \conf{4}{f}{1}\,\conf{5}{g}{1}\,\conf{6}{s}{1}, \conb{4}{f}{1}\,\conb{6}{s}{1}\,\conb{8}{d}{1}, \conf{4}{f}{1}\,\conf{6}{p}{1}\,\conf{7}{s}{1}, \conb{4}{f}{1}\,\conb{6}{s}{1}\,\conb{9}{s}{1}, \conf{4}{f}{1}\,\conf{6}{p}{1}\,\conf{6}{d}{1}, \conb{4}{f}{1}\,\conb{6}{s}{1}\,\conb{9}{p}{1}, \conf{5}{d}{1}\,\conf{6}{s}{1}\,\conf{6}{p}{1}, \conb{4}{f}{1}\,\conb{6}{s}{1}\,\conb{9}{d}{1}, \conf{4}{f}{1}\,\conf{6}{p}{1}\,\conf{7}{p}{1}, \conb{4}{f}{3}, \conf{4}{f}{1}\,\conf{5}{f}{1}\,\conf{6}{p}{1}, \conb{5}{d}{2}\,\conb{7}{s}{1}, \conf{4}{f}{1}\,\conf{6}{p}{1}\,\conf{8}{s}{1}, \conb{5}{d}{2}\,\conb{6}{d}{1}, \conf{4}{f}{1}\,\conf{6}{p}{1}\,\conf{7}{d}{1}, \conb{4}{f}{1}\,\conb{6}{p}{1}\,\conb{8}{p}{1}, \conf{4}{f}{1}\,\conf{6}{p}{1}\,\conf{6}{f}{1}, \conb{4}{f}{1}\,\conb{5}{g}{1}\,\conb{6}{p}{1}, \conf{5}{d}{2}\,\conf{7}{p}{1}, \conb{4}{f}{1}\,\conb{6}{p}{1}\,\conb{8}{d}{1}, \conf{5}{d}{2}\,\conf{5}{f}{1}, \conb{5}{d}{2}\,\conb{8}{s}{1}, \conf{5}{d}{2}\,\conf{7}{d}{1}, \conb{4}{f}{1}\,\conb{6}{d}{1}\,\conb{7}{s}{1}, \conf{5}{d}{2}\,\conf{8}{p}{1}, \conb{4}{f}{1}\,\conb{6}{d}{2}, \conf{5}{d}{2}\,\conf{6}{f}{1}, \conb{5}{d}{1}\,\conb{6}{s}{1}\,\conb{7}{s}{1}, \conf{5}{d}{2}\,\conf{5}{g}{1}, \conb{6}{p}{1}, \conf{5}{d}{1}\,\conf{6}{p}{2}, \conb{4}{f}{1}\,\conb{7}{s}{1}\,\conb{7}{p}{1}, \conf{5}{d}{1}\,\conf{6}{s}{1}\,\conf{6}{d}{1}, \conb{5}{d}{2}\,\conb{8}{d}{1}, \conf{4}{f}{1}\,\conf{6}{d}{1}\,\conf{7}{p}{1}, \conb{4}{f}{1}\,\conb{5}{f}{1}\,\conb{7}{s}{1}, \conf{4}{f}{1}\,\conf{5}{f}{1}\,\conf{6}{d}{1}, \conb{5}{d}{1}\,\conb{6}{s}{1}\,\conb{7}{p}{1}, \conf{5}{d}{1}\,\conf{5}{f}{1}\,\conf{6}{s}{1}, \conb{4}{f}{1}\,\conb{7}{p}{2}, \conf{4}{f}{1}\,\conf{5}{f}{1}\,\conf{7}{p}{1}, \conb{4}{f}{1}\,\conb{7}{s}{1}\,\conb{8}{s}{1}, \conf{4}{f}{1}\,\conf{5}{f}{2}, \conb{4}{f}{1}\,\conb{6}{d}{1}\,\conb{8}{s}{1}, \conf{4}{f}{1}\,\conf{7}{s}{1}\,\conf{7}{d}{1}, \conb{5}{d}{1}\,\conb{6}{s}{1}\,\conb{8}{s}{1}, \conf{4}{f}{1}\,\conf{6}{d}{1}\,\conf{7}{d}{1}, \conb{6}{s}{1}\,\conb{6}{p}{2}, \conf{4}{f}{1}\,\conf{7}{s}{1}\,\conf{8}{p}{1}, \conb{4}{f}{1}\,\conb{6}{d}{1}\,\conb{8}{p}{1}, \conf{5}{d}{1}\,\conf{6}{s}{1}\,\conf{7}{d}{1}, \conb{4}{f}{1}\,\conb{5}{g}{1}\,\conb{7}{s}{1}, \conf{4}{f}{1}\,\conf{6}{f}{1}\,\conf{7}{s}{1}, \conb{5}{d}{1}\,\conb{6}{s}{1}\,\conb{8}{p}{1}, \conf{4}{f}{1}\,\conf{6}{d}{1}\,\conf{6}{f}{1}, \conb{4}{f}{1}\,\conb{5}{g}{1}\,\conb{6}{d}{1}, \conf{5}{d}{1}\,\conf{6}{s}{1}\,\conf{6}{f}{1}, \conb{5}{d}{1}\,\conb{5}{g}{1}\,\conb{6}{s}{1}, \conf{4}{f}{1}\,\conf{7}{s}{1}\,\conf{8}{d}{1}, \conb{4}{f}{1}\,\conb{7}{p}{1}\,\conb{8}{s}{1}, \conf{4}{f}{1}\,\conf{5}{f}{1}\,\conf{8}{s}{1}, \conb{4}{f}{1}\,\conb{6}{d}{1}\,\conb{8}{d}{1}, \conf{5}{d}{1}\,\conf{6}{s}{1}\,\conf{8}{d}{1}, \conb{4}{f}{1}\,\conb{7}{p}{1}\,\conb{7}{d}{1}, \conf{4}{f}{1}\,\conf{5}{f}{1}\,\conf{7}{d}{1}, \conb{4}{f}{1}\,\conb{5}{f}{1}\,\conb{8}{p}{1}, \conf{4}{f}{1}\,\conf{7}{p}{1}\,\conf{8}{p}{1}, \conb{4}{f}{1}\,\conb{5}{f}{1}\,\conb{5}{g}{1}, \conf{5}{d}{1}\,\conf{6}{p}{1}\,\conf{6}{d}{1}, \conb{4}{f}{1}\,\conb{5}{f}{1}\,\conb{6}{f}{1}, \conf{4}{f}{1}\,\conf{5}{g}{1}\,\conf{7}{p}{1}, \conb{4}{f}{1}\,\conb{6}{f}{1}\,\conb{7}{p}{1}, \conf{6}{d}{1}, \conb{7}{s}{1}, \conf{4}{f}{1}\,\conf{5}{f}{1}\,\conf{8}{d}{1}, \conb{4}{f}{1}\,\conb{7}{p}{1}\,\conb{8}{d}{1}, \conf{5}{d}{1}\,\conf{6}{p}{1}\,\conf{7}{s}{1}, \conb{5}{d}{1}\,\conb{6}{s}{1}\,\conb{9}{s}{1}, \conf{5}{d}{1}\,\conf{6}{s}{1}\,\conf{9}{p}{1}, \conb{4}{f}{1}\,\conb{7}{d}{1}\,\conb{8}{s}{1}, \conf{7}{p}{1}, \conb{4}{f}{1}\,\conb{7}{d}{2}, \conf{5}{d}{1}\,\conf{6}{s}{1}\,\conf{9}{d}{1}, \conb{4}{f}{1}\,\conb{8}{s}{1}\,\conb{8}{p}{1}, \conf{4}{f}{1}\,\conf{7}{d}{1}\,\conf{8}{p}{1}, \conb{4}{f}{1}\,\conb{6}{f}{1}\,\conb{8}{s}{1}, \conf{5}{d}{1}\,\conf{6}{p}{1}\,\conf{7}{p}{1}, \conb{4}{f}{1}\,\conb{5}{g}{1}\,\conb{8}{s}{1}, \conf{4}{f}{1}\,\conf{6}{f}{1}\,\conf{7}{d}{1}, \conb{4}{f}{1}\,\conb{5}{g}{1}\,\conb{7}{d}{1}, \conf{5}{f}{1}, \conb{4}{f}{1}\,\conb{6}{f}{1}\,\conb{8}{p}{1}, \conf{4}{f}{1}\,\conf{5}{g}{1}\,\conf{6}{f}{1}, \conb{4}{f}{1}\,\conb{5}{g}{1}\,\conb{8}{p}{1}, \conf{5}{d}{1}\,\conf{5}{f}{1}\,\conf{6}{p}{1}, \conb{4}{f}{1}\,\conb{8}{p}{2}, \conf{4}{f}{1}\,\conf{5}{g}{2}, \conb{4}{f}{1}\,\conb{6}{f}{2}, \conf{4}{f}{1}\,\conf{8}{s}{1}\,\conf{8}{d}{1}, \conb{4}{f}{1}\,\conb{7}{d}{1}\,\conb{8}{d}{1}, \conf{8}{s}{1}, \conb{4}{f}{1}\,\conb{8}{d}{2}, \conf{7}{d}{1}, \conb{5}{d}{1}\,\conb{6}{p}{1}\,\conb{8}{s}{1}, \conf{8}{p}{1}, \conb{4}{f}{1}\,\conb{5}{g}{1}\,\conb{8}{d}{1}, \conf{6}{s}{1}\,\conf{6}{p}{1}\,\conf{7}{s}{1}, \conb{5}{d}{1}\,\conb{6}{p}{1}\,\conb{7}{d}{1}, \conf{4}{f}{1}\,\conf{8}{p}{1}\,\conf{8}{d}{1}, \conb{4}{f}{1}\,\conb{6}{f}{1}\,\conb{8}{d}{1}, \conf{6}{f}{1}, \conb{5}{g}{1}, \conf{5}{d}{1}\,\conf{6}{p}{1}\,\conf{8}{p}{1}, \conb{5}{d}{1}\,\conb{6}{p}{1}\,\conb{6}{f}{1}, \conf{8}{d}{1}, \conb{5}{d}{1}\,\conb{5}{g}{1}\,\conb{6}{p}{1}, \conf{6}{s}{1}\,\conf{6}{p}{1}\,\conf{6}{d}{1}, \conb{5}{d}{1}\,\conb{6}{p}{1}\,\conb{8}{d}{1}, \conf{6}{s}{1}\,\conf{6}{p}{1}\,\conf{7}{p}{1}, \conb{5}{d}{1}\,\conb{6}{d}{1}\,\conb{7}{s}{1}, \conf{5}{d}{1}\,\conf{6}{d}{2}, \conb{5}{f}{1}\,\conb{6}{s}{1}\,\conb{6}{p}{1}, \conf{5}{d}{1}\,\conf{6}{d}{1}\,\conf{7}{p}{1}, \conb{6}{s}{1}\,\conb{6}{p}{1}\,\conb{8}{s}{1}, \conf{5}{d}{1}\,\conf{5}{f}{1}\,\conf{6}{d}{1}, \conb{6}{s}{1}\,\conb{6}{p}{1}\,\conb{7}{d}{1}, \conf{6}{s}{1}\,\conf{6}{p}{1}\,\conf{8}{p}{1}, \conb{6}{s}{1}\,\conb{6}{p}{1}\,\conb{6}{f}{1}, \conf{5}{d}{1}\,\conf{7}{s}{1}\,\conf{7}{p}{1}, \conb{5}{g}{1}\,\conb{6}{s}{1}\,\conb{6}{p}{1}, \conf{6}{s}{1}\,\conf{6}{p}{1}\,\conf{8}{d}{1}, \conb{5}{d}{1}\,\conb{5}{f}{1}\,\conb{7}{s}{1}, \conf{5}{d}{1}\,\conf{6}{d}{1}\,\conf{8}{s}{1}, \conb{5}{d}{1}\,\conb{6}{d}{1}\,\conb{7}{d}{1}, \conf{5}{d}{1}\,\conf{6}{d}{1}\,\conf{8}{p}{1}, \conb{5}{d}{1}\,\conb{6}{d}{1}\,\conb{6}{f}{1}, \conf{5}{d}{1}\,\conf{5}{g}{1}\,\conf{6}{d}{1}, \conb{5}{d}{1}\,\conb{7}{p}{2}, \conf{5}{d}{1}\,\conf{7}{s}{1}\,\conf{8}{s}{1}, \conb{5}{d}{1}\,\conb{7}{s}{1}\,\conb{7}{d}{1}, \conf{5}{d}{1}\,\conf{5}{f}{1}\,\conf{7}{p}{1}, \conb{5}{d}{1}\,\conb{6}{d}{1}\,\conb{8}{d}{1}, \conf{5}{d}{1}\,\conf{7}{s}{1}\,\conf{8}{p}{1}, \conb{6}{s}{1}, \conf{6}{s}{1}\,\conf{6}{d}{1}\,\conf{7}{s}{1}, \conb{5}{d}{1}\,\conb{5}{f}{2}, \conf{5}{d}{1}\,\conf{6}{f}{1}\,\conf{7}{s}{1}, \conb{5}{d}{1}\,\conb{5}{g}{1}\,\conb{7}{s}{1}, \conf{5}{d}{1}\,\conf{7}{s}{1}\,\conf{8}{d}{1}, \conb{6}{s}{1}\,\conb{7}{s}{1}\,\conb{7}{p}{1}, \conf{5}{d}{1}\,\conf{7}{p}{1}\,\conf{8}{s}{1}, \conb{5}{d}{1}\,\conb{5}{f}{1}\,\conb{8}{s}{1}, \conf{5}{d}{1}\,\conf{7}{p}{1}\,\conf{7}{d}{1}, \conb{5}{d}{1}\,\conb{5}{f}{1}\,\conb{7}{d}{1}, \conf{5}{f}{1}\,\conf{6}{s}{1}\,\conf{7}{s}{1}, \conb{6}{s}{1}\,\conb{6}{d}{2}, \conf{5}{d}{1}\,\conf{5}{f}{1}\,\conf{8}{p}{1}, \conb{5}{d}{1}\,\conb{7}{p}{1}\,\conb{8}{p}{1}, \conf{5}{d}{1}\,\conf{6}{f}{1}\,\conf{7}{p}{1}, \conb{5}{d}{1}\,\conb{5}{g}{1}\,\conb{7}{p}{1}, \conf{5}{d}{1}\,\conf{5}{f}{1}\,\conf{5}{g}{1}, \conb{5}{d}{1}\,\conb{5}{f}{1}\,\conb{6}{f}{1}, \conf{6}{s}{1}\,\conf{6}{d}{1}\,\conf{7}{p}{1}, \conb{5}{f}{1}\,\conb{6}{s}{1}\,\conb{6}{d}{1}, \conf{5}{d}{1}\,\conf{7}{p}{1}\,\conf{8}{d}{1}, \conb{5}{d}{1}\,\conb{5}{f}{1}\,\conb{8}{d}{1}, \conf{6}{s}{1}\,\conf{7}{s}{1}\,\conf{8}{s}{1}, \conb{6}{s}{1}\,\conb{7}{s}{1}\,\conb{7}{d}{1}, \conf{5}{d}{1}\,\conf{7}{d}{1}\,\conf{8}{s}{1}, \conb{6}{s}{1}\,\conb{7}{p}{2}, \conf{6}{s}{1}\,\conf{7}{s}{1}\,\conf{8}{p}{1}, \conb{5}{f}{1}\,\conb{6}{s}{1}\,\conb{7}{p}{1}, \conf{5}{d}{1}\,\conf{7}{d}{2}, \conb{5}{g}{1}\,\conb{6}{s}{1}\,\conb{7}{s}{1}, \conf{6}{s}{1}\,\conf{6}{f}{1}\,\conf{7}{s}{1}, \conb{5}{d}{1}\,\conb{8}{s}{1}\,\conb{8}{p}{1}, \conf{5}{d}{1}\,\conf{7}{d}{1}\,\conf{8}{p}{1}, \conb{6}{s}{1}\,\conb{6}{d}{1}\,\conb{8}{s}{1}, \conf{5}{d}{1}\,\conf{6}{f}{1}\,\conf{8}{s}{1}, \conb{5}{d}{1}\,\conb{5}{g}{1}\,\conb{8}{s}{1}, \conf{5}{d}{1}\,\conf{6}{f}{1}\,\conf{7}{d}{1}, \conb{5}{f}{2}\,\conb{6}{s}{1}, \conf{5}{d}{1}\,\conf{5}{g}{1}\,\conf{7}{d}{1}, \conb{6}{s}{1}\,\conb{6}{d}{1}\,\conb{7}{d}{1}, \conf{6}{s}{1}\,\conf{7}{s}{1}\,\conf{8}{d}{1}, \conb{5}{d}{1}\,\conb{5}{g}{1}\,\conb{8}{p}{1}, \conf{6}{s}{1}\,\conf{6}{d}{1}\,\conf{8}{p}{1}, \conb{5}{d}{1}\,\conb{8}{p}{2}, \conf{5}{d}{1}\,\conf{5}{g}{1}\,\conf{6}{f}{1}, \conb{5}{d}{1}\,\conb{5}{g}{2}, \conf{6}{s}{1}\,\conf{6}{d}{1}\,\conf{6}{f}{1}, \conb{5}{d}{1}\,\conb{6}{f}{2}, \conf{5}{d}{1}\,\conf{6}{f}{1}\,\conf{8}{p}{1}, \conb{5}{d}{1}\,\conb{8}{s}{1}\,\conb{8}{d}{1}, \conf{5}{d}{1}\,\conf{7}{d}{1}\,\conf{8}{d}{1}, \conb{5}{g}{1}\,\conb{6}{s}{1}\,\conb{6}{d}{1}, \conf{6}{s}{1}\,\conf{7}{p}{1}\,\conf{8}{s}{1}, \conb{6}{s}{1}\,\conb{7}{p}{1}\,\conb{7}{d}{1}, \conf{5}{f}{1}\,\conf{6}{s}{1}\,\conf{8}{s}{1}, \conb{6}{s}{1}\,\conb{6}{d}{1}\,\conb{8}{d}{1}, \conf{5}{f}{1}\,\conf{6}{s}{1}\,\conf{7}{d}{1}, \conb{5}{d}{1}\,\conb{5}{g}{1}\,\conb{8}{d}{1}, \conf{5}{d}{1}\,\conf{8}{p}{1}\,\conf{8}{d}{1}, \conb{5}{f}{1}\,\conb{5}{g}{1}\,\conb{6}{s}{1}, \conf{5}{d}{1}\,\conf{6}{f}{1}\,\conf{8}{d}{1}, \conb{6}{s}{1}\,\conb{7}{p}{1}\,\conb{8}{p}{1}, \conf{5}{f}{1}\,\conf{6}{s}{1}\,\conf{8}{p}{1}, \conb{6}{s}{1}\,\conb{6}{f}{1}\,\conb{7}{p}{1}, \conf{5}{g}{1}\,\conf{6}{s}{1}\,\conf{7}{p}{1}, \conb{5}{f}{1}\,\conb{6}{s}{1}\,\conb{6}{f}{1}, \conf{5}{d}{1}\,\conf{8}{d}{2}, \conb{6}{s}{1}\,\conb{7}{p}{1}\,\conb{8}{d}{1}, \conf{5}{f}{1}\,\conf{6}{s}{1}\,\conf{8}{d}{1}, \conb{6}{s}{1}\,\conb{7}{d}{1}\,\conb{8}{s}{1}, \conf{6}{s}{1}\,\conf{7}{d}{2}, \conb{6}{s}{1}\,\conb{8}{s}{1}\,\conb{8}{p}{1}, \conf{6}{s}{1}\,\conf{7}{d}{1}\,\conf{8}{p}{1}, \conb{5}{g}{1}\,\conb{6}{s}{1}\,\conb{8}{s}{1}, \conf{6}{s}{1}\,\conf{6}{f}{1}\,\conf{8}{s}{1}, \conb{6}{s}{1}\,\conb{6}{f}{1}\,\conb{7}{d}{1}, \conf{5}{g}{1}\,\conf{6}{s}{1}\,\conf{7}{d}{1}, \conb{5}{g}{1}\,\conb{6}{s}{1}\,\conb{8}{p}{1}, \conf{6}{s}{1}\,\conf{8}{p}{2}, \conb{5}{g}{1}\,\conb{6}{s}{1}\,\conb{6}{f}{1}, \conf{6}{s}{1}\,\conf{6}{f}{2}, \conb{5}{g}{2}\,\conb{6}{s}{1}, \conf{6}{s}{1}\,\conf{6}{f}{1}\,\conf{8}{p}{1}, \conb{6}{s}{1}\,\conb{8}{s}{1}\,\conb{8}{d}{1}, \conf{6}{s}{1}\,\conf{7}{d}{1}\,\conf{8}{d}{1}, \conb{5}{g}{1}\,\conb{6}{s}{1}\,\conb{8}{d}{1}, \conf{6}{s}{1}\,\conf{8}{p}{1}\,\conf{8}{d}{1}, \conb{6}{s}{1}\,\conb{6}{f}{1}\,\conb{8}{d}{1}, \conf{6}{s}{1}\,\conf{8}{d}{2}
}\\
	\end{tabular}
    \end{ruledtabular}
\end{table*}
\begin{table*}
	\caption*{\justifying \textsc{TABLE I CONTINUED:} Configurations included in the \FAC\  calculations for lanthanide ions. The presented configurations are sorted by energy. The FMC weights used for the radial optimization of the potential are given for each ion. Configurations are displayed in alternating black and blue to improve visual clarity and facilitate distinction between successive entries.}
	\label{tab:FAC_configs2}
    \begin{ruledtabular}
	\begin{tabular}{@{}cccl@{}}
		Ion & FMC weights & \#\footnote{Number of included configurations} & Configurations \\ 
		\midrule
        \multirow{1}{*}{\ion{Ce}{iii}} &
        \multirow{1}{*}{4\textit{f}$^{2.00}$} &
        111  &
        \parbox[t]{0.70\textwidth}{\justifying\noindent
        \conf{4}{f}{2}, \conb{4}{f}{1}\,\conb{5}{d}{1}, \conf{4}{f}{1}\,\conf{6}{s}{1}, \conb{4}{f}{1}\,\conb{6}{p}{1}, \conf{5}{d}{2}, \conb{5}{d}{1}\,\conb{6}{s}{1}, \conf{4}{f}{1}\,\conf{7}{s}{1}, \conb{4}{f}{1}\,\conb{6}{d}{1}, \conf{4}{f}{1}\,\conf{5}{f}{1}, \conb{4}{f}{1}\,\conb{7}{p}{1}, \conf{5}{d}{1}\,\conf{6}{p}{1}, \conb{4}{f}{1}\,\conb{7}{d}{1}, \conf{4}{f}{1}\,\conf{8}{s}{1}, \conb{6}{s}{2}, \conf{4}{f}{1}\,\conf{6}{f}{1}, \conb{4}{f}{1}\,\conb{5}{g}{1}, \conf{4}{f}{1}\,\conf{8}{p}{1}, \conb{4}{f}{1}\,\conb{8}{d}{1}, \conf{4}{f}{1}\,\conf{9}{s}{1}, \conb{4}{f}{1}\,\conb{9}{p}{1}, \conf{4}{f}{1}\,\conf{9}{d}{1}, \conb{6}{s}{1}\,\conb{6}{p}{1}, \conf{5}{d}{1}\,\conf{6}{d}{1}, \conb{5}{d}{1}\,\conb{7}{s}{1}, \conf{5}{d}{1}\,\conf{5}{f}{1}, \conb{5}{d}{1}\,\conb{7}{p}{1}, \conf{6}{p}{2}, \conb{5}{d}{1}\,\conb{8}{s}{1}, \conf{5}{d}{1}\,\conf{7}{d}{1}, \conb{5}{d}{1}\,\conb{6}{f}{1}, \conf{5}{d}{1}\,\conf{5}{g}{1}, \conb{5}{d}{1}\,\conb{8}{p}{1}, \conf{6}{s}{1}\,\conf{6}{d}{1}, \conb{6}{s}{1}\,\conb{7}{s}{1}, \conf{5}{d}{1}\,\conf{8}{d}{1}, \conb{5}{f}{1}\,\conb{6}{s}{1}, \conf{6}{s}{1}\,\conf{7}{p}{1}, \conb{5}{d}{1}\,\conb{9}{s}{1}, \conf{5}{d}{1}\,\conf{9}{p}{1}, \conb{5}{d}{1}\,\conb{9}{d}{1}, \conf{6}{s}{1}\,\conf{8}{s}{1}, \conb{6}{s}{1}\,\conb{7}{d}{1}, \conf{6}{p}{1}\,\conf{6}{d}{1}, \conb{5}{g}{1}\,\conb{6}{s}{1}, \conf{6}{s}{1}\,\conf{8}{p}{1}, \conb{6}{p}{1}\,\conb{7}{s}{1}, \conf{6}{s}{1}\,\conf{6}{f}{1}, \conb{5}{f}{1}\,\conb{6}{p}{1}, \conf{6}{s}{1}\,\conf{8}{d}{1}, \conb{6}{p}{1}\,\conb{7}{p}{1}, \conf{6}{p}{1}\,\conf{7}{d}{1}, \conb{6}{p}{1}\,\conb{8}{s}{1}, \conf{6}{p}{1}\,\conf{6}{f}{1}, \conb{5}{g}{1}\,\conb{6}{p}{1}, \conf{6}{p}{1}\,\conf{8}{p}{1}, \conb{6}{p}{1}\,\conb{8}{d}{1}, \conf{6}{d}{2}, \conb{6}{d}{1}\,\conb{7}{s}{1}, \conf{7}{s}{2}, \conb{5}{f}{1}\,\conb{6}{d}{1}, \conf{5}{f}{1}\,\conf{7}{s}{1}, \conb{6}{d}{1}\,\conb{7}{p}{1}, \conf{7}{s}{1}\,\conf{7}{p}{1}, \conb{5}{f}{2}, \conf{5}{f}{1}\,\conf{7}{p}{1}, \conb{7}{p}{2}, \conf{6}{d}{1}\,\conf{7}{d}{1}, \conb{6}{d}{1}\,\conb{8}{s}{1}, \conf{7}{s}{1}\,\conf{8}{s}{1}, \conb{6}{d}{1}\,\conb{6}{f}{1}, \conf{5}{g}{1}\,\conf{6}{d}{1}, \conb{7}{s}{1}\,\conb{7}{d}{1}, \conf{6}{d}{1}\,\conf{8}{p}{1}, \conb{6}{f}{1}\,\conb{7}{s}{1}, \conf{5}{f}{1}\,\conf{7}{d}{1}, \conb{7}{s}{1}\,\conb{8}{p}{1}, \conf{5}{g}{1}\,\conf{7}{s}{1}, \conb{5}{f}{1}\,\conb{5}{g}{1}, \conf{5}{f}{1}\,\conf{8}{s}{1}, \conb{5}{f}{1}\,\conb{8}{p}{1}, \conf{5}{f}{1}\,\conf{6}{f}{1}, \conb{6}{d}{1}\,\conb{8}{d}{1}, \conf{7}{p}{1}\,\conf{7}{d}{1}, \conb{7}{p}{1}\,\conb{8}{s}{1}, \conf{7}{s}{1}\,\conf{8}{d}{1}, \conb{5}{g}{1}\,\conb{7}{p}{1}, \conf{6}{f}{1}\,\conf{7}{p}{1}, \conb{7}{p}{1}\,\conb{8}{p}{1}, \conf{5}{f}{1}\,\conf{8}{d}{1}, \conb{7}{p}{1}\,\conb{8}{d}{1}, \conf{7}{d}{2}, \conb{7}{d}{1}\,\conb{8}{s}{1}, \conf{8}{s}{2}, \conb{6}{f}{1}\,\conb{7}{d}{1}, \conf{5}{g}{1}\,\conf{7}{d}{1}, \conb{5}{g}{1}\,\conb{8}{s}{1}, \conf{6}{f}{1}\,\conf{8}{s}{1}, \conb{5}{g}{1}\,\conb{6}{f}{1}, \conf{5}{g}{2}, \conb{8}{s}{1}\,\conb{8}{p}{1}, \conf{7}{d}{1}\,\conf{8}{p}{1}, \conb{6}{f}{2}, \conf{5}{g}{1}\,\conf{8}{p}{1}, \conb{6}{f}{1}\,\conb{8}{p}{1}, \conf{8}{p}{2}, \conb{7}{d}{1}\,\conb{8}{d}{1}, \conf{5}{g}{1}\,\conf{8}{d}{1}, \conb{8}{s}{1}\,\conb{8}{d}{1}, \conf{6}{f}{1}\,\conf{8}{d}{1}, \conb{8}{p}{1}\,\conb{8}{d}{1}, \conf{8}{d}{2}
        }\\
        \midrule
        
        \multirow{1}{*}{\ion{Pr}{ii}} &
        \multirow{1}{*}{4\textit{f}$^{2.98}$ 5\textit{d}$^{\,0.52}$  6\textit{s}$^{0.50}$} &
        27  &
        \parbox[t]{0.70\textwidth}{\justifying\noindent
        \conf{4}{f}{3}\,\conf{6}{s}{1}, \conb{4}{f}{3}\,\conb{5}{d}{1}, \conf{4}{f}{2}\,\conf{5}{d}{2}, \conb{4}{f}{2}\,\conb{5}{d}{1}\,\conb{6}{s}{1}, \conf{4}{f}{2}, \conb{4}{f}{3}\,\conb{6}{p}{1}, \conf{4}{f}{2}\,\conf{5}{d}{1}\,\conf{6}{p}{1}, \conb{4}{f}{2}\,\conb{6}{s}{1}\,\conb{6}{p}{1}, \conf{4}{f}{3}\,\conf{7}{s}{1}, \conb{4}{f}{3}\,\conb{6}{d}{1}, \conf{4}{f}{3}\,\conf{7}{p}{1}, \conb{4}{f}{3}\,\conb{5}{f}{1}, \conf{4}{f}{3}\,\conf{6}{f}{1}, \conb{4}{f}{3}\,\conb{5}{g}{1}, \conf{4}{f}{2}\,\conf{6}{s}{1}\,\conf{7}{s}{1}, \conb{4}{f}{2}\,\conb{6}{p}{2}, \conf{4}{f}{2}\,\conf{6}{s}{1}\,\conf{6}{d}{1}, \conb{4}{f}{2}\,\conb{6}{s}{1}\,\conb{7}{p}{1}, \conf{4}{f}{2}\,\conf{5}{f}{1}\,\conf{6}{s}{1}, \conb{4}{f}{4}, \conf{4}{f}{1}\,\conf{5}{d}{2}\,\conf{6}{s}{1}, \conb{4}{f}{2}\,\conb{6}{s}{1}\,\conb{6}{f}{1}, \conf{4}{f}{2}\,\conf{5}{g}{1}\,\conf{6}{s}{1}, \conb{4}{f}{1}\,\conb{5}{d}{1}, \conf{4}{f}{1}\,\conf{5}{d}{1}\,\conf{6}{s}{1}\,\conf{6}{p}{1}, \conb{4}{f}{1}\,\conb{6}{p}{1}, \conf{4}{f}{1}\,\conf{6}{s}{1}\,\conf{6}{p}{2}
        }\\
        \midrule

        \multirow{1}{*}{\ion{Pr}{iii}} &
        \multirow{1}{*}{4\textit{f}$^{3.00}$} &
        17  &
        \parbox[t]{0.70\textwidth}{\justifying\noindent
        \conf{4}{f}{3}, \conb{4}{f}{2}\,\conb{5}{d}{1}, \conf{4}{f}{2}\,\conf{6}{s}{1}, \conb{4}{f}{2}\,\conb{6}{p}{1}, \conf{4}{f}{1}\,\conf{5}{d}{2}, \conb{4}{f}{2}\,\conb{7}{s}{1}, \conf{4}{f}{2}\,\conf{6}{d}{1}, \conb{4}{f}{2}\,\conb{5}{f}{1}, \conf{4}{f}{2}\,\conf{7}{p}{1}, \conb{4}{f}{1}\,\conb{5}{d}{1}\,\conb{6}{s}{1}, \conf{4}{f}{2}\,\conf{8}{s}{1}, \conb{4}{f}{2}\,\conb{6}{f}{1}, \conf{4}{f}{2}\,\conf{5}{g}{1}, \conb{4}{f}{1}\,\conb{5}{d}{1}\,\conb{6}{p}{1}, \conf{4}{f}{1}, \conb{4}{f}{1}\,\conb{6}{s}{1}\,\conb{6}{p}{1}, \conf{4}{f}{1}\,\conf{6}{p}{2}
        }\\
        \midrule

        \multirow{1}{*}{\ion{Nd}{ii}} &
        \multirow{1}{*}{4\textit{f}$^{4.00}$\,6\textit{s}$^{0.35}$\,5\textit{d}$^{0.65}$} &
        30  &
        \parbox[t]{0.70\textwidth}{\justifying\noindent
        \conf{4}{f}{4}\,\conf{6}{s}{1}, \conb{4}{f}{4}\,\conb{5}{d}{1}, \conf{4}{f}{3}\,\conf{5}{d}{2}, \conb{4}{f}{3}\,\conb{5}{d}{1}\,\conb{6}{s}{1}, \conf{4}{f}{4}\,\conf{6}{p}{1}, \conb{4}{f}{3}, \conf{4}{f}{3}\,\conf{5}{d}{1}\,\conf{6}{p}{1}, \conb{4}{f}{4}\,\conb{7}{s}{1}, \conf{4}{f}{4}\,\conf{6}{d}{1}, \conb{4}{f}{3}\,\conb{6}{s}{1}\,\conb{6}{p}{1}, \conf{4}{f}{4}\,\conf{7}{p}{1}, \conb{4}{f}{4}\,\conb{5}{f}{1}, \conf{4}{f}{4}\,\conf{8}{s}{1}, \conb{4}{f}{4}\,\conb{7}{d}{1}, \conf{4}{f}{4}\,\conf{8}{p}{1}, \conb{4}{f}{4}\,\conb{6}{f}{1}, \conf{4}{f}{4}\,\conf{5}{g}{1}, \conb{4}{f}{4}\,\conb{8}{d}{1}, \conf{4}{f}{3}\,\conf{5}{d}{1}\,\conf{6}{d}{1}, \conb{4}{f}{4}\,\conb{7}{f}{1}, \conf{4}{f}{4}\,\conf{6}{g}{1}, \conb{4}{f}{4}\,\conb{8}{f}{1}, \conf{4}{f}{4}\,\conf{7}{g}{1}, \conb{4}{f}{4}\,\conb{8}{g}{1}, \conf{4}{f}{3}\,\conf{6}{p}{2}, \conb{4}{f}{3}\,\conb{5}{d}{1}\,\conb{5}{f}{1}, \conf{4}{f}{3}\,\conf{6}{s}{1}\,\conf{7}{s}{1}, \conb{4}{f}{3}\,\conb{6}{s}{1}\,\conb{6}{d}{1}, \conf{4}{f}{3}\,\conf{5}{d}{1}\,\conf{6}{f}{1}, \conb{4}{f}{3}\,\conb{5}{d}{1}\,\conb{5}{g}{1}
        }\\
        \midrule

        \multirow{1}{*}{\ion{Nd}{iii}} &
        \multirow{1}{*}{4\textit{f}$^{4.00}$} &
        30  &
        \parbox[t]{0.70\textwidth}{\justifying\noindent
        \conf{4}{f}{4}, \conb{4}{f}{3}\,\conb{5}{d}{1}, \conf{4}{f}{3}\,\conf{6}{s}{1}, \conb{4}{f}{3}\,\conb{6}{p}{1}, \conf{4}{f}{2}\,\conf{5}{d}{2}, \conb{4}{f}{3}\,\conb{6}{d}{1}, \conf{4}{f}{3}\,\conf{7}{s}{1}, \conb{4}{f}{3}\,\conb{5}{f}{1}, \conf{4}{f}{3}\,\conf{7}{p}{1}, \conb{4}{f}{3}\,\conb{8}{s}{1}, \conf{4}{f}{3}\,\conf{7}{d}{1}, \conb{4}{f}{3}\,\conb{6}{f}{1}, \conf{4}{f}{3}\,\conf{8}{p}{1}, \conb{4}{f}{3}\,\conb{5}{g}{1}, \conf{4}{f}{2}\,\conf{5}{d}{1}\,\conf{6}{s}{1}, \conb{4}{f}{3}\,\conb{8}{d}{1}, \conf{4}{f}{3}\,\conf{7}{f}{1}, \conb{4}{f}{3}\,\conb{6}{g}{1}, \conf{4}{f}{3}\,\conf{8}{f}{1}, \conb{4}{f}{3}\,\conb{7}{g}{1}, \conf{4}{f}{3}\,\conf{8}{g}{1}, \conb{4}{f}{2}\,\conb{5}{d}{1}\,\conb{6}{p}{1}, \conf{4}{f}{2}, \conb{4}{f}{2}\,\conb{6}{s}{1}\,\conb{6}{p}{1}, \conf{4}{f}{2}\,\conf{5}{d}{1}\,\conf{6}{d}{1}, \conb{4}{f}{2}\,\conb{5}{d}{1}\,\conb{5}{f}{1}, \conf{4}{f}{2}\,\conf{6}{p}{2}, \conb{4}{f}{2}\,\conb{5}{d}{1}\,\conb{6}{f}{1}, \conf{4}{f}{2}\,\conf{5}{d}{1}\,\conf{5}{g}{1}, \conb{4}{f}{2}\,\conb{5}{d}{1}\,\conb{6}{g}{1}
        }\\
        \midrule

        \multirow{1}{*}{\ion{Pm}{ii}} &
        \multirow{1}{*}{4\textit{f}$^{4.91}$\,6\textit{s}$^{1.09}$} &
        11  &
        \parbox[t]{0.70\textwidth}{\justifying\noindent
        \conf{4}{f}{5}\,\conf{6}{s}{1}, \conb{4}{f}{5}\,\conb{5}{d}{1}, \conf{4}{f}{4}\,\conf{5}{d}{1}\,\conf{6}{s}{1}, \conb{4}{f}{5}\,\conb{6}{p}{1}, \conf{4}{f}{4}, \conb{4}{f}{4}\,\conb{6}{s}{1}\,\conb{6}{p}{1}, \conf{4}{f}{4}\,\conf{6}{s}{1}\,\conf{6}{d}{1}, \conb{4}{f}{4}\,\conb{5}{f}{1}\,\conb{6}{s}{1}, \conf{4}{f}{4}\,\conf{6}{s}{1}\,\conf{7}{d}{1}, \conb{4}{f}{4}\,\conb{6}{s}{1}\,\conb{6}{f}{1}, \conf{4}{f}{4}\,\conf{5}{g}{1}\,\conf{6}{s}{1}
        }\\
        \midrule

        \multirow{1}{*}{\ion{Pm}{iii}} &
        \multirow{1}{*}{4\textit{f}$^{5.00}$} &
        21  &
        \parbox[t]{0.70\textwidth}{\justifying\noindent
        \conf{4}{f}{5}, \conb{4}{f}{4}\,\conb{5}{d}{1}, \conf{4}{f}{4}\,\conf{6}{s}{1}, \conb{4}{f}{4}\,\conb{6}{p}{1}, \conf{4}{f}{4}\,\conf{7}{s}{1}, \conb{4}{f}{4}\,\conb{6}{d}{1}, \conf{4}{f}{3}\,\conf{5}{d}{2}, \conb{4}{f}{4}\,\conb{7}{p}{1}, \conf{4}{f}{4}\,\conf{8}{s}{1}, \conb{4}{f}{3}\,\conb{5}{d}{1}\,\conb{6}{s}{1}, \conf{4}{f}{4}\,\conf{7}{d}{1}, \conb{4}{f}{4}\,\conb{8}{p}{1}, \conf{4}{f}{4}\,\conf{8}{d}{1}, \conb{4}{f}{3}\,\conb{5}{d}{1}\,\conb{6}{p}{1}, \conf{4}{f}{3}, \conb{4}{f}{3}\,\conb{6}{s}{1}\,\conb{6}{p}{1}, \conf{4}{f}{3}\,\conf{5}{d}{1}\,\conf{6}{d}{1}, \conb{4}{f}{3}\,\conb{6}{p}{2}, \conf{4}{f}{3}\,\conf{6}{s}{1}\,\conf{6}{d}{1}, \conb{4}{f}{3}\,\conb{6}{p}{1}\,\conb{6}{d}{1}, \conf{4}{f}{3}\,\conf{6}{d}{2}
        }\\
        \midrule

        \multirow{1}{*}{\ion{Sm}{ii}} &
        \multirow{1}{*}{4\textit{f}$^{5.65}$\,5\textit{d}$^{0.87}$\,6\textit{s}$^{0.43}$\,6\textit{p}$^{0.05}$} &
        7  &
        \parbox[t]{0.70\textwidth}{\justifying\noindent
        \conf{4}{f}{6}\,\conf{6}{s}{1}, \conb{4}{f}{6}\,\conb{5}{d}{1}, \conf{4}{f}{5}\,\conf{5}{d}{1}\,\conf{6}{s}{1}, \conb{4}{f}{6}\,\conb{6}{p}{1}, \conf{4}{f}{5}, \conb{4}{f}{5}\,\conb{6}{s}{1}\,\conb{6}{p}{1}, \conf{4}{f}{7}
        }\\
        \midrule

        \multirow{1}{*}{\ion{Sm}{iii}} &
        \multirow{1}{*}{4\textit{f}$^{5.88}$\,6\textit{p}$^{1.12}$} &
        34  &
        \parbox[t]{0.70\textwidth}{\justifying\noindent
        \conf{4}{f}{6}, \conb{4}{f}{5}\,\conb{5}{d}{1}, \conf{4}{f}{5}\,\conf{6}{s}{1}, \conb{4}{f}{5}\,\conb{6}{p}{1}, \conf{4}{f}{5}\,\conf{7}{s}{1}, \conb{4}{f}{5}\,\conb{6}{d}{1}, \conf{4}{f}{5}\,\conf{7}{p}{1}, \conb{4}{f}{5}\,\conb{5}{f}{1}, \conf{4}{f}{4}\,\conf{5}{d}{2}, \conb{4}{f}{5}\,\conb{8}{s}{1}, \conf{4}{f}{5}\,\conf{7}{d}{1}, \conb{4}{f}{5}\,\conb{8}{p}{1}, \conf{4}{f}{5}\,\conf{5}{g}{1}, \conb{4}{f}{5}\,\conb{6}{f}{1}, \conf{4}{f}{5}\,\conf{8}{d}{1}, \conb{4}{f}{4}\,\conb{5}{d}{1}\,\conb{6}{s}{1}, \conf{4}{f}{4}\,\conf{5}{d}{1}\,\conf{6}{p}{1}, \conb{4}{f}{4}, \conf{4}{f}{4}\,\conf{6}{s}{1}\,\conf{6}{p}{1}, \conb{4}{f}{4}\,\conb{5}{d}{1}\,\conb{7}{s}{1}, \conf{4}{f}{4}\,\conf{5}{d}{1}\,\conf{7}{p}{1}, \conb{4}{f}{4}\,\conb{6}{p}{2}, \conf{4}{f}{4}\,\conf{5}{d}{1}\,\conf{7}{d}{1}, \conb{4}{f}{4}\,\conb{6}{s}{1}\,\conb{7}{s}{1}, \conf{4}{f}{4}\,\conf{6}{s}{1}\,\conf{7}{p}{1}, \conb{4}{f}{4}\,\conb{6}{s}{1}\,\conb{7}{d}{1}, \conf{4}{f}{4}\,\conf{6}{p}{1}\,\conf{7}{s}{1}, \conb{4}{f}{4}\,\conb{6}{p}{1}\,\conb{7}{p}{1}, \conf{4}{f}{4}\,\conf{6}{p}{1}\,\conf{7}{d}{1}, \conb{4}{f}{4}\,\conb{7}{s}{1}\,\conb{7}{p}{1}, \conf{4}{f}{4}\,\conf{7}{p}{2}, \conb{4}{f}{4}\,\conb{7}{s}{1}\,\conb{7}{d}{1}, \conf{4}{f}{4}\,\conf{7}{p}{1}\,\conf{7}{d}{1}, \conb{4}{f}{4}\,\conb{7}{d}{2}
        }\\
        \midrule

        \multirow{1}{*}{\ion{Eu}{ii}} &
        \multirow{1}{*}{4\textit{f}$^{6.81}$\,5\textit{d}$^{0.18}$\,6\textit{s}$^{1.01}$} &
        10  &
        \parbox[t]{0.70\textwidth}{\justifying\noindent
        \conf{4}{f}{7}\,\conf{6}{s}{1}, \conb{4}{f}{7}\,\conb{5}{d}{1}, \conf{4}{f}{7}\,\conf{6}{p}{1}, \conb{4}{f}{6}\,\conb{5}{d}{2}, \conf{4}{f}{6}\,\conf{5}{d}{1}\,\conf{6}{s}{1}, \conb{4}{f}{7}\,\conb{7}{s}{1}, \conf{4}{f}{7}\,\conf{6}{d}{1}, \conb{4}{f}{6}\,\conb{5}{d}{1}\,\conb{6}{p}{1}, \conf{4}{f}{7}\,\conf{8}{s}{1}, \conb{4}{f}{8}
        }\\
        \midrule

        \multirow{1}{*}{\ion{Eu}{iii}} &
        \multirow{1}{*}{4\textit{f}$^{6.72}$\,5\textit{d}$^{0.18}$\,6\textit{s}$^{0.09}$\,6\textit{p}$^{0.01}$} &
        13  &
        \parbox[t]{0.70\textwidth}{\justifying\noindent
        \conf{4}{f}{7}, \conb{4}{f}{6}\,\conb{5}{d}{1}, \conf{4}{f}{6}\,\conf{6}{s}{1}, \conb{4}{f}{6}\,\conb{6}{p}{1}, \conf{4}{f}{6}\,\conf{7}{s}{1}, \conb{4}{f}{6}\,\conb{6}{d}{1}, \conf{4}{f}{6}\,\conf{7}{p}{1}, \conb{4}{f}{6}\,\conb{8}{s}{1}, \conf{4}{f}{6}\,\conf{7}{d}{1}, \conb{4}{f}{6}\,\conb{8}{p}{1}, \conf{4}{f}{5}\,\conf{5}{d}{1}\,\conf{6}{s}{1}, \conb{4}{f}{5}\,\conb{5}{d}{1}\,\conb{6}{p}{1}, \conf{4}{f}{5}\,\conf{6}{s}{1}\,\conf{6}{p}{1}
        }\\
        \midrule

        \multirow{1}{*}{\ion{Gd}{ii}} &
        \multirow{1}{*}{4\textit{f}$^{7.94}$\,5\textit{d}$^{0.26}$\,6\textit{p}$^{0.80}$} &
        8  &
        \parbox[t]{0.70\textwidth}{\justifying\noindent
        \conf{4}{f}{7}\,\conf{5}{d}{1}\,\conf{6}{s}{1}, \conb{4}{f}{7}\,\conb{5}{d}{2}, \conf{4}{f}{8}\,\conf{6}{s}{1}, \conb{4}{f}{7}, \conf{4}{f}{8}\,\conf{5}{d}{1}, \conb{4}{f}{8}\,\conb{6}{p}{1}, \conf{4}{f}{7}\,\conf{5}{d}{1}\,\conf{6}{p}{1}, \conb{4}{f}{7}\,\conb{6}{s}{1}\,\conb{6}{p}{1}
        }\\
        \midrule

        \multirow{1}{*}{\ion{Gd}{iii}} &
        \multirow{1}{*}{4\textit{f}$^{7.85}$\,6\textit{s}$^{0.15}$} &
        10  &
        \parbox[t]{0.70\textwidth}{\justifying\noindent
        \conf{4}{f}{7}\,\conf{5}{d}{1}, \conb{4}{f}{8}, \conf{4}{f}{7}\,\conf{6}{s}{1}, \conb{4}{f}{7}\,\conb{6}{p}{1}, \conf{4}{f}{7}\,\conf{7}{s}{1}, \conb{4}{f}{7}\,\conb{6}{d}{1}, \conf{4}{f}{7}\,\conf{7}{p}{1}, \conb{4}{f}{7}\,\conb{5}{f}{1}, \conf{4}{f}{7}\,\conf{7}{d}{1}, \conb{4}{f}{7}\,\conb{6}{f}{1}
        }\\
        \midrule

        \multirow{1}{*}{\ion{Tb}{ii}} &
        \multirow{1}{*}{4\textit{f}$^{8.89}$\,6\textit{s}$^{0.05}$\,6\textit{p}$^{1.06}$} &
        8  &
        \parbox[t]{0.70\textwidth}{\justifying\noindent
        \conf{4}{f}{9}\,\conf{6}{s}{1}, \conb{4}{f}{8}\,\conb{5}{d}{2}, \conf{4}{f}{8}\,\conf{5}{d}{1}\,\conf{6}{s}{1}, \conb{4}{f}{8}, \conf{4}{f}{9}\,\conf{5}{d}{1}, \conb{4}{f}{9}\,\conb{6}{p}{1}, \conf{4}{f}{8}\,\conf{5}{d}{1}\,\conf{6}{p}{1}, \conb{4}{f}{8}\,\conb{6}{s}{1}\,\conb{6}{p}{1}
        }\\
        \midrule

        \multirow{1}{*}{\ion{Tb}{iii}} &
        \multirow{1}{*}{4\textit{f}$^{8.79}$\,6\textit{s}$^{0.21}$} &
        9  &
        \parbox[t]{0.70\textwidth}{\justifying\noindent
        \conf{4}{f}{9}, \conb{4}{f}{8}\,\conb{5}{d}{1}, \conf{4}{f}{8}\,\conf{6}{s}{1}, \conb{4}{f}{8}\,\conb{6}{p}{1}, \conf{4}{f}{8}\,\conf{7}{s}{1}, \conb{4}{f}{8}\,\conb{6}{d}{1}, \conf{4}{f}{8}\,\conf{5}{f}{1}, \conb{4}{f}{8}\,\conb{8}{p}{1}, \conf{4}{f}{8}\,\conf{6}{f}{1}
        }\\
	\end{tabular}
    \end{ruledtabular}
\end{table*}
\begin{table*}
	\caption*{\justifying \textsc{TABLE I CONTINUED:} Configurations included in the \FAC\  calculations for lanthanide ions. The presented configurations are sorted by energy. The FMC weights used for the radial optimization of the potential are given for each ion. Configurations are displayed in alternating black and blue to improve visual clarity and facilitate distinction between successive entries.}
	\label{tab:FAC_configs3}
    \begin{ruledtabular}
	\begin{tabular}{@{}cccl@{}}
		Ion & FMC weights & \#\footnote{Number of included configurations} & Configurations \\ 
		\midrule

        \multirow{1}{*}{\ion{Dy}{ii}} &
        \multirow{1}{*}{4\textit{f}$^{11.00}$} &
        10  &
        \parbox[t]{0.70\textwidth}{\justifying\noindent
        \conf{4}{f}{10}\,\conf{6}{s}{1}, \conb{4}{f}{9}\,\conb{5}{d}{1}\,\conb{6}{s}{1}, \conf{4}{f}{9}, \conb{4}{f}{10}\,\conb{5}{d}{1}, \conf{4}{f}{9}\,\conf{5}{d}{2}, \conb{4}{f}{10}\,\conb{6}{p}{1}, \conf{4}{f}{9}\,\conf{6}{s}{1}\,\conf{6}{p}{1}, \conb{4}{f}{9}\,\conb{5}{d}{1}\,\conb{6}{p}{1}, \conf{4}{f}{9}\,\conf{6}{p}{2}, \conb{4}{f}{11}
        }\\
        \midrule

        \multirow{1}{*}{\ion{Dy}{iii}} &
        \multirow{1}{*}{4\textit{f}$^{9.73}$\,5\textit{d}$^{0.27}$} &
        9  &
        \parbox[t]{0.70\textwidth}{\justifying\noindent
        \conf{4}{f}{10}, \conb{4}{f}{9}\,\conb{5}{d}{1}, \conf{4}{f}{9}\,\conf{6}{s}{1}, \conb{4}{f}{9}\,\conb{6}{p}{1}, \conf{4}{f}{8}\,\conf{5}{d}{2}, \conb{4}{f}{8}\,\conb{5}{d}{1}\,\conb{6}{s}{1}, \conf{4}{f}{8}\,\conf{5}{d}{1}\,\conf{6}{p}{1}, \conb{4}{f}{8}\,\conb{6}{s}{1}\,\conb{6}{p}{1}, \conf{4}{f}{8}\,\conf{6}{p}{2}
        }\\
        \midrule

        \multirow{1}{*}{\ion{Ho}{ii}} &
        \multirow{1}{*}{4\textit{f}$^{10.92}$\,6\textit{s}$^{1.08}$} &
        15  &
        \parbox[t]{0.70\textwidth}{\justifying\noindent
        \conf{4}{f}{11}\,\conf{6}{s}{1}, \conb{4}{f}{11}\,\conb{5}{d}{1}, \conf{4}{f}{11}\,\conf{6}{p}{1}, \conb{4}{f}{10}, \conf{4}{f}{10}\,\conf{5}{d}{1}\,\conf{6}{s}{1}, \conb{4}{f}{10}\,\conb{5}{d}{2}, \conf{4}{f}{10}\,\conf{6}{s}{1}\,\conf{6}{p}{1}, \conb{4}{f}{10}\,\conb{5}{d}{1}\,\conb{6}{p}{1}, \conf{4}{f}{10}\,\conf{6}{p}{2}, \conb{4}{f}{12}, \conf{4}{f}{9}\,\conf{5}{d}{2}\,\conf{6}{s}{1}, \conb{4}{f}{9}\,\conb{5}{d}{1}, \conf{4}{f}{9}\,\conf{5}{d}{1}\,\conf{6}{s}{1}\,\conf{6}{p}{1}, \conb{4}{f}{9}\,\conb{6}{p}{1}, \conf{4}{f}{9}\,\conf{6}{s}{1}\,\conf{6}{p}{2}
        }\\
        \midrule

        \multirow{1}{*}{\ion{Ho}{iii}} &
        \multirow{1}{*}{4\textit{f}$^{10.74}$\,6\textit{s}$^{0.26}$} &
        15  &
        \parbox[t]{0.70\textwidth}{\justifying\noindent
       \conf{4}{f}{11}, \conb{4}{f}{10}\,\conb{5}{d}{1}, \conf{4}{f}{10}\,\conf{6}{s}{1}, \conb{4}{f}{10}\,\conb{6}{p}{1}, \conf{4}{f}{10}\,\conf{7}{s}{1}, \conb{4}{f}{10}\,\conb{6}{d}{1}, \conf{4}{f}{10}\,\conf{7}{p}{1}, \conb{4}{f}{10}\,\conb{5}{f}{1}, \conf{4}{f}{10}\,\conf{8}{s}{1}, \conb{4}{f}{10}\,\conb{7}{d}{1}, \conf{4}{f}{10}\,\conf{8}{p}{1}, \conb{4}{f}{10}\,\conb{9}{s}{1}, \conf{4}{f}{10}\,\conf{8}{d}{1}, \conb{4}{f}{10}\,\conb{9}{p}{1}, \conf{4}{f}{9}\,\conf{5}{d}{1}\,\conf{6}{d}{1}
        }\\
        \midrule

        \multirow{1}{*}{\ion{Er}{ii}} &
        \multirow{1}{*}{4\textit{f}$^{11.72}$\,5\textit{d}$^{0.42}$\,6\textit{s}$^{0.86}$} &
        19  &
        \parbox[t]{0.70\textwidth}{\justifying\noindent
        \conf{4}{f}{12}\,\conf{6}{s}{1}, \conb{4}{f}{11}\,\conb{5}{d}{1}\,\conb{6}{s}{1}, \conf{4}{f}{11}, \conb{4}{f}{11}\,\conb{5}{d}{2}, \conf{4}{f}{12}\,\conf{5}{d}{1}, \conb{4}{f}{12}\,\conb{6}{p}{1}, \conf{4}{f}{11}\,\conf{6}{s}{1}\,\conf{6}{p}{1}, \conb{4}{f}{11}\,\conb{5}{d}{1}\,\conb{6}{p}{1}, \conf{4}{f}{12}\,\conf{7}{s}{1}, \conb{4}{f}{12}\,\conb{7}{p}{1}, \conf{4}{f}{11}\,\conf{6}{s}{1}\,\conf{7}{s}{1}, \conb{4}{f}{11}\,\conb{5}{d}{1}\,\conb{7}{s}{1}, \conf{4}{f}{11}\,\conf{5}{d}{1}\,\conf{6}{d}{1}, \conb{4}{f}{11}\,\conb{6}{s}{1}\,\conb{6}{d}{1}, \conf{4}{f}{11}\,\conf{5}{d}{1}\,\conf{7}{p}{1}, \conb{4}{f}{11}\,\conb{6}{s}{1}\,\conb{7}{p}{1}, \conf{4}{f}{13}, \conb{4}{f}{11}\,\conb{6}{d}{2}, \conf{4}{f}{11}\,\conf{7}{p}{2}
        }\\
        \midrule

        \multirow{1}{*}{\ion{Er}{iii}} &
        \multirow{1}{*}{4\textit{f}$^{11.73}$\,5\textit{d}$^{0.27}$} &
        15  &
        \parbox[t]{0.70\textwidth}{\justifying\noindent
      \conf{4}{f}{12}, \conb{4}{f}{11}\,\conb{5}{d}{1}, \conf{4}{f}{11}\,\conf{6}{s}{1}, \conb{4}{f}{11}\,\conb{6}{p}{1}, \conf{4}{f}{11}\,\conf{7}{s}{1}, \conb{4}{f}{11}\,\conb{6}{d}{1}, \conf{4}{f}{11}\,\conf{7}{p}{1}, \conb{4}{f}{11}\,\conb{5}{f}{1}, \conf{4}{f}{11}\,\conf{8}{s}{1}, \conb{4}{f}{11}\,\conb{7}{d}{1}, \conf{4}{f}{11}\,\conf{8}{p}{1}, \conb{4}{f}{11}\,\conb{6}{f}{1}, \conf{4}{f}{10}\,\conf{5}{d}{2}, \conb{4}{f}{11}\,\conb{9}{s}{1}, \conf{4}{f}{11}\,\conf{8}{d}{1}, \conb{4}{f}{11}\,\conb{9}{p}{1}, \conf{4}{f}{11}\,\conf{6}{g}{1}, \conb{4}{f}{10}\,\conb{5}{d}{1}\,\conb{6}{s}{1}, \conf{4}{f}{10}, \conb{4}{f}{10}\,\conb{5}{d}{1}\,\conb{6}{p}{1}, \conf{4}{f}{10}\,\conf{6}{s}{1}\,\conf{6}{p}{1}, \conb{4}{f}{10}\,\conb{5}{d}{1}\,\conb{6}{d}{1}, \conf{4}{f}{10}\,\conf{5}{d}{1}\,\conf{7}{s}{1}, \conb{4}{f}{10}\,\conb{5}{d}{1}\,\conb{7}{p}{1}, \conf{4}{f}{10}\,\conf{6}{s}{1}\,\conf{7}{s}{1}, \conb{4}{f}{10}\,\conb{6}{s}{1}\,\conb{6}{d}{1}, \conf{4}{f}{10}\,\conf{5}{d}{1}\,\conf{8}{s}{1}, \conb{4}{f}{10}\,\conb{5}{d}{1}\,\conb{8}{p}{1}, \conf{4}{f}{10}\,\conf{6}{s}{1}\,\conf{7}{p}{1}, \conb{4}{f}{10}\,\conb{6}{s}{1}\,\conb{8}{s}{1}, \conf{4}{f}{10}\,\conf{6}{s}{1}\,\conf{8}{p}{1}
        }\\
        \midrule

        \multirow{1}{*}{\ion{Tm}{ii}} &
        \multirow{1}{*}{4\textit{f}$^{11.87}$\,6\textit{p}$^{0.13}$} &
        25  &
        \parbox[t]{0.70\textwidth}{\justifying\noindent
        \conf{4}{f}{13}\,\conf{6}{s}{1}, \conb{4}{f}{12}\,\conb{5}{d}{1}\,\conb{6}{s}{1}, \conf{4}{f}{12}, \conb{4}{f}{13}\,\conb{5}{d}{1}, \conf{4}{f}{13}\,\conf{6}{p}{1}, \conb{4}{f}{12}\,\conb{5}{d}{2}, \conf{4}{f}{12}\,\conf{6}{s}{1}\,\conf{6}{p}{1}, \conb{4}{f}{12}\,\conb{5}{d}{1}\,\conb{6}{p}{1}, \conf{4}{f}{13}\,\conf{7}{s}{1}, \conb{4}{f}{13}\,\conb{7}{p}{1}, \conf{4}{f}{13}\,\conf{6}{d}{1}, \conb{4}{f}{13}\,\conb{8}{s}{1}, \conf{4}{f}{13}\,\conf{8}{p}{1}, \conb{4}{f}{12}\,\conb{6}{s}{1}\,\conb{7}{s}{1}, \conf{4}{f}{12}\,\conf{6}{p}{2}, \conb{4}{f}{12}\,\conb{6}{s}{1}\,\conb{6}{d}{1}, \conf{4}{f}{12}\,\conf{6}{s}{1}\,\conf{7}{p}{1}, \conb{4}{f}{12}\,\conb{6}{s}{1}\,\conb{8}{s}{1}, \conf{4}{f}{12}\,\conf{6}{s}{1}\,\conf{8}{p}{1}, \conb{5}{p}{6}, \conf{4}{f}{11}\,\conf{5}{d}{2}\,\conf{6}{s}{1}, \conb{4}{f}{11}\,\conb{5}{d}{1}, \conf{4}{f}{11}\,\conf{5}{d}{1}\,\conf{6}{s}{1}\,\conf{6}{p}{1}, \conb{4}{f}{11}\,\conb{6}{p}{1}, \conf{4}{f}{11}\,\conf{6}{s}{1}\,\conf{6}{p}{2}
        }\\
        \midrule

        \multirow{1}{*}{\ion{Tm}{iii}} &
        \multirow{1}{*}{4\textit{f}$^{10.94}$\,5\textit{d}$^{0.49}$\,6\textit{s}$^{1.01}$\,6\textit{p}$^{0.56}$} &
        15  &
        \parbox[t]{0.70\textwidth}{\justifying\noindent
        \conf{4}{f}{13}, \conb{4}{f}{12}\,\conb{5}{d}{1}, \conf{4}{f}{12}\,\conf{6}{s}{1}, \conb{4}{f}{12}\,\conb{6}{p}{1}, \conf{4}{f}{12}\,\conf{7}{s}{1}, \conb{4}{f}{12}\,\conb{6}{d}{1}, \conf{4}{f}{12}\,\conf{7}{p}{1}, \conb{4}{f}{12}\,\conb{8}{s}{1}, \conf{4}{f}{12}\,\conf{8}{p}{1}, \conb{4}{f}{11}\,\conb{5}{d}{2}, \conf{4}{f}{11}\,\conf{5}{d}{1}\,\conf{6}{s}{1}, \conb{4}{f}{11}, \conf{4}{f}{11}\,\conf{5}{d}{1}\,\conf{6}{p}{1}, \conb{4}{f}{11}\,\conb{6}{s}{1}\,\conb{6}{p}{1}, \conf{4}{f}{11}\,\conf{6}{p}{2}
        }\\
        \midrule

        \multirow{1}{*}{\ion{Yb}{ii}} &
        \multirow{1}{*}{4\textit{f}$^{13.51}$\,5\textit{d}$^{0.16}$\,6\textit{s}$^{1.18}$\,6\textit{p}$^{0.15}$} &
        40  &
        \parbox[t]{0.70\textwidth}{\justifying\noindent
        \conf{6}{s}{1}, \conb{4}{f}{13}, \conf{4}{f}{13}\,\conf{5}{d}{1}\,\conf{6}{s}{1}, \conb{6}{p}{1}, \conf{5}{d}{1}, \conb{4}{f}{13}\,\conb{5}{d}{2}, \conf{4}{f}{13}\,\conf{6}{s}{1}\,\conf{6}{p}{1}, \conb{4}{f}{13}\,\conb{5}{d}{1}\,\conb{6}{p}{1}, \conf{7}{s}{1}, \conb{6}{d}{1}, \conf{7}{p}{1}, \conb{5}{f}{1}, \conf{8}{s}{1}, \conb{8}{p}{1}, \conf{6}{f}{1}, \conb{9}{s}{1}, \conf{4}{f}{13}\,\conf{6}{s}{1}\,\conf{7}{s}{1}, \conb{8}{d}{1}, \conf{7}{f}{1}, \conb{10}{s}{1}, \conf{9}{d}{1}, \conb{8}{f}{1}, \conf{4}{f}{13}\,\conf{6}{p}{2}, \conb{4}{f}{13}\,\conb{6}{s}{1}\,\conb{6}{d}{1}, \conf{10}{d}{1}, \conb{9}{f}{1}, \conf{11}{d}{1}, \conb{10}{f}{1}, \conf{11}{f}{1}, \conb{12}{f}{1}, \conf{4}{f}{13}\,\conf{6}{s}{1}\,\conf{7}{p}{1}, \conb{13}{f}{1}, \conf{14}{f}{1}, \conb{4}{f}{13}\,\conb{6}{s}{1}\,\conb{8}{s}{1}, \conf{4}{f}{13}\,\conf{6}{s}{1}\,\conf{8}{p}{1}, \conb{4}{f}{12}\,\conb{5}{d}{2}\,\conb{6}{s}{1}, \conf{4}{f}{12}\,\conf{5}{d}{1}, \conb{4}{f}{12}\,\conb{5}{d}{1}\,\conb{6}{s}{1}\,\conb{6}{p}{1}, \conf{4}{f}{12}\,\conf{6}{p}{1}, \conb{4}{f}{12}\,\conb{6}{s}{1}\,\conb{6}{p}{2}
        }\\
        \midrule

        \multirow{1}{*}{\ion{Yb}{iii}} &
        \multirow{1}{*}{4\textit{f}$^{13.65}$\,5\textit{d}$^{0.35}$} &
        43  &
        \parbox[t]{0.70\textwidth}{\justifying\noindent
        \conf{5}{p}{6}, \conb{4}{f}{13}\,\conb{5}{d}{1}, \conf{4}{f}{13}\,\conf{6}{s}{1}, \conb{4}{f}{13}\,\conb{6}{p}{1}, \conf{4}{f}{13}\,\conf{7}{s}{1}, \conb{4}{f}{13}\,\conb{6}{d}{1}, \conf{4}{f}{13}\,\conf{7}{p}{1}, \conb{4}{f}{13}\,\conb{8}{s}{1}, \conf{4}{f}{13}\,\conf{8}{p}{1}, \conb{4}{f}{12}\,\conb{5}{d}{2}, \conf{4}{f}{12}\,\conf{5}{d}{1}\,\conf{6}{s}{1}, \conb{4}{f}{12}, \conf{4}{f}{12}\,\conf{5}{d}{1}\,\conf{6}{p}{1}, \conb{4}{f}{12}\,\conb{6}{s}{1}\,\conb{6}{p}{1}, \conf{4}{f}{12}\,\conf{5}{d}{1}\,\conf{7}{s}{1}, \conb{4}{f}{12}\,\conb{5}{d}{1}\,\conb{6}{d}{1}, \conf{4}{f}{12}\,\conf{5}{d}{1}\,\conf{7}{p}{1}, \conb{4}{f}{12}\,\conb{6}{p}{2}, \conf{4}{f}{12}\,\conf{6}{s}{1}\,\conf{7}{s}{1}, \conb{4}{f}{12}\,\conb{5}{d}{1}\,\conb{8}{s}{1}, \conf{4}{f}{12}\,\conf{6}{s}{1}\,\conf{6}{d}{1}, \conb{4}{f}{12}\,\conb{5}{d}{1}\,\conb{8}{p}{1}, \conf{4}{f}{12}\,\conf{6}{s}{1}\,\conf{7}{p}{1}, \conb{4}{f}{12}\,\conb{6}{s}{1}\,\conb{8}{s}{1}, \conf{4}{f}{12}\,\conf{6}{s}{1}\,\conf{8}{p}{1}, \conb{4}{f}{12}\,\conb{6}{p}{1}\,\conb{7}{s}{1}, \conf{4}{f}{12}\,\conf{6}{p}{1}\,\conf{6}{d}{1}, \conb{4}{f}{12}\,\conb{6}{p}{1}\,\conb{7}{p}{1}, \conf{4}{f}{12}\,\conf{6}{p}{1}\,\conf{8}{s}{1}, \conb{4}{f}{12}\,\conb{6}{p}{1}\,\conb{8}{p}{1}, \conf{4}{f}{12}\,\conf{6}{d}{1}\,\conf{7}{s}{1}, \conb{4}{f}{12}\,\conb{6}{d}{2}, \conf{4}{f}{12}\,\conf{7}{s}{1}\,\conf{7}{p}{1}, \conb{4}{f}{12}\,\conb{6}{d}{1}\,\conb{7}{p}{1}, \conf{4}{f}{12}\,\conf{7}{p}{2}, \conb{4}{f}{12}\,\conb{7}{s}{1}\,\conb{8}{s}{1}, \conf{4}{f}{12}\,\conf{6}{d}{1}\,\conf{8}{s}{1}, \conb{4}{f}{12}\,\conb{7}{s}{1}\,\conb{8}{p}{1}, \conf{4}{f}{12}\,\conf{6}{d}{1}\,\conf{8}{p}{1}, \conb{4}{f}{12}\,\conb{7}{p}{1}\,\conb{8}{s}{1}, \conf{4}{f}{12}\,\conf{7}{p}{1}\,\conf{8}{p}{1}, \conb{4}{f}{12}\,\conb{8}{s}{1}\,\conb{8}{p}{1}, \conf{4}{f}{12}\,\conf{8}{p}{2}
        }\\

	\end{tabular}
    \end{ruledtabular}
\end{table*}

In this work, we present atomic structure calculations performed with the \FAC\ code for all singly and doubly ionized lanthanide elements, focusing on level energies and electric dipole (E1) transitions. Magnetic dipole (M1) and electric quadrupole (E2) transitions are beyond the scope of this study and will be addressed in future work. The computed results are systematically compared against available experimental and semi-empirical datasets to evaluate their accuracy. Specifically, we benchmark our level energies and E1 transitions against five data sources: the NIST ASD \citep{NIST_ASD}, the DREAM database \citep{2020Atoms...8...18Q}, and recent experimental measurements reported by \citet[][]{2024MNRAS.527.4440F}, \citet[][]{2024ApJS..274....9D}, and \citet[][]{2025ApJS..278....7V}. While the NIST ASD is the standard reference for atomic structure data due to its stringent inclusion criteria, its coverage of lanthanide ions is limited, with only 705 validated transitions reported across 11 singly and doubly ionized species.

The DREAM database provides a much larger collection of spectroscopic parameters for lanthanides in low ionization stages, derived using the pseudo-relativistic Hartree–Fock approach implemented in the \HFR\ code. It includes 60\,160 E1 transitions for singly and doubly ionized lanthanides, although the data are concentrated among a few well-characterized ions. Several ions in our sample either lack representation entirely or have only a limited number of transitions available in DREAM.

We obtained the correct ground state configuration for all ions except \ion{Gd}{ii}. An overview of experimental and calculated ground states for all ions in this study is given in table~\ref{tab:ground_states}. 

\begin{table}[hbt]
\centering
	\caption{\justifying Overview of the ground state configurations in our \FAC\ atomic structure calculations and from experiments. Checkmarks in the NIST columns indicate that the calculated ground state configuration matches the one reported on the NIST ASD.}
	\label{tab:ground_states}
      \begin{ruledtabular}
	\begin{tabular}{@{}lcccc}
        & \multicolumn{2}{c}{Singly Ionised} & \multicolumn{2}{c}{Doubly Ionised}  \\            \cline{2-3}\cline{4-5}
Element & \FAC & NIST     & \FAC & NIST   \\ \hline
La      & \conf{5}{d}{2} & \ding{51}   & \conf{5}{d}{1} & \ding{51} \\
Ce      & \conf{4}{f}{1}\,\conf{5}{d}{2} & \ding{51}   & \conf{4}{f}{2} & \ding{51} \\
Pr      & \conf{4}{f}{3}\,\conf{6}{s}{1} & \ding{51}   & \conf{4}{f}{3} & \ding{51} \\
Nd      & \conf{4}{f}{4}\,\conf{6}{s}{1} & \ding{51}   & \conf{4}{f}{4} & \ding{51} \\
Pm      & \conf{4}{f}{5}\,\conf{6}{s}{1} & \ding{51}   & \conf{4}{f}{5}& \ding{51} \\
Sm      & \conf{4}{f}{6}\,\conf{6}{s}{1} & \ding{51}   & \conf{4}{f}{6} & \ding{51} \\
Eu      & \conf{4}{f}{7}\,\conf{6}{s}{1} & \ding{51}   & \conf{4}{f}{7} & \ding{51} \\
Gd      & \conf{4}{f}{7}\,\conf{5}{d}{2} & \conf{4}{f}{7}\,\conf{5}{d}{1}\,\conf{6}{s}{1} & \conf{4}{f}{7}\,\conf{5}{d}{1} & \ding{51} \\
Tb      & \conf{4}{f}{9}\,\conf{6}{s}{1} & \ding{51}   & \conf{4}{f}{9} & \ding{51} \\
Dy      & \conf{4}{f}{10}\,\conf{6}{s}{1} & \ding{51}   & \conf{4}{f}{8}\,\conf{6}{s}{2} & \ding{51} \\
Ho      & \conf{4}{f}{11}\,\conf{6}{s}{1} & \ding{51}   & \conf{4}{f}{11}\,\conf{6}{s}{1} & \ding{51} \\
Er      & \conf{4}{f}{12}\,\conf{6}{s}{1} & \ding{51}   & \conf{4}{f}{12} & \ding{51} \\
Tm      & \conf{4}{f}{13}\,\conf{6}{s}{1} & \ding{51}   & \conf{4}{f}{13} & \ding{51} \\
Yb      & \conf{4}{f}{14}\,\conf{6}{s}{1} & \ding{51}   & \conf{4}{f}{14} & \ding{51} \\
	\end{tabular}
      \end{ruledtabular}

\end{table}

\subsection{Atomic Datasets}
\subsubsection{Lanthanum (Z=57)}
\ion{La}{iii}, and to a lesser extent \ion{La}{ii}, exhibit the most straightforward spectra among the lanthanides. Initial experimental investigations of \ion{La}{ii} were carried out in early studies \cite{Meggers_1927, Russel_Meggers_1932}, with the latter providing comprehensive energy level assignments. Zeeman spectroscopy was later employed to refine the $g$-factors of previously identified levels \cite[][]{1945JOSA...35..658H}. More recently, updated energy levels were measured via Fourier transform (FT) spectroscopy \cite[][]{2018JQSRT.211..188G}.

After a first analysis of the \ion{La}{iii} spectrum \cite[][]{1926PNAS...12..551G, 1931PPS....43...53B} and subsequent refinements \cite[][]{Russel_Meggers_1932, 1935CJRes..13A...1L, 1965JOSA...55.1283S}, a major re-evaluation was carried out in 1967, utilizing a sliding spark source to record lines from 2\,100 to 11\,000\,\AA\ \cite[][]{1967JOSA...57.1459O}.

In total, 115 and 41 energy levels arising from 13 and 23 configurations have been reported for \ion{La}{ii} and \ion{La}{iii}, respectively. The compilation by \citet[][]{1978aelr.rept.....M} revised nearly all known \ion{La}{ii} levels and associated configurations. For \ion{La}{iii}, energy levels reconstructed from DREAM transitions are in close agreement with values in the NIST database.

Given the comparatively simple electronic structure of lanthanum ions, analogous to caesium and barium, with one or two valence electrons, we constructed atomic models comprising 111 configurations (including single and double excitations up to {\conf{8}{s}{}, \conf{8}{p}{}, \conf{8}{d}{}, \conf{6}{f}{}}) for \ion{La}{ii}, and 23 configurations with single excitations up to {\conf{10}{s}{}, \conf{9}{p}{}, \conf{9}{d}{}, \conf{8}{f}{}, \conf{8}{g}{}} for \ion{La}{iii} (see table~\ref{tab:FAC_configs1}).

We successfully identified all levels of \ion{La}{iii} and 108 out of 115 levels of \ion{La}{ii}. The wide spacing of energy levels simplified level identification and reduced the dependence on detailed experimental cross-matching. Most level identifications were made using LS term assignments. The computed level structures are shown in figure~\ref{fig:LaII_levels} and figure~\ref{fig:LaIII_levels}. Median calibration corrections of $2763$\,cm$^{-1}$ and $821$\,cm$^{-1}$ were applied to \ion{La}{ii} and \ion{La}{iii}, respectively (see figures~\ref{fig:xmatch_accuracy_singly} and \ref{fig:xmatch_accuracy_doubly}).

\subsubsection{Cerium (Z=58)}

The classification of the \ion{Ce}{ii} spectrum was initiated in early work \cite[][]{1937PhRv...52.1209A, 1941JOSA...31..439H}, which identified approximately 300 levels along with their associated $g$-factors and configurations. Further measurements were carried out in the 1970s, leading to the cataloging of 192 odd- and 288 even-parity levels originating from five and seven configurations, respectively \cite[][]{Corliss_1973}.

The spectrum of \ion{Ce}{iii} was initially investigated in early studies \cite[][]{Bruin_1937, PhysRev.52.456}, which reported energy levels corresponding to seven configurations. A definitive identification of the ground-state configuration \conf{4}{f}{2} was achieved only later \cite[][]{1965JOSA...55.1283S}, along with an almost complete set of \ion{Ce}{iii} levels for 14 of the 17 experimentally confirmed configurations. Subsequent measurements completed the level structure \cite[][]{1972PhyS....6..139J}.

Both \ion{Ce}{ii} and \ion{Ce}{iii} are experimentally well characterized up to 50\,000\,cm$^{-1}$. The NIST ASD \cite[][]{NIST_ASD} includes 480 levels from 12 \ion{Ce}{ii} configurations and 214 levels from 17 \ion{Ce}{iii} configurations, each accompanied by configuration information. Additional \ion{Ce}{iii} levels are available in the DREAM database \cite{2020Atoms...8...18Q}. However, due to the lack of configuration assignments and the high level density near 50\,000\,cm$^{-1}$, these levels are not suitable for calibration purposes.

Our atomic structure models include 111 and 282 configurations for \ion{Ce}{ii} and \ion{Ce}{iii}, respectively, with single and double excitations from the ground state up to {\conf{8}{s}{}, \conf{8}{p}{}, \conf{8}{d}{}, \conf{6}{f}{}, \conf{5}{g}{}}. All configurations used are listed in table~\ref{tab:FAC_configs1}. The resulting energy levels, separated by parity, are shown in figure~\ref{fig:CeII_levels} and figure~\ref{fig:CeIII_levels}.

For \ion{Ce}{ii}, odd-parity levels match experimental data closely, but even-parity levels exhibit a systematic offset of approximately 8\,000\,cm$^{-1}$. Attempts to optimize the central potential to reduce this discrepancy while maintaining agreement for the odd levels were unsuccessful. In \ion{Ce}{iii}, the even-parity levels show better agreement, while the odd levels are subject to deviations both at low and high excitation. Nonetheless, the offset in \ion{Ce}{iii} odd-parity levels is less severe than that seen in \ion{Ce}{ii} even-parity levels. We successfully identified 441 \ion{Ce}{ii} and 209 \ion{Ce}{iii} levels. Calibration corrections of $6221$\,cm$^{-1}$ for \ion{Ce}{ii} and $1163$\,cm$^{-1}$ for \ion{Ce}{iii} were applied.

\subsubsection{Praseodymium (Z=59)}

Praseodymium is among the most extensively studied lanthanides. Early determinations of \ion{Pr}{ii} energy levels appeared in foundational work \cite[][]{1941PhRv...60..722R}, and were subsequently expanded using Fourier transform spectroscopy \cite[][]{1989PhyS...39..694G, 1989PhyS...39..710G}. These efforts were followed by re-measurements of 39 levels \cite[][]{2001PhyS...64..455I} and further additions obtained through laser-induced fluorescence techniques \cite[][]{2001EPJD...17..275F, 2005PhyS...72..300F, 2007ADNDT..93..127F}.

For \ion{Pr}{iii}, a total of 234 odd and 167 even levels have been identified \cite[][]{1963JOSA...53..831S, PMID:31929635}, with supplementary data available from subsequent studies \cite[][]{1968PhRv..174...89C, 1974_Sugar, 1974PhyS....9..325W}.

Experimental energy levels for \ion{Pr}{ii} and \ion{Pr}{iii} were compiled by \citet[][]{1978aelr.book.....M} and are included in the NIST ASD \citep{NIST_ASD}. We make use of 111 \ion{Pr}{ii} and 400 \ion{Pr}{iii} levels with identified electronic configurations (see table~\ref{tab:NIST_DREAM_available_data}). For comparative purposes, we also include levels reconstructed from transitions in the DREAM database \citep{2020Atoms...8...18Q}, although these lack configuration details and are therefore excluded from calibration.

Our atomic structure models are built from 27 and 17 configurations for \ion{Pr}{ii} and \ion{Pr}{iii}, respectively, including all configurations listed in the NIST ASD. We also include single and double excitations up to {\conf{7}{s}{}, \conf{7}{p}{}, \conf{7}{d}{}, \conf{5}{f}{}, \conf{5}{g}{}} yielding additional levels below the ionization potential, which are important for opacity.

The resulting theoretical energy levels for both parity classes are shown in figure~\ref{fig:PrII_levels} and figure~\ref{fig:PrIII_levels}. The agreement between our uncalibrated results and experimental data is generally good. We successfully identified 97 \ion{Pr}{ii} and 370 \ion{Pr}{iii} levels. Calibration corrections of $2145$\,cm$^{-1}$ for \ion{Pr}{ii} and $5227$\,cm$^{-1}$ for \ion{Pr}{iii} were applied (see figure~\ref{fig:xmatch_accuracy_singly} and figure~\ref{fig:xmatch_accuracy_doubly}).

\subsubsection{Neodymium (Z=60)}

The interpretation of the \ion{Nd}{ii} spectrum began with a comprehensive reference volume \cite[][]{1978aelr.book.....M}, which incorporated foundational measurements of 20 low-lying even levels and 57 upper odd levels \cite[][]{1942PhRv...61..167A}. Subsequent extensive experimental data were primarily derived from doctoral theses \cite[][]{Wyart_thesis_1968, Hoekstra_thesis_1969}, and were further supplemented by spectroscopic analyses \cite[][]{1970AcSpB..25..333B, 1971JOSA...61.1335B}, leading to the identification of approximately 500 odd and nearly 200 even levels. Although updates and corrections to these datasets have been incorporated into the NIST energy database, the most recent revision from 2002 remains unpublished \cite[][]{Blaise2002unpublished}.

For \ion{Nd}{iii}, the earlier compilation \cite[][]{1978aelr.book.....M} includes only 29 spectroscopic levels, based on unpublished work \cite[][]{Crosswhite1976unpublished} that drew on experimental investigations using grating spectroscopy of spark discharges \cite[][]{1961JOSA...51..820D, 1963ApOpt...2..675D}. These levels correspond to the even parity ground configuration \conf{4}{f}{4} and the odd parity excited configuration\conf{4}{f}{3}\,\conf{5}{d}{1}. An additional 11 levels from the same excited configuration were later identified \cite[][]{Aldenius_Thesis2001} and supported by stellar spectroscopy \cite[][]{2006A&A...456..329R}. More recently, a significantly extended set of 144 experimentally determined levels has been published \cite[][]{2024AA...684A.149D}, covering the \conf{4}{f}{4},\,\conf{4}{f}{3}\,\conf{5}{d}{1}, \conf{4}{f}{3}\,\conf{6}{s}{1}, and \conf{4}{f}{3}\,\conf{6}{p}{1} configurations, based on emission spectra from Penning and hollow cathode discharge lamps.

In this work, we include a comprehensive dataset comprising 744 \ion{Nd}{ii} levels, most accompanied by detailed electronic structure and LS term information. For \ion{Nd}{iii}, we incorporate the new experimental dataset from \citet{2024AA...684A.149D}, resulting in a total of 264 identified energy levels.

The \FAC\ calculations presented here follow an established methodology \cite[][]{2023MNRAS.524.3083F}, employing a set of 30 electronic configurations (table~\ref{tab:FAC_configs1}) that includes single-electron excitations up to a principal quantum number of $n=8$ and double excitations up to $n=6$. This configuration set has been demonstrated to achieve opacity convergence, as the inclusion of additional configurations neither increases the overall opacity nor improves the accuracy of energy levels most relevant for opacity calculations.

Using the newly published \ion{Nd}{iii} experimental data as a calibration basis, we identify 26 of the 29 (NIST) and 30 (DREAM) known levels within our calculations. The majority correspond to odd parity states, with only five even parity levels belonging to the \conf{4}{f}{4} $5I$ multiplet identified. The scarcity of high-lying even parity experimental levels introduces systematic uncertainties in our models, potentially resulting in deviations of several electron volts for these states.

\begin{figure*}
	\includegraphics[width=\linewidth]{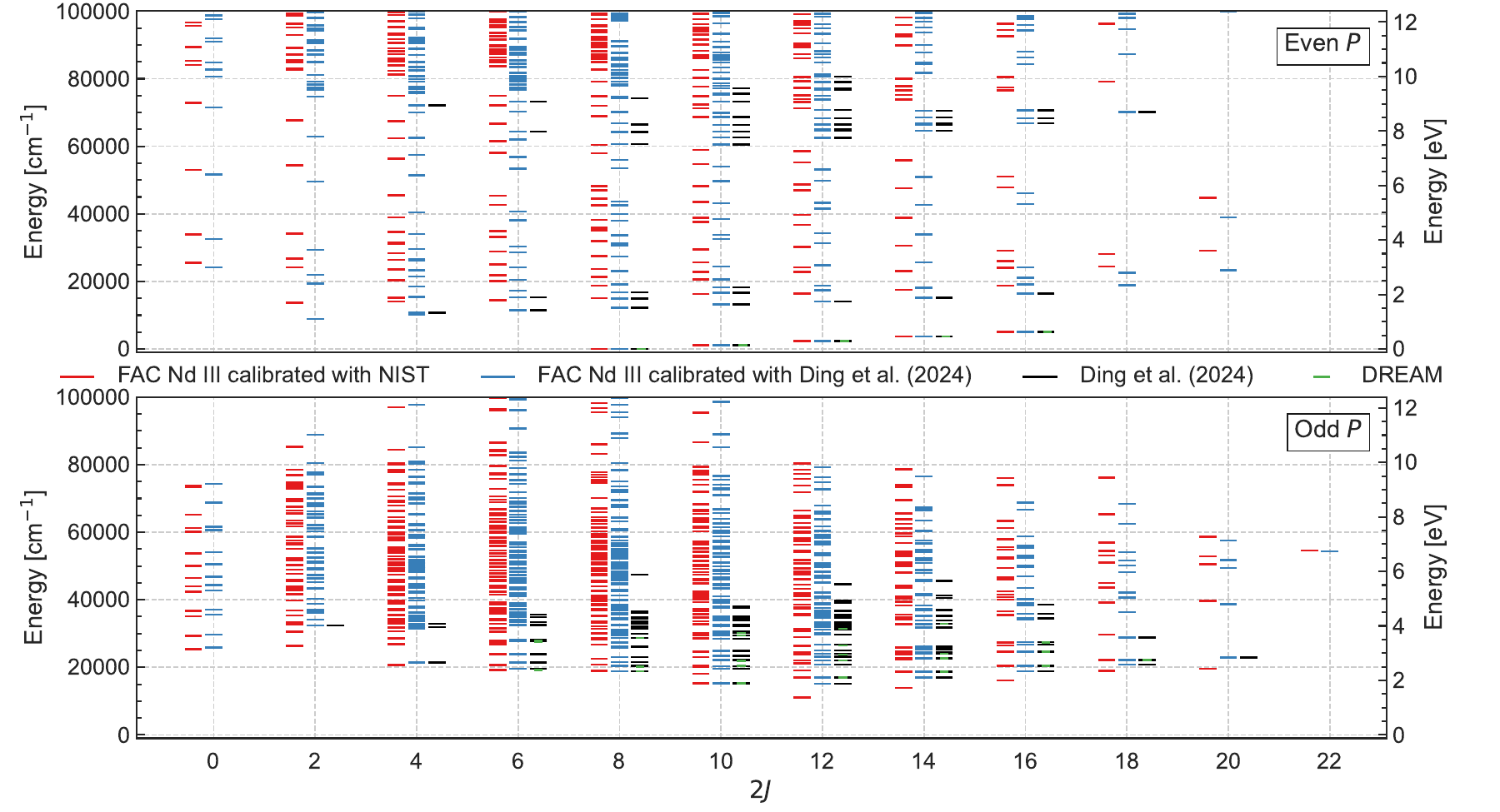}
	\caption{\justifying Comparison of the energy levels for \ion{Nd}{iii} for even (top) and odd (bottom) parity using the \FAC\ code. Red horizontal lines indicate the calculation, which was calibrated on experimental NIST/DREAM data, while blue lines indicate the level energies calibrated on the extensive dataset from \citet[][black horizontal bars]{2024AA...684A.149D}. Identification of higher lying states, for which the agreement with measurements is generally worse, leads to an increase in the applied correction (see figure~\ref{fig:xmatch_accuracy_doubly}).}  
	\label{fig:NdIII_levels_comparison} 
\end{figure*}

Including the dataset of \citet{2024AA...684A.149D} increases the number of identified \ion{Nd}{iii} levels to 231, enhancing the identification by an order of magnitude. figure~\ref{fig:NdIII_levels_comparison} compares \FAC\ calculations calibrated with NIST/DREAM data and those incorporating the new dataset. The updated calibration modifies energies by approximately 0.5\,eV near 20\,000\,cm$^{-1}$ and up to 2\,eV for even parity levels above 60\,000\,cm$^{-1}$. These changes affect both experimentally identified and unobserved levels (see Section~\ref{sec:calibration}). Given the sensitivity of opacity calculations to level energy accuracies, the impact of these data on opacity is discussed in Section~\ref{sec:opacities}.
We apply a calibration correction of $3801$\,cm$^{-1}$ to \ion{Nd}{ii} and $7785$\,cm$^{-1}$ to \ion{Nd}{iii}. The latter is a substantial increase relative to the correction based on NIST/DREAM alone ($3376$\,cm$^{-1}$), reflecting the poorer agreement at higher excitation energies absent in earlier data.

\subsubsection{Promethium (Z=61)}

The analysis of promethium spectra presents distinct experimental challenges due to the absence of stable isotopes. Initial observations of \ion{Pm}{ii} identified levels within the \conf{4}{f}{5}\,\conf{6}{s}{1} ground configuration \cite{1950PhRv...78..628W}, followed by an extended classification of transitions \cite{1958JOSA...48..389G}. These early efforts were later summarized in a comprehensive compilation \citet[][]{1978aelr.book.....M}. Subsequent work identified five additional energy levels associated with the \conf{4}{f}{5}\,\conf{5}{d}{1} configuration \cite[][]{1995JPhB...28.3615O}.

Experimental data for \ion{Pm}{iii} remain extremely limited. An initial study confirmed only the ground level \cite[][]{1971JOSA...61.1682M}, with further spectroscopic investigations hindered by the radioactive nature of promethium, which complicates the preparation of stable and sufficiently pure samples.

Our \FAC\ atomic structure models include seven electronic configurations for \ion{Pm}{ii} and fourteen for \ion{Pm}{iii}, as listed in table~\ref{tab:FAC_configs1}. All experimentally verified configurations are included, along with single and double excitations up to {\conf{7}{s}{}, \conf{7}{p}{}, \conf{7}{d}{}, \conf{6}{f}{}} for both ionization stages.

The computed energy levels are shown in figure~\ref{fig:PmII_levels} and figure~\ref{fig:PmIII_levels}. For \ion{Pm}{ii}, we identify 23 experimental levels, allowing the application of a calibration correction of $617$\,cm$^{-1}$. These correspond exclusively to odd-parity levels from the ground-state multiplet. No even-parity levels could be confidently matched due to the relatively high excitation energies of known experimental levels (20\,000 to 30\,000\,cm$^{-1}$) and the absence of spectroscopic term designations. The even-parity region in our \FAC\ calculations is highly congested, precluding unambiguous identification.

For \ion{Pm}{iii}, the absence of experimental reference levels beyond the ground state precludes meaningful calibration. Nevertheless, our calculations correctly recover the ground-state configurations \conf{4}{f}{5}\,\conf{6}{s}{1} for \ion{Pm}{ii} and \conf{4}{f}{5} for \ion{Pm}{iii}, in agreement with the NIST ASD \citep{NIST_ASD}.

\subsubsection{Samarium (Z=62)}

The spectrum of \ion{Sm}{ii} was initially explored in early work that reported 40 even-parity and over 200 odd-parity levels \cite[][]{1936ApJ....84...26A}. This dataset was later expanded through analysis of Zeeman splitting, leading to the identification of more than 70 even and 300 odd levels \cite[][]{1969AcSpB..24..405B}.

For \ion{Sm}{iii}, only 43 energy levels are currently available, derived primarily from spark emission spectroscopy \cite[][]{1967JOSA...57..867D}. A small number of additional \conf{4}{f}{6} $^5D$ levels have been determined in solid-state environments; however, these values may differ significantly from those of free ions, with discrepancies reaching several hundred cm$^{-1}$ \cite[e.g.][]{1967JOSA...57..867D, 1978aelr.book.....M}.

We use the dataset compiled by \citet[][]{1978aelr.book.....M} and hosted by the NASA ADS, which lists 376 and 43 levels for \ion{Sm}{ii} and \ion{Sm}{iii}, respectively. While all \ion{Sm}{iii} levels include configuration assignments, only 127 of the \ion{Sm}{ii} levels have associated data. Additionally, the experimental data for even-parity states of \ion{Sm}{iii} is sparse, with just eight levels identified across seven distinct $J$--$\pi$ groups. DREAM-derived levels are also included but duplicate those found in NIST ASD.

Our \FAC\ atomic structure models comprise seven configurations for \ion{Sm}{ii} and 35 for \ion{Sm}{iii}, including all experimentally confirmed configurations. We also include single and double excitations up to {\conf{6}{s}{}, \conf{6}{p}{}, \conf{5}{d}{}, \conf{4}{f}{}} for the singly ionized ion and {\conf{8}{s}{}, \conf{8}{p}{}, \conf{7}{d}{}, \conf{6}{f}{}} for the doubly ionized species (see table~\ref{tab:FAC_configs1}).

With the exception of even-parity levels in \ion{Sm}{ii}, our calculated level structure agrees well with the experimental datasets (see figure~\ref{fig:SmII_levels} and figure~\ref{fig:SmIII_levels}). In many cases, the deviation from experimental levels is less than 3000\,cm$^{-1}$. However, the even-parity levels of \ion{Sm}{ii} exhibit a systematic offset of approximately 12\,000\,cm$^{-1}$. We identify 122 levels in \ion{Sm}{ii}, with a calibration correction of $2540$\,cm$^{-1}$, and 39 levels in \ion{Sm}{iii}, requiring a correction of $1767$\,cm$^{-1}$.

\subsubsection{Europium (Z=63)}

The earliest investigations of the \ion{Eu}{ii} spectrum identified nine levels associated with the \conf{4}{f}{7}\,\conf{6}{s}{1}, \conf{4}{f}{7}\,\conf{5}{d}{1}, \conf{4}{f}{7}\,\conf{6}{p}{1} configurations \cite[][]{1934PhRv...45..499A}. This was followed by a more comprehensive study based on spark spectra and Zeeman measurements, which assigned LS terms to 156 levels from 1861 observed transitions \cite[][]{1941PhRv...60..641R}. Subsequent measurements provided isotope shifts and identified 13 additional levels \cite[][]{2013EPJST.222.2279F}. Some of the term assignments in the earlier compilation \cite[][]{1941PhRv...60..641R} remain tentative and are therefore excluded from the reference set adopted in this work.

The \ion{Eu}{iii} spectrum was first examined in detail through the identification of 105 energy levels \cite[][]{1974JOSA...64.1484S}. A more recent reanalysis of previously unidentified lines led to the addition of 39 levels, with energies spanning from 34\,816 to 86\,209\,cm$^{-1}$ \cite[][]{2008A&A...483..339W}.

A revised compilation by \citet[][]{1978aelr.book.....M} includes 141 \ion{Eu}{ii} levels from ten configurations and 105 \ion{Eu}{iii} levels from four configurations. Many identified levels lie above 30\,000\,cm$^{-1}$, where dense level spacing complicates assignment. Nevertheless, a sufficient number of low-lying \ion{Eu}{iii} and odd-parity \ion{Eu}{ii} levels are experimentally constrained.

Our atomic structure model for \ion{Eu}{ii} includes all ten confirmed configurations and incorporates single and double excitations up to {\conf{8}{s}{}, \conf{6}{p}{}, \conf{6}{d}{}, \conf{4}{f}{}}. For \ion{Eu}{iii}, we consider the \conf{4}{f}{7}, \conf{4}{f}{6}\,\conf{6}{s}{1}, \conf{4}{f}{6}\,\conf{6}{p}{1}, and \conf{4}{f}{7}\,\conf{5}{d}{1} configurations. The identified levels in our calculations match well with experimental data.

We apply calibration corrections of $2194$\,cm$^{-1}$ for \ion{Eu}{ii} and $2437$\,cm$^{-1}$ for \ion{Eu}{iii}. The large upper bound in the correction for \ion{Eu}{ii} (see fig.~\ref{fig:xmatch_accuracy_singly}) arises from significant offsets in levels involving the \conf{4}{f}{6}\,\conf{5}{d}{2}, \conf{4}{f}{6}\,\conf{5}{d}{1}\,\conf{6}{s}{1}, and \conf{4}{f}{7}\,\conf{6}{p}{1} configurations.

\subsubsection{Gadolinium (Z=64)}

The spectrum of \ion{Gd}{ii} was initially examined using Zeeman pattern analysis, leading to the identification of 9 odd and 11 even levels \cite[][]{1940PhRv...57..292A}. From over 2600 arc spectral lines, 137 energy levels were subsequently derived \cite[][]{1950JOSA...40..550R}, followed by the identification of an additional 178 levels through spark spectroscopy \cite[][]{1971AcSpB..26....1B}. High-energy states above 35\,000\,cm$^{-1}$ were later added to the dataset \cite[][]{1998AcSpB..53..633V}. A comprehensive compilation includes 164 odd and 150 even levels \cite[][]{1978aelr.book.....M}, with configuration assignments provided in related spectroscopic analyses \cite[][]{1970JOSA...60..763S}.

The \ion{Gd}{iii} spectrum was investigated using spark discharge spectroscopy in a series of early studies \cite[][]{1962PhDT........18C, 1963JOSA...53..695C, Kielkopf_thesis_1969}, although line blending among different ionization stages posed substantial challenges. A re-evaluation of these data, including previously unpublished results, yielded 25 confirmed energy levels \cite[][]{1978aelr.book.....M}.

In our \FAC\ atomic structure model, gadolinium is treated with a minimal configuration set to ensure a manageable CI space due to its complex electronic structure. This results in 46\,733 and 23\,053 computed energy levels for \ion{Gd}{ii} and \ion{Gd}{iii}, respectively. For calibration, we use 239 \ion{Gd}{ii} and 25 \ion{Gd}{iii} experimental levels with assigned configurations.

Despite the limited CI expansion, the agreement for \ion{Gd}{ii} is satisfactory. Our calculations recover a ground state configuration of \conf{4}{f}{7}\,\conf{5}{d}{1}\,\conf{6}{s}{1}, which differs from the experimental ground state \conf{4}{f}{7}\,\conf{5}{d}{2} by less than 2500\,cm$^{-1}$ (see table~\ref{tab:ground_states}). This discrepancy arises from the difficulty in balancing configuration mixing.

Calibration corrections of $3927$\,cm$^{-1}$ for \ion{Gd}{ii} and $3870$\,cm$^{-1}$ for \ion{Gd}{iii} are applied.

\subsubsection{Terbium (Z=65)}

The \ion{Tb}{ii} spectrum was initially characterized through hyperfine structure measurements and optical Zeeman spectroscopy, revealing transitions associated with two configurations \cite[][]{osti_4775737}. Due to its mid-shell electronic structure, terbium displays a dense level spectrum with numerous overlapping lines, generally of low to moderate intensity. The observed transitions were categorized into two systems: system A, involving odd-parity lower levels (primarily from the \conf{4}{f}{9}\,\conf{6}{s}{1} configuration), and system B, involving even-parity lower levels. At the time of the initial study, the ground-state configuration had not been definitively assigned to either system, and only relative energies were provided. Additional levels were later identified using similar experimental techniques \cite[][]{klinkenberg_1977}, with further revisions and compilations included in subsequent analyses \cite[][]{Blaise_1976}. Despite these efforts, the experimental characterization of \ion{Tb}{ii} remains incomplete, with only a limited number of low-lying states firmly established.

The \ion{Tb}{iii} spectrum was examined using hollow-cathode discharge spectroscopy \cite[][]{1972Phy....61..443M}. From over 2\,500 observed lines, a curated set of 485 lines was used to reconstruct 111 energy levels, which are considered complete up to approximately 20\,000\,cm$^{-1}$.

Our atomic structure models include all experimentally confirmed configurations (table~\ref{tab:FAC_configs1}). Due to the complexity of terbium's mid-lanthanide electronic structure, configuration interaction expansion is limited. Despite this, our \FAC\ calculations yield 40\,502 and 17\,440 levels for \ion{Tb}{ii} and \ion{Tb}{iii}, respectively.

Despite the high level density above 20\,000\,cm$^{-1}$, we identify 60 experimental levels in \ion{Tb}{ii} and 103 in \ion{Tb}{iii}. Calibration corrections of $1639$\,cm$^{-1}$ and $2845$\,cm$^{-1}$ are applied to \ion{Tb}{ii} and \ion{Tb}{iii}, respectively. 

\subsubsection{Dysprosium (Z=66)}

Early analyses of the \ion{Dy}{ii} spectrum identified several recurring energy differences \cite{1914ApJ....40..298P}, with further patterns revealed through the identification of six constant differences between spectral lines \cite{1942RvMP...14...96M}. Additional structural insights were provided through Zeeman-effect studies that attributed specific transitions to the \conf{4}{f}{10}\,\conf{6}{s}{1} configuration \cite{1952JOSA...42..117B}. Substantial progress in understanding the structure of \ion{Dy}{ii} was made through absorption spectra obtained from oxyacetylene flames \cite{1964AcSpB..20.1117M}. Zeeman structure analyses of 129 transitions further contributed to the dataset \cite{1970JOSA...60.1209V}, alongside the documentation of over 22\,000 optical and near-infrared lines \cite{1971JOSA...61..704C}. Building upon this extensive line list, additional levels were identified through new Zeeman measurements \cite{1972AcSpB..27..443W, 1972AcSpB..27..616W, 1976Phy....89..361W}, resulting in most of the currently known 563 energy levels. More recently, 24 previously unreported levels were revealed using Fourier transform spectroscopy \cite{2000PhyS...62..463N}.

Spectroscopic data for \ion{Dy}{iii} remain considerably limited. Initial investigations identified several energy levels, though a comprehensive analysis was not achieved \cite{1963JOSA...53..202C}. Additional levels were later documented through semi-empirical calculations \cite{1997JOSAB..14..511S}.

Our \FAC\ model incorporates 10 electronic configurations for \ion{Dy}{ii} and 9 for \ion{Dy}{iii} (see table~\ref{tab:FAC_configs1}). These include all experimentally verified configurations, with additional SD-excitations ensuring comprehensive coverage of relevant atomic transitions. We identified 341 experimental levels for \ion{Dy}{ii} within our calculations, with calibration corrections of $1618\,$cm$^{-1}$. For \ion{Dy}{iii} we use level information from \citet[][]{1997JOSAB..14..511S}, as the NIST ASD only contains information on the ground state. We identified a total of 106 experimental levels for \ion{Dy}{iii}. The calibration corrections amount to $2293\,$cm$^{-1}$.

\subsubsection{Holmium (Z=67)}

Initial spectroscopic investigations of singly ionized holmium focused on the splitting between the two lowest levels of the \conf{4}{f}{11}\,\conf{6}{s}{1} configuration, along with two excited levels, determined through hyperfine structure analysis \cite{1968JOSA...58.1519S}. This work was later expanded through the analysis of 547 spectral lines, leading to the identification of eight additional excited levels \cite{1971JOSA...61.1429L}. Further structural insights into the \conf{4}{f}{11}\,\conf{6}{p}{1} electronic configuration were obtained using 13 known energy levels, resulting in the discovery of four previously unreported levels \cite{1974JPhB....7.1111W}. A more extensive study employed ultraviolet Fourier transform spectroscopy to analyze 303 spectral lines and determine energy levels and hyperfine constants for 100 states, including 41 not documented in earlier datasets \cite{2009PhyS...79c5306G}.

The \ion{Ho}{iii} spectrum was initially analyzed through the identification of 42 energy levels, which included the \conf{4}{f}{11} ground configuration as well as levels associated with three additional known configurations \cite{1966PhDT........18M, 1967JOSA...57..870M}. The assignment of these levels was supported in part by theoretical predictions \cite{1965PhDT........85B}. Additional levels were later reported in unpublished work \cite{1973PhDT........17H}, and a more comprehensive analysis identified 80 of the currently known 120 levels \cite{1976PhyBC..85..386W}. Most of the newly characterized levels were attributed to the \conf{4}{f}{10}\,\conf{5}{d}{1} configuration, with further additions to the \conf{4}{f}{11} configuration including the $^4$F term.

Our \FAC\ atomic structure model includes 15 electronic configurations for \ion{Ho}{ii} and 15 for \ion{Ho}{iii}, which encompass all experimentally confirmed configurations plus SD-excitations as detailed in table~\ref{tab:FAC_configs1}. We successfully identified 42 levels for \ion{Ho}{ii} with calibration corrections of $1901\,$cm$^{-1}$, and 119 levels for \ion{Ho}{iii} with corrections of $3136\,$cm$^{-1}$.

\subsubsection{Erbium (Z=68)}
The spectroscopic analysis of \ion{Er}{ii} began with the identification of eight low-lying levels belonging to the \conf{4}{f}{12}\,\conf{6}{s}{1} configuration, based on Zeeman measurements and wavelength data \cite{1959JOSA...49..200M, 1958JOSA...48..542L, 1945SpecErd......G}. These levels were subsequently reanalyzed and grouped into four $J_1j$ terms according to their energy separations \cite{1962JOSA...52..504J}. Theoretical calculations employing electrostatic parameters provided further insight into the \conf{4}{f}{12}\,\conf{6}{s}{1} configuration \cite{1963ApJ...138..272C}, while additional spectroscopic data expanded the known set of low-lying levels \cite{1965JOSA...55..471M, 1970JOSA...60...94S, 1971ApJ...167..205S}. Infrared spectral observations later led to the identification of four levels within the \conf{4}{f}{12}\,\conf{5}{d}{1} configuration \cite{1971JOSA...61.1495S}. A more recent and comprehensive reassessment of the \ion{Er}{ii} spectrum combined theoretical modeling with new experimental measurements to assign 130 even-parity levels to the \conf{4}{f}{12}\,\{\conf{6}{s}{1}, \conf{5}{d}{1}\} and \conf{4}{f}{11}\,\{\conf{6}{s}{1}\,\conf{6}{p}{1}, \conf{5}{d}{1}\,\conf{6}{p}{1}\} configurations, and 230 odd-parity levels to the \conf{4}{f}{12}\,\conf{6}{p}{1} and \conf{4}{f}{11}\,\{\conf{5}{d}{1}\,\conf{6}{s}{1}, \conf{5}{d}{2}, \conf{6}{s}{2}\} configurations \cite{2009PhyS...79c5301W}.

Initial investigations of the \ion{Er}{iii} spectrum were based on early spectroscopic studies \cite{1961JOSA...51..337D, 1965PhDT........85B}. A more comprehensive analysis was subsequently carried out using spark discharge spectra \cite{1973JOSA...63..358S}. The set of known energy levels was later re-evaluated and expanded from 45 to 115 through the application of semi-empirical parametric methods \citep{1974PhyS....9..325W, 1974Phy....77..159W, 1980PhyS...22..583W, 1997PhyS...56..446W}.

In our calculations, we utilized 19 configurations for \ion{Er}{ii} and 31 for \ion{Er}{iii}, as detailed in table~\ref{tab:FAC_configs1}. We identified 153 experimental levels for \ion{Er}{ii} with calibration corrections of $2584\,$cm$^{-1}$, and 46 levels for \ion{Er}{iii} with corrections of $2644\,$cm$^{-1}$.

\subsubsection{Thulium (Z=69)}

Early spectroscopic investigations of \ion{Tm}{ii} identified four levels of the ground state using Zeeman effect measurements \cite{1942RvMP...14...96M}. Additional levels were observed through absorption spectroscopy in fuel-rich oxyacetylene flames \cite{1964AcSpB..20.1117M}, and further expanded using spectra obtained from electrode-less discharge tubes \cite{1965JPh....26..605B, 1969AcSpB..24..367C}. Level assignments to the \conf{4}{f}{13}\,\conf{6}{p}{1} and \conf{4}{f}{12}\,\{\conf{5}{d}{1}\,\conf{6}{s}{1}, \conf{6}{s}{2}, \conf{6}{s}{1}\,\conf{6}{p}{1}\} configurations were introduced in subsequent analyses \cite{1967JOSA...57.1358S}, and continued in later spectroscopic studies  \cite{1971PhyS....3..231C, 1974JPhB....7.1112W}. More recently, new odd- and even-parity levels were presented, along with revised JJ-values for several states using the Racah–Slater method \cite{2011CaJPh..89..451W}. The three lowest terms of the spectrum were also computed using the Configuration Interaction with Perturbation Theory (CIPT) method \cite{2020JPhB...53c5004L}.

For \ion{Tm}{iii}, spectroscopic data were initially compiled and subsequently expanded in successive works \cite{1969JOSA...59.1383S, 1970JOSA...60..454S}, with additional theoretical analysis contributing further refinements \cite{1971PhyS....4...53W}.

We obtained 361 levels of 9 configurations for \ion{Tm}{ii} and 120 levels from 4 configurations for \ion{Tm}{iii}. Adopting level energies from the DREAM database \citep{2020Atoms...8...18Q} yields fewer levels in the case of \ion{Tm}{ii}, whereas for \ion{Tm}{iii} all levels from the NIST ASD \citep[][]{NIST_ASD} are reconstructed. 

Our \FAC\ atomic structure model incorporates 25 configurations for \ion{Tm}{ii} and 15 for \ion{Tm}{iii}, including all experimentally confirmed configurations as detailed in table~\ref{tab:FAC_configs1} and correctly replicating the experimental ground levels. We identified 289 experimental levels for \ion{Tm}{ii} from 9 configurations with calibration corrections of $1789\,$cm$^{-1}$, and 114 levels for \ion{Tm}{iii} from 4 configurations with corrections of $2695\,$cm$^{-1}$.

\subsubsection{Ytterbium (Z=70)}

A comprehensive spectroscopic analysis of \ion{Yb}{ii} documented 315 levels spanning even- and odd-parity configurations, with level assignments informed by Zeeman pattern analysis \cite{1967JOSA...57..396M}. Subsequent work assigned six additional levels to the \conf{4}{f}{13}\,\conf{5}{d}{2} configuration \cite{1968JOSA...58..837S}. The emission spectrum was later re-examined, accompanied by theoretical investigations of configurations such as \conf{4}{f}{13}\,\conf{5}{d}\,\conf{6}{p}{1}, \conf{4}{f}{13}\,\conf{5}{d}{2}, and others not included in the NIST database \cite{1979JQSRT...21..115W}.

The \ion{Yb}{iii} spectrum was previously analyzed in foundational studies \cite[][]{Bryant1961, Bryant1965}, with one additional level of the \conf{4}{f}{13}\,\conf{5}{d}{1} configuration added in later work \cite{1970JOSA...60..571S}. Owing to the relative simplicity of its electronic structure, the spectrum of \ion{Yb}{iii} is comparatively well understood.

Our \FAC\ computations include 40 configurations for \ion{Yb}{ii} and 15 for \ion{Yb}{iii}, as detailed in table~\ref{tab:FAC_configs1}. We successfully identified 216 experimental levels from 29 configurations for \ion{Yb}{ii} with calibration corrections of $9047\,$cm$^{-1}$, and 46 levels from 6 configurations for \ion{Yb}{iii} with corrections of $4363\,$cm$^{-1}$. Both experimental ground state levels are correctly reproduced in our calculations.

\subsection{Results of the atomic structure calculations}
The results of our \FAC\ computations pertaining to singly and doubly ionized lanthanides are encapsulated in table~\ref{tab:NIST_DREAM_available_data}. Leveraging experimental data for the calibration procedure, we have identified 66\,591 E1 transitions with precise wavelengths (where both upper and lower levels are identified and calibrated) within our computed data. For only a single ion (\ion{Pm}{iii}), among the 28 ions investigated in this study, we were unable to identify any levels beyond the ground state; consequently, we cannot provide precise wavelengths for this ion. In summary, we identified 28\,690\,443 transitions below the ionization threshold, stemming from 146\,856 energy levels. The total calculated transitions with precise wavelengths are significantly fewer than those experimentally confirmed for certain ions, notably \ion{Ce}{ii}, \ion{Pr}{iii}, \ion{Dy}{iii}, \ion{Er}{iii}, and \ion{Yb}{ii}. In these cases, crucial information for identifying their upper and lower energy levels among the experimentally measured energy levels is absent. DREAM solely offers energies and angular momentum for the levels pertinent to the transition. Unless these levels are also available in the NIST ASD and details on the electronic configuration and/or the principal contribution to the spectroscopic term are available, we cannot use the upper and lower levels from DREAM transitions in our calibration procedure. Furthermore, due to significant mixing between configurations at energies beyond $\sim$30\,000 cm$^{-1}$, the identification of experimental levels within our calculations is infeasible in numerous instances.

We emphasize that the number of transitions with accurate wavelengths in our data, despite being lower than those available in the DREAM database, is still extremely important for future line identification in KNe. Experimentally confirmed transitions alone cannot be used in radiative transfer simulations, as the bulk of the opacity comes from lines without experimental information, requiring theoretical atomic structure calculations to fill the gaps. 

\begin{figure*}
	\includegraphics[width=0.97\textwidth]{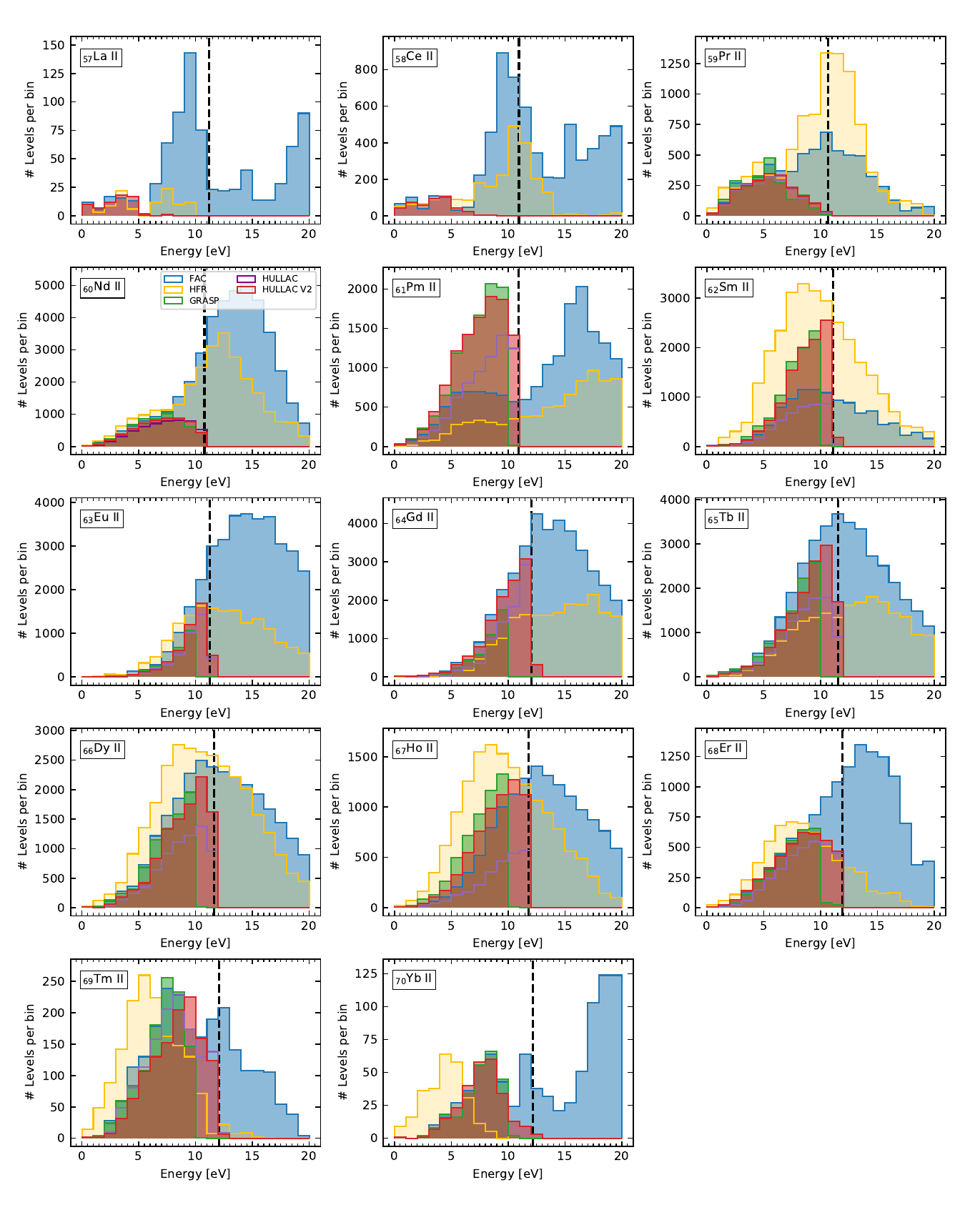}
	\caption{\justifying Distributions of energy levels for singly ionized lanthanide ions. Different colors represent datasets computed with various atomic structure codes: \FAC\ (this work; calibrated data) shown in blue, \HULLAC\ shown in purple for version V1 \cite{2020MNRAS.496.1369T} and red for version V2 \cite{2024MNRAS.535.2670K}, \HFR\ in yellow \cite[][]{2025A&A...696A..32D}, and \GRASP\ in green \cite[][]{2019ApJS..240...29G, 2020ApJS..248...17R, 2021ApJS..257...29R}. Vertical dashed lines indicate the ionization thresholds of the respective ions.
	}
	\label{fig:Level_densities_singly}
\end{figure*}

\begin{figure*}
	\includegraphics[width=0.97\textwidth]{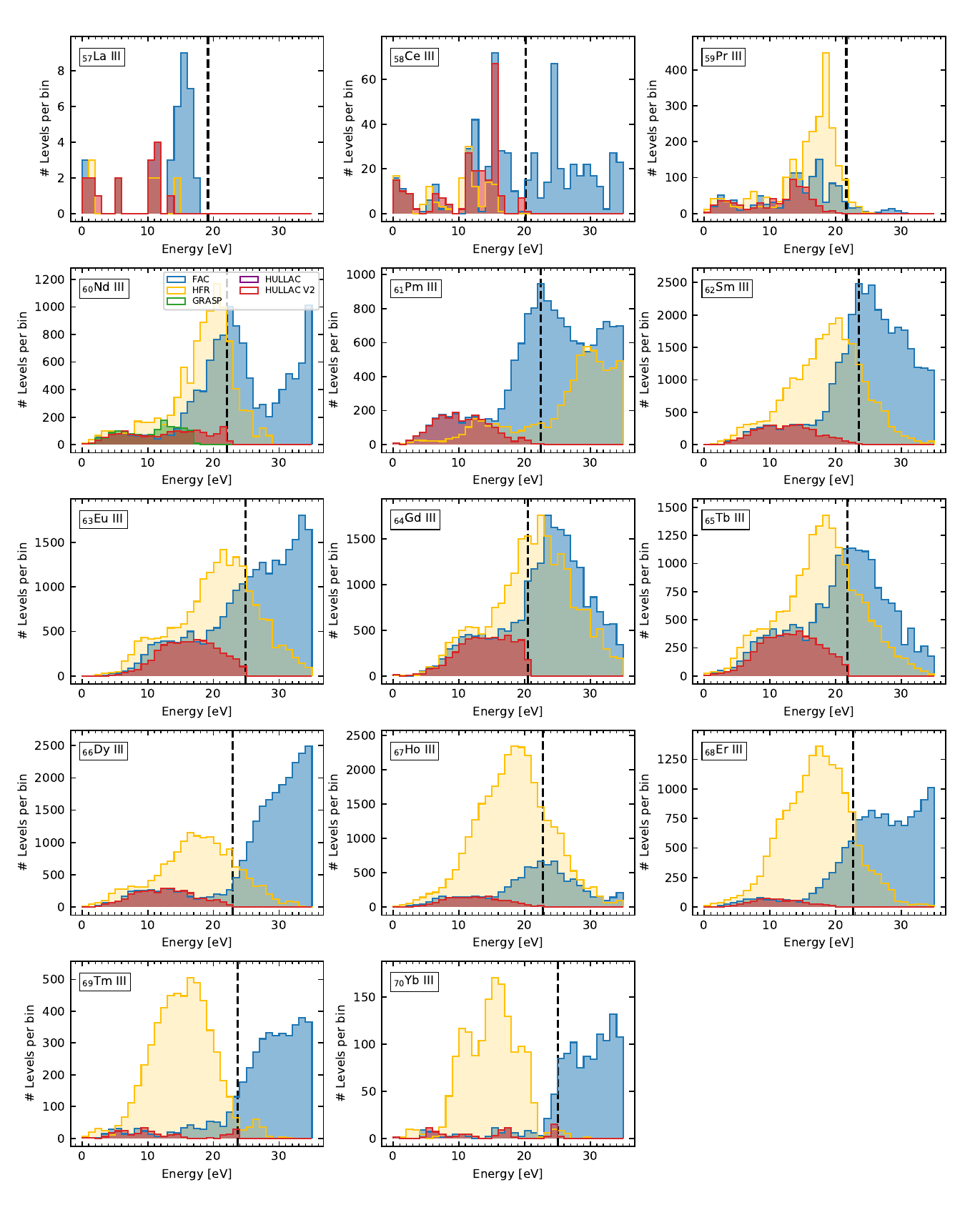}
	\caption{\justifying Distributions of energy levels for doubly ionized lanthanide ions. Different colors represent datasets computed with various atomic structure codes: \FAC\ (this work; calibrated data) shown in blue, \HULLAC\ shown in purple for version V1 \cite{2020MNRAS.496.1369T} and red for version V2 \cite{2024MNRAS.535.2670K}, \HFR\ in yellow \cite[][]{2025A&A...696A..32D}, and \GRASP\ in green \cite[][]{2019ApJS..240...29G}. Vertical dashed lines indicate the ionization thresholds of the respective ions.
	}
	\label{fig:Level_densities_doubly}
\end{figure*}

We show a comparison of all our calculated energy levels with published theoretical calculations using the \HULLAC\ (V1: \citep{2020MNRAS.496.1369T}; V2: \citep{2024MNRAS.535.2670K}), \HFR\ \citep{2025A&A...696A..32D} and \GRASP\ \citep[][]{2019ApJS..240...29G, 2020ApJS..248...17R, 2021ApJS..257...29R} codes in figure~\ref{fig:Level_densities_singly} and figure~\ref{fig:Level_densities_doubly} for singly and doubly ionized ions, respectively. 


For the first ionization stage, our results exhibit excellent agreement with the aforementioned published atomic structure datasets. Variations in the density of levels primarily stem from the calibration procedure applied to low-lying energy levels. The inclusion of additional electronic configurations beyond those considered in earlier studies \cite[][]{2024MNRAS.535.2670K, 2020ApJS..248...17R, 2021ApJS..257...29R} often results in a marked increase in the density of energy levels near the ionization threshold. This enhancement arises directly from the broader range of configuration state functions incorporated in the present \FAC\ calculations.

For \ion{Sm}{ii}, computational limitations prevented the inclusion of as many configurations as those employed in previous studies \cite[][]{2024MNRAS.535.2670K, 2020ApJS..248...17R}, leading to an underestimation of the level density above 7\,eV for this ion. In addition to bound levels, our models also include a substantial number of auto-ionizing states above the ionization potential. As illustrated in figure~\ref{fig:Level_densities_singly}, the overall distribution of energy levels tends to peak at excitation energies exceeding the ionization threshold.

For singly ionized species in the second half of the lanthanide series, moderate disagreement arises in comparisons with \HFR\ calculations. This is particularly evident for energy levels close to the ground state, where our \FAC\ levels have been calibrated against experimental data. In contrast, the \HFR\ results predict a significantly higher density of low-lying levels, which is inconsistent with existing experimental constraints. This artificial over-density of near-ground levels in \HFR\ models contributes to an increase in predicted opacity, as discussed in Section~\ref{sec:expansion_opacity}.

For doubly ionized lanthanides, we observe more pronounced discrepancies between our results and those from \HFR\ and \HULLAC. While \HULLAC\ calculations tend to show a decline in level density approaching the ionization threshold, both our calibrated \FAC\ results and the \HFR\ data exhibit a marked increase. In our case, the number of auto-ionizing levels surpasses that of bound states by roughly an order of magnitude, an outcome directly attributable to the broader configuration state function space included in the \FAC\ calculations.

A notable discrepancy emerges when comparing the present results with those reported in a recent study \cite[][]{2025A&A...696A..32D}, whose \HFR\ calculations exhibit a pronounced increase in the number of levels at intermediate excitation energies, beginning at approximately 15\,eV (see figure~\ref{fig:Level_densities_doubly}). While this elevated level density is consistent with \HULLAC\ and \FAC\ trends in the first half of the lanthanide series, it far exceeds the values for the second half, surpassing our bound level counts by more than an order of magnitude. For many ions in the latter group, where our \FAC\ level structure has been calibrated to experimental data and reproduces observed level densities, the \HFR\ results are in clear disagreement with empirical evidence. The implications of these discrepancies for radiative transfer and opacity calculations are further explored in Section~\ref{sec:expansion_opacity}.

\begin{table*}
\centering
	\caption{\justifying Summary of available energy levels and transitions from the NIST and DREAM databases for singly and doubly ionised lanthanides. Also shown are the number of configurations, levels, transitions, and calibrated transitions below the ionisation energy for each ion in the calibrated \FAC\ data set.}
	\label{tab:NIST_DREAM_available_data}
\begin{ruledtabular}
\begin{tabular}{lcccccccccc}
                      & \multicolumn{3}{c}{NIST}             & \multicolumn{2}{c}{DREAM} & Ferrara+24 & \multicolumn{4}{c}{FAC calibrated below E$_{\text{ion}}$} \\
                      \cline{2-4}\cline{5-6}\cline{7-7}\cline{8-11}
	    Ion           & \# config    & \# levels   & \# lines  & \# levels  & \# lines & \# lines & \# config  & \# levels & \# lines & \# lines calib.     \\  \hline
        \ion{La}{ii}  & 14 & 115 & 84  & --  & --    & -- & 44  & 472     & 17743       & 1239  \\
        \ion{La}{iii} & 23 & 41  & 0   & 40  & 131   & -- & 23  & 41      & 219           & 219     \\
        \ion{Ce}{ii}  & 12 & 480 & 283 & 501 & 15989 & -- & 40  & 2829  & 408639      & 14592 \\
        \ion{Ce}{iii} & 17 & 214 & 255 & 214 & 2935  & -- & 24  & 295     & 7083        & 3680  \\
        \ion{Pr}{ii}  & 9  & 192 & 0   & 58  & 144   & 157& 18  & 3689  & 571453      & 881     \\
        \ion{Pr}{iii} & 10 & 400 & 0   & 539 & 18331 & -- & 15  & 1105  & 63935       & 9436  \\
        \ion{Nd}{ii}  & 8  & 744 & 0   & 61  & 106   & 681& 27  & 9994  & 3336077   & 5434  \\
        \ion{Nd}{iii} \footnote{We used data from \cite{2024AA...684A.149D} for the calibration of Nd\,\textsc{iii} instead of data from DREAM/NIST ADS.} & 2  & 29  & 0   & 30  & 52    & -- & 21  & 4580  & 667955      & 3675      \\
        \ion{Pm}{ii}  & 3  & 171 & 0   & --  & --    & -- & 7   & 4990  & 913018      & 0       \\
        \ion{Pm}{iii} & 1  & 1   & 0   & --  & --    & -- & 14  & 5526  & 797829      & 0       \\
        \ion{Sm}{ii}  & 6  & 376 & 7   & 71  & 162   & 585& 6   & 6145  & 1111875   & 1106  \\
        \ion{Sm}{iii} & 2  & 43  & 0   & 39  & 81    & -- & 16  & 11177 & 1911972   & 95      \\
        \ion{Eu}{ii}  & 10 & 141 & 13  & --  & --    & 45 & 10  & 6781  & 888646      & 224     \\
        \ion{Eu}{iii} & 4  & 105 & 0   & 106 & 893   & -- & 11  & 8470  & 1308810   & 893     \\
        \ion{Gd}{ii}  & 8  & 314 & 0   & --  & --    & -- & 8   & 12394 & 3070318   & 2909  \\
        \ion{Gd}{iii} & 5  & 25  & 0   & 19  & 44    & -- & 10  & 6298  & 814002      & 85      \\
        \ion{Tb}{ii}  & 8  & 140 & 8   & --  & --    & -- & 8   & 16092 & 5245495   & 445     \\
        \ion{Tb}{iii} & 4  & 111 & 0   & 106 & 913   & -- & 9   & 8385  & 1335501   & 928     \\
        \ion{Dy}{ii}  & 9  & 563 & 17  & --  & --    & 275& 9   & 12493 & 3580925   & 4617  \\
        \ion{Dy}{iii}\footnote{A total of 106 experimental energy levels from 4 configurations of \citet[][]{1997JOSAB..14..511S} are used for the calibration of \ion{Dy}{iii} instead of data from the NIST ADS.} & 1  & 1   & 0   & 107 & 1304  & -- & 6   & 3910  & 355148      & 1246       \\
        \ion{Ho}{ii}  & 4  & 43  & 4   & --  & --    & -- & 8   & 5192  & 646913       & 106     \\
        \ion{Ho}{iii} & 4  & 121 & 0   & 122 & 1325  & -- & 14  & 4658  & 462551      & 1162  \\
        \ion{Er}{ii}  & 8  & 360 & 11  & 17  & 19    & 359& 16  & 5072  & 838067      & 983     \\
        \ion{Er}{iii} & 4  & 48  & 0   & 115 & 1308  & -- & 18  & 2837  & 195156      & 216     \\
        \ion{Tm}{ii}  & 9  & 361 & 13  & 297 & 7881  & -- & 17  & 1517  & 114318      & 7307  \\
        \ion{Tm}{iii} & 4  & 120 & 0   & 120 & 1479  & -- & 11  & 650     & 16361       & 1200  \\
        \ion{Yb}{ii}  & 29 & 340 & 10  & 269 & 6792  & -- & 34  & 348     & 8849        & 3711  \\
        \ion{Yb}{iii} & 6  & 47  & 0   & 53  & 271   & -- & 14  & 916     & 1585       & 202     \\
		\midrule
        total         &    &     &     &     &       &     & & 146856 & 28690443  & 66591 \\
	\end{tabular}
    \end{ruledtabular}
\end{table*}

\section{Opacities}
\label{sec:opacity}
\subsection{Oscillator Strengths}
\label{sec:oscillator_strengths}

In this section, we present electric dipole (E1) transitions, along with corresponding oscillator strengths and radiative lifetimes, for all 28 singly and doubly ionized lanthanide ions considered in this study. The transitions were computed in the Babushkin (length) gauge, and all relevant quantities for radiative transfer applications are provided in machine-readable format in table~\ref{tab:transitions_all} in the Appendix. The level indices used in the transition data correspond to those listed in table~\ref{tab:level_energies_all}. 

To complement the more general databases introduced in Section~\ref{sec:calculations}, we also benchmark our results against transition data obtained from high-resolution experiments reported in recent dedicated studies. Hollow cathode lamp spectroscopy was carried out using the Catania Astrophysical Observatory Spectropolarimeter \cite[][]{2024MNRAS.527.4440F}. In parallel, branching ratios were derived from emission spectra recorded with two different high-resolution spectrometers \cite[][]{2024ApJS..274....9D, 2025ApJS..278....7V}. A detailed overview of the experimental levels and transitions used for comparison with our calculated dataset is provided in table~\ref{tab:NIST_DREAM_available_data}.

\begin{figure*}
	\includegraphics[width=\textwidth]{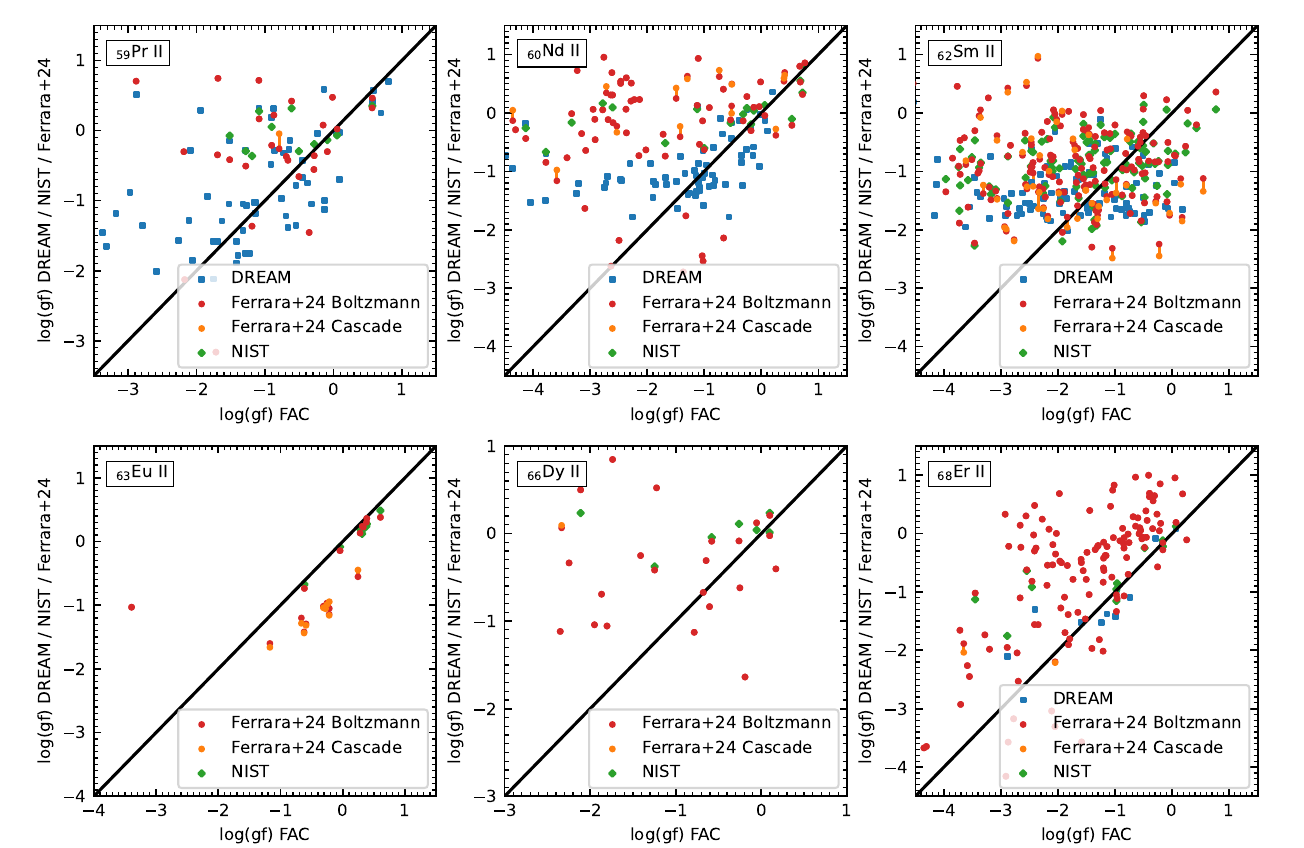}
	\caption{\justifying Oscillator strengths derived from experimental measurements by \citet[][]{2024MNRAS.527.4440F} and the NIST Atomic Spectra Database \citep{NIST_ASD}, as well as from theoretical calculations available in the DREAM database \citep{2020Atoms...8...18Q}, are compared with those obtained in this work using the \FAC\ code for transitions in which both upper and lower energy levels have been identified. In both the DREAM dataset and our \FAC\ calculations, energy levels were calibrated to experimental values, and oscillator strengths were computed based on these calibrated energies.}
	\label{fig:ferrara_transitions}
\end{figure*}

\begin{figure*}
	\includegraphics[width=0.65\linewidth]{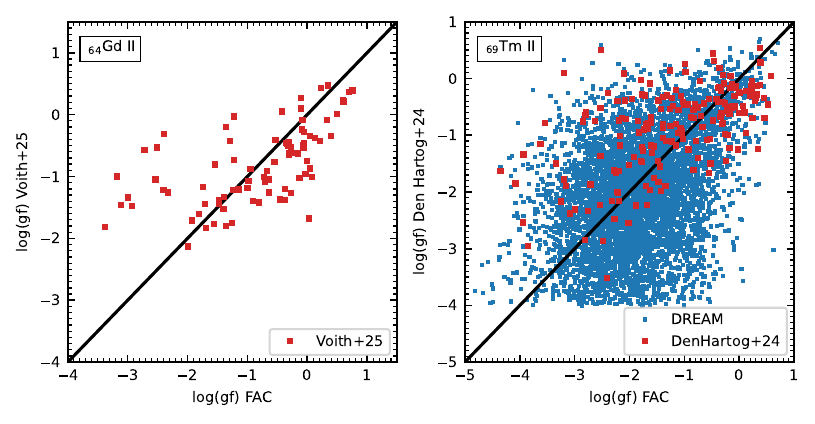}
	\caption{\justifying Oscillator strengths obtained from experimental measurements by \citet[][]{2025ApJS..278....7V} and  \citet[][]{2024ApJS..274....9D}, along with theoretical values from the DREAM database \citep{2020Atoms...8...18Q}, are compared with results from our \FAC\ calculations for transitions in which both the upper and lower energy levels have been identified. In both the DREAM dataset and our \FAC\ computations, energy levels were calibrated to experimental data, and oscillator strengths were derived using these calibrated values.}
	\label{fig:DenHartog_transitions}
\end{figure*}

Figure~\ref{fig:ferrara_transitions} compares transition data from the NIST Atomic Spectra Database and from recent measurements for six singly ionized lanthanide elements \cite[][]{2024MNRAS.527.4440F}, alongside atomic structure calculations performed in this work using the \FAC\ code. Similarly, figure~\ref{fig:DenHartog_transitions} presents a comparison of our results for \ion{Gd}{ii} and \ion{Tm}{ii} with experimental data obtained in high-resolution spectroscopic studies \cite[][]{2024ApJS..274....9D, 2025ApJS..278....7V}. In all cases, we include only those transitions for which both the upper and lower energy levels have been experimentally calibrated. Where available, theoretical oscillator strengths from the DREAM database are also included for comparison.

For strong transitions, defined here as those with $\log(gf) > -1$, we find that our \FAC\ calculations are generally consistent with experimental data, yielding a typical scatter of approximately 0.5 dex around the one-to-one relation $\log(gf)_{\mathrm{FAC}} = \log(gf)_{\mathrm{exp}}$. However, for weaker transitions with $\log(gf) < -1$, we observe a systematic discrepancy: the experimentally derived oscillator strengths are frequently higher than those predicted by our calculations. This behavior is partly expected, as experimental techniques are often limited to detecting transitions with at least moderate strength (typically down to $\log(gf) \sim -2$), inherently biasing the comparison against the full set of theoretically predicted weak lines. Nevertheless, one would anticipate a comparable number of transitions where experiments yield smaller oscillator strengths than theory, yet such cases appear under-represented. Most weak transitions in our \FAC\ dataset exhibit significantly suppressed $\log(gf)$ values relative to experimental findings.

This trend is not unique to our results. A similar pattern was noted in previous studies, such as figure~4 of \cite{2024ApJS..274....9D}, where experimental data were compared with \GRASP\ calculations \citep{2021ApJS..257...29R}. The fact that this discrepancy arises across different atomic structure codes -- \textit{ab initio} \GRASP\ and semi-empirical \FAC\ -- as well as across three distinct experimental setups (used in figures~\ref{fig:ferrara_transitions} and \ref{fig:DenHartog_transitions}), suggests that it is not attributable to a specific theoretical model or measurement technique. If such systematic underestimation of oscillator strengths by theoretical methods is widespread, the implications for KN opacity modeling and, particularly, mass estimations based on synthetic spectra, could be significant.

\begin{figure*}
	\includegraphics[width=\textwidth]{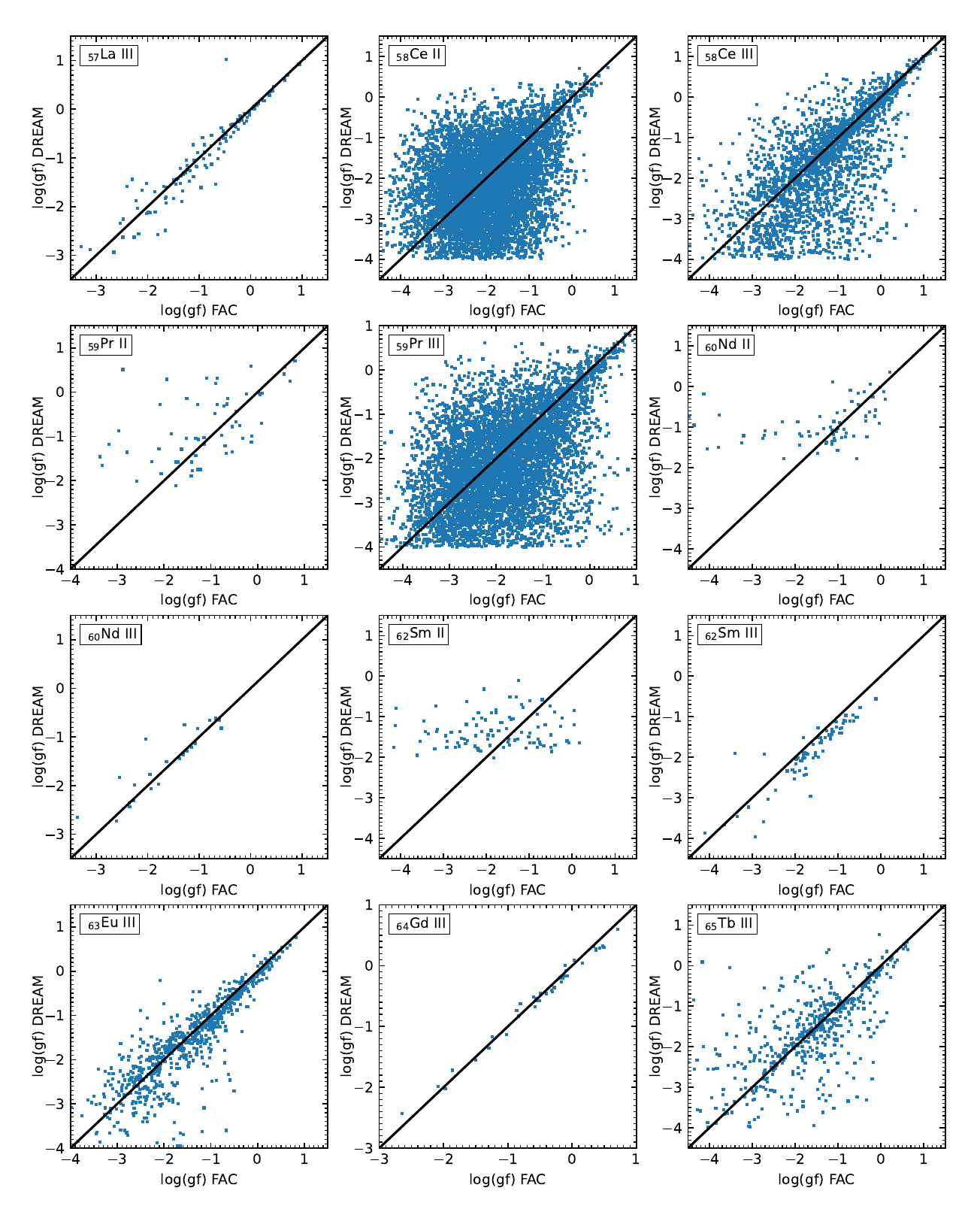}
	\caption{\justifying Same as figure~\ref{fig:ferrara_transitions} but for ions for which only oscillator strengths from the DREAM database \citep{2020Atoms...8...18Q} are available. Shown are ions from $_{57}$\ion{La}{iii} to $_{65}$\ion{Tb}{iii}.}
	\label{fig:dream_transitions_1}
\end{figure*}

\begin{figure*}
	\includegraphics[width=\textwidth]{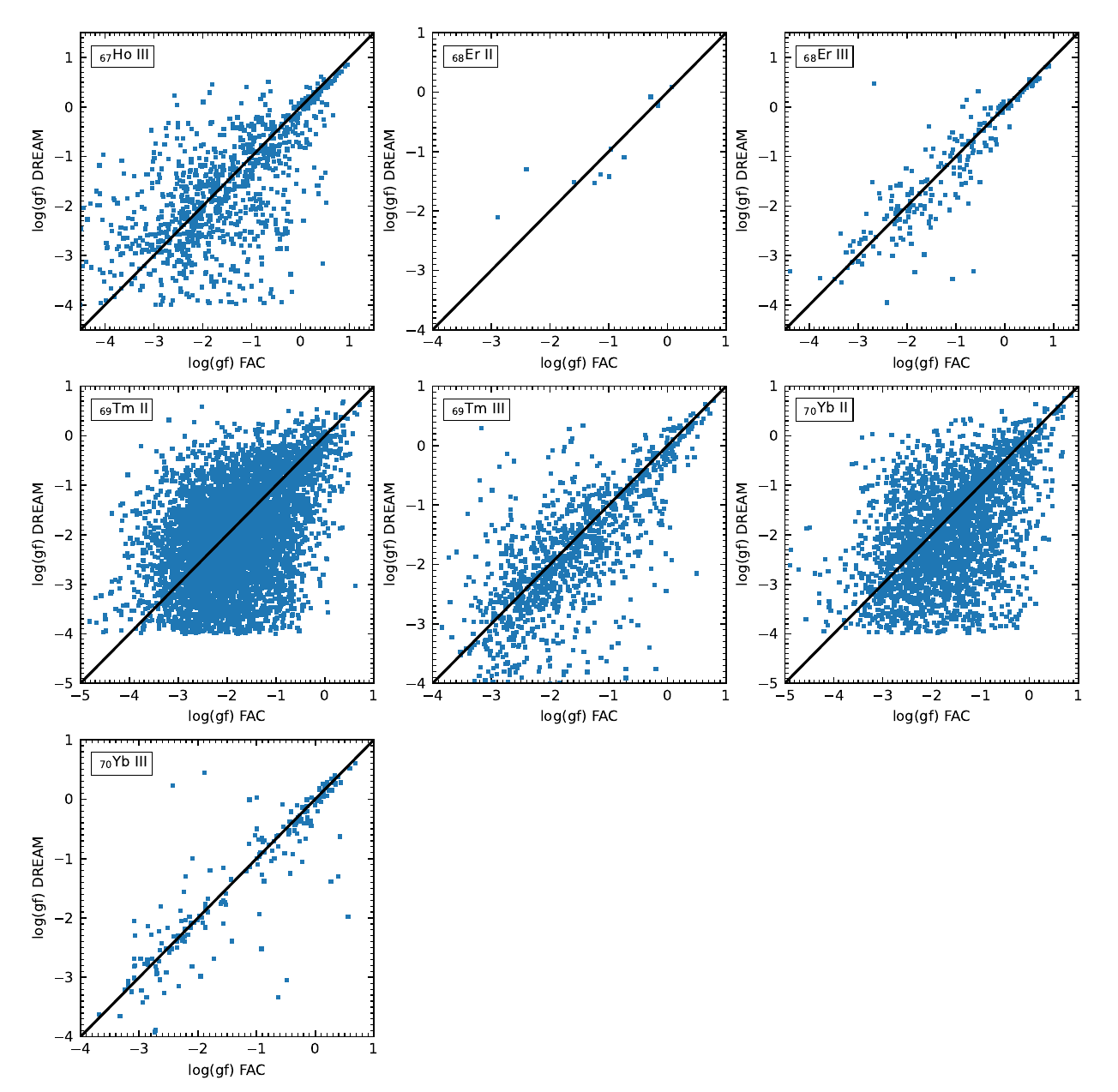}
	\caption{\justifying Same as figure~\ref{fig:ferrara_transitions} but for ions for which only oscillator strengths from the DREAM database \citep{2020Atoms...8...18Q} are available. Shown are ions from $_{67}$\ion{Ho}{iii} to $_{70}$\ion{Yb}{iii}.}
	\label{fig:dream_transitions_2}
\end{figure*}

In figures~\ref{fig:dream_transitions_1} and \ref{fig:dream_transitions_2}, we further compare our \FAC\ oscillator strengths with calibrated theoretical data from the DREAM database, which are based on \HFR\ calculations \citep{2020Atoms...8...18Q}. Although the large number of transitions (see table~\ref{tab:NIST_DREAM_available_data}) naturally results in considerable scatter, especially among weak lines with $\log(gf) < -1$, we generally find good agreement between the two theoretical datasets for most ions. Notable exceptions include \ion{Nd}{ii}, for which a similar trend to the experimental comparisons is observed: weak lines are substantially weaker in our \FAC\ calculations compared to the \HFR\ results. In contrast, \ion{Sm}{ii} exhibits a flat distribution of oscillator strengths within the range $\log(gf) \sim [-2, 0]$, in line with experimental observations.

For the remaining ions, \FAC\ and \HFR\ oscillator strengths are generally distributed around the one-to-one correspondence in figures~\ref{fig:dream_transitions_1} and \ref{fig:dream_transitions_2}. It is worth noting that the degree of scatter correlates with the number of experimentally known levels: ions with many experimentally calibrated levels tend to exhibit larger scatter due to the increased complexity of higher-lying electronic configurations, whereas ions with only a few low-lying measured states, typically involving simpler near-ground configurations, show tighter agreement.


\subsection{Bound-Bound Opacities}
\label{sec:opacities}
This study aims to deliver precise bound-bound opacities for conditions representative of kilonova ejecta, approximately a few days post-merger. The opacities presented here have been calculated purely for demonstration rather than serving as direct inputs for radiative transfer simulations. We offer a complete set of level energies and transitions for radiative transfer inputs instead of using pre-computed opacities, which ionization balance approximations may significantly influence in mono-atomic plasmas. 

In this work, we adopt the Sobolev formalism, which is widely used in transient radiative transfer. In the Sobolev optical depth
\begin{equation}
	\tau_{i,k} = \frac{\pi e^2}{m_e c } f_{k} n_{i,k} \lambda_{k} t .
	\label{equ:sobolev_optical_depth}
\end{equation}
$f_{k}$ is the oscillator strength of the line with transition wavelength $\lambda_{k}$ and the corresponding lower level number density $n_{i,k}$, where the indices $i$ and $k$ indicate ionization states and level numbers, respectively. The number densities of the lower levels can be computed by assuming LTE holds, using the Saha ionization
\begin{equation}
	\frac{n_{i}}{n_{i-1}} = \frac{Z_i(T)g_e}{Z_{i-1}(T)n_e}e^{-E_{\mathrm{ion}}/k_B T}
\end{equation}
and Boltzmann excitation equations
\begin{equation}
	n_{i,k} = \frac{g_{k}}{Z_i(T)}e^{-E_k/k_B T}n_i, 
	\label{equ:boltzmann_excitation}
\end{equation}
where $g_{k}$ is the multiplicity of the level $(2J_{k}+1)$, $E_k$ is the level excitation energy, $E_{\mathrm{ion}}$ are the ionization energies and $Z_i(T)$ are the partition functions
\begin{equation}
	Z_i(T) = \sum_k g_{k} e^{-E_k/k_B T}.
	\label{equ:partition_function}
\end{equation}

By applying the Sobolev optical depth, expansion opacities can be calculated 
\citep[refer to][]{1993ApJ...412..731E, 2013ApJ...774...25K, 2013ApJ...775..113T}
\begin{equation}
	\kappa_{\mathrm{exp}} (\lambda) = \frac{1}{ct\rho}\sum_{i,k} \frac{\lambda_k}{\Delta\lambda}\left(1-e^{-\tau_{i,k}}\right),
	\label{equ:expansion_opacity}
\end{equation}
where $t$ indicates the time after the merger, $\rho$ represents the ejecta density, $\Delta\lambda$ is the width of the bin, and $\lambda_l$ is the wavelength of the transition. 

The absolute scale of the opacity curves is unaffected by the choice of bin width, as demonstrated in prior studies \cite[][]{2020MNRAS.496.1369T, 2023MNRAS.524.3083F}. In the limit of very narrow bin widths and weak transitions, the expansion opacity reduces to the line-binned form \cite{2020MNRAS.493.4143F}. The combination of large velocity gradients typical of kilonova ejecta and the wavelength separation between prominent spectral lines fulfills the conditions required for the Sobolev approximation \cite{2020MNRAS.496.1369T}. For a general discussion of the Sobolev approximation in the context of expanding atmospheres, see \cite{1960mes..book.....S} and the treatment provided in the standard reference text \cite{1999isw..book.....L}.

In this work, we calculate all opacities presuming LTE conditions within the ejecta (see refs.~\cite{2022MNRAS.513.5174P, 2021MNRAS.506.5863H} for a discussion on non-LTE influences on kilonova opacities). We assume LTE is a suitable approximation in the initial days of the KN evolution. LTE opacities solely depend on the plasma state of the ejecta -- temperature, density and time post-merger -- as well as its composition (see equations~\ref{equ:sobolev_optical_depth} through \ref{equ:partition_function}). 

We adopt $T=5000$\,K, $t=1$\,day post-merger, and $\rho=10^{-13}$\,g\,cm$^{-3}$, in line with prior research on KN opacities \citep[see ][ among others]{2017Natur.551...80K, 2019ApJS..240...29G, 2020MNRAS.496.1369T, 2023MNRAS.524.3083F}. These values are based on simple blackbody fits to the observed 1.4-day spectrum of AT2017gfo and estimates of ejecta mass from light curve modeling. In LTE, temperatures do not differentiate between plasma and radiation. 

We resolve the ionization balance using ionization energies from the NIST ASD \citep{NIST_ASD}. In constructing the Saha equations, partition functions for all ion stages are ideally necessary, yet, given the physical conditions ($T<10000\,$K), only ions up to three times ionized are required. As calculated level energies for neutral and triply ionized species are not provided in this work, we source these values from the NIST ASD. While there is ample data for neutral species, triply ionized ions frequently lack excited states beyond the ground state. Thus, we caution that using our computed opacities near the transition to a triply ionized plasma might entail uncertainties in the ionization balance due to underestimated partition functions. For lanthanide ions, this transition is anticipated around $8000\,$K.

\begin{figure}
	\includegraphics[width=\linewidth]{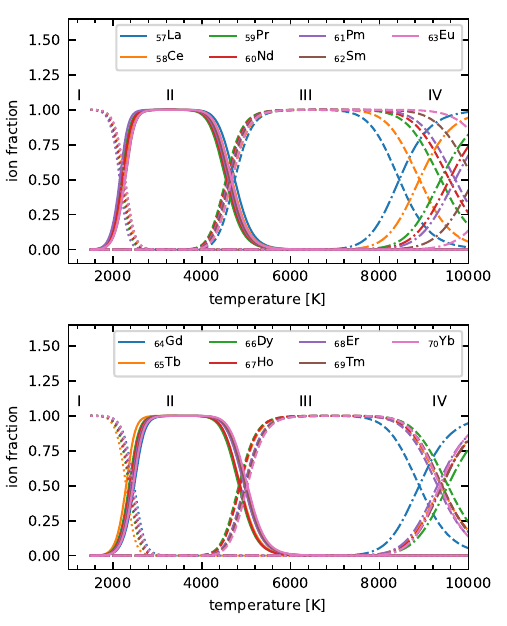}
	\caption{\justifying Ion fractions of neutral to triply ionized lanthanides as a function of temperature. Data of the singly (\textsc{ii}) and doubly (\textsc{iii}) ionized species was computed with the \FAC\ code. Data of the other two ionization stages (\textsc{i} and \textsc{iv}) was substituted from Kurucz' GFALL. Top panel: ions from $_{57}$La to $_{63}$Eu. Bottom panel: ions from $_{64}$Gd to $_{70}$Yb.}
	\label{fig:ionization_balance}
\end{figure}
In the temperature range inferred from continuum fitting of AT2017gfo ($2000$\,--\,$5000$\,K), only singly and doubly ionized species are expected to be significantly populated. For ions with available experimental data, uncertainties in partition functions are estimated to be below 5\,\%. Overall, the impact of these uncertainties on the Saha equilibrium is comparatively modest when set against the larger variations in level energies and transition strengths observed across different atomic datasets for lanthanides. Although highly excited states often show the greatest discrepancies between atomic structure models, their contribution to the ionization balance is negligible due to suppression by the exponential factor in Equation~\eqref{equ:partition_function}. Partition functions exert the strongest influence near the temperature regimes where the dominant ionization stage changes.

Figure~\ref{fig:ionization_balance} shows ion fractions for neutral through triply ionized species, based on our calibrated atomic data. Ionization transitions occur over relatively narrow temperature intervals, with most lanthanide species transitioning within $\sim$400\,K. An exception is the $\mathrm{III} \longleftrightarrow \mathrm{IV}$ transition, which is expected to span a broader temperature range. Missing atomic levels can lead to underestimates in partition functions and shift the temperature at which ionization transitions occur, depending on the ratio between the partition functions of successive ionization stages. It is important to note that the present analysis is valid only within the range $2\,500$\,–\,$8\,500$\,K, where singly and doubly ionized species are dominant.
\subsubsection{Lanthanide Expansion Opacity}
\label{sec:expansion_opacity}

\begin{figure*}
	\includegraphics[width=0.94\textwidth]{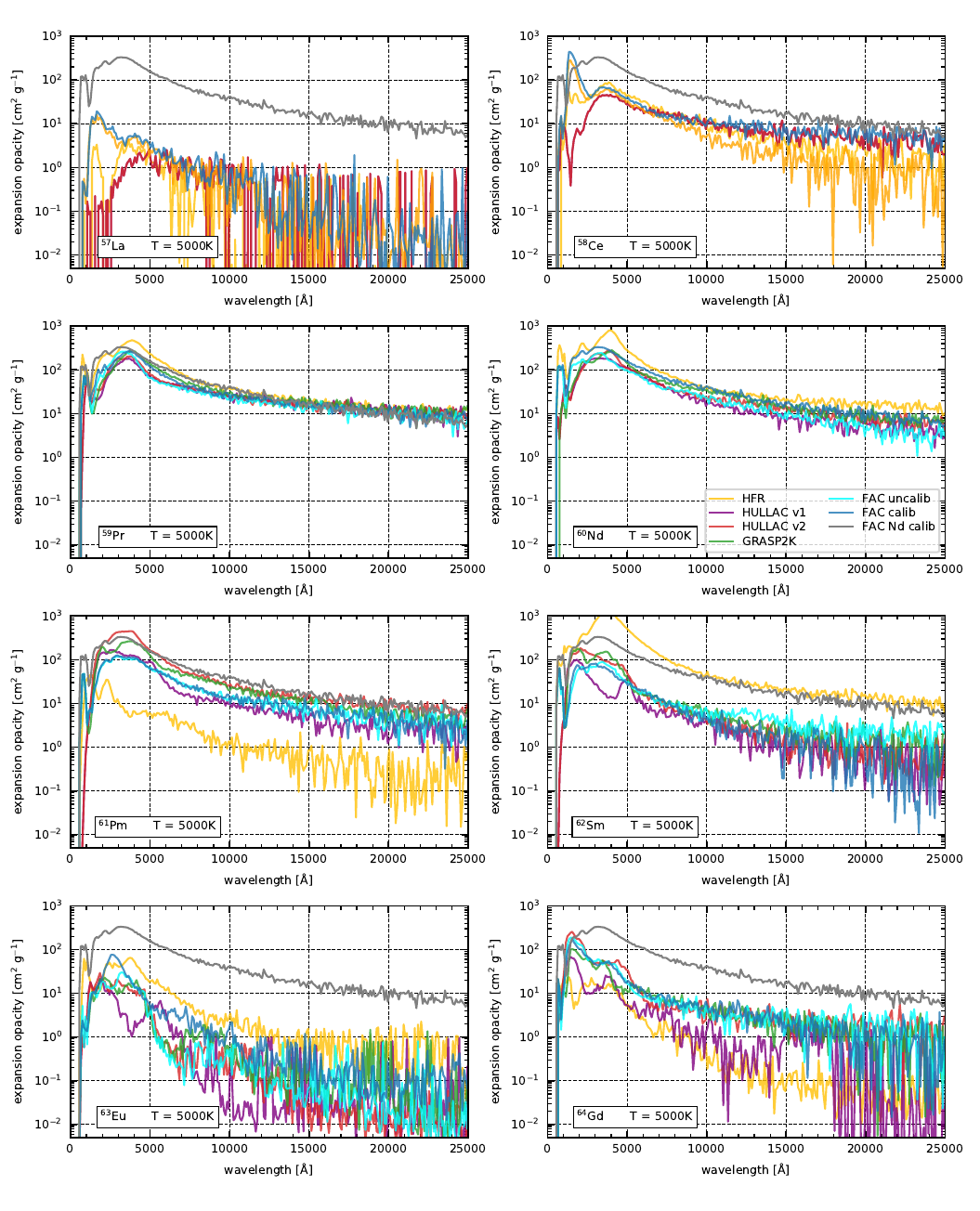}
	\caption{\justifying Expansion opacity results for ions from Z=57 (La) to Z=64 (Gd) are presented at $T=5000$\,K, $\rho=10^{-13}$\,g\,cm$^{-3}$ and $t=1$\,day, using a bin width of 100\,\AA. The comparison includes uncalibrated (orange) and calibrated (blue) \FAC\ calculations, as well as \HULLAC\ data from two versions: V1 (purple) \cite{2020MNRAS.496.1369T} and V2 (red) \cite{2024MNRAS.535.2670K}. Results from \GRASP\ calculations (green) are also shown, based on datasets from previous studies \cite{2019ApJS..240...29G, 2020ApJS..248...17R, 2021ApJS..257...29R}. For \GRASP, missing data for doubly ionized species were substituted with calibrated \FAC\ results. For ease of comparison across species, the expansion opacity of Nd derived from \FAC\ calculations is included in gray in all panels.}
	\label{fig:expansion_opacity_La-Gd}
\end{figure*}
\begin{figure*}
	\includegraphics[width=0.94\textwidth]{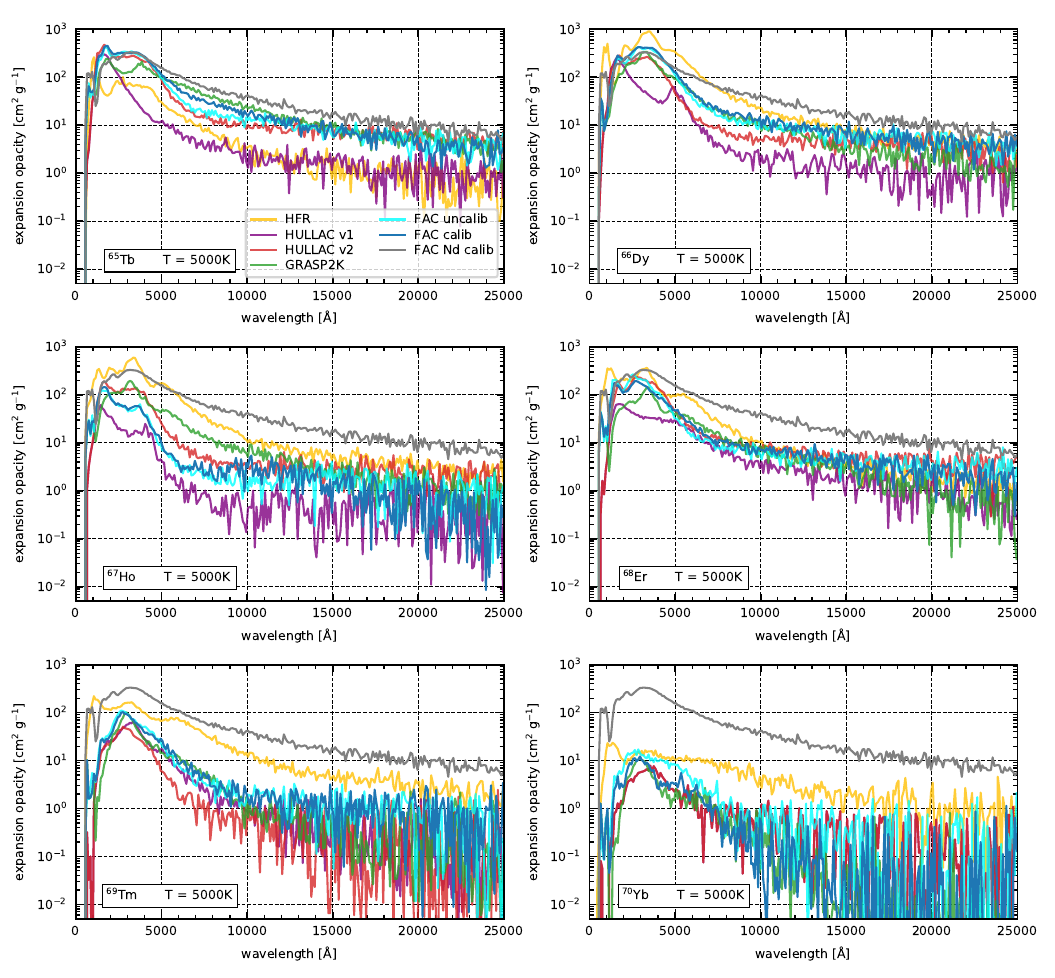}
	\caption{\justifying Expansion opacity results for ions from Z=65 (Tb) to Z=70 (Yb) are presented at $T=5000$\,K, $\rho=10^{-13}$\,g\,cm$^{-3}$ and $t=1$\,day, using a bin width of 100\,\AA. The comparison includes uncalibrated (orange) and calibrated (blue) \FAC\ calculations, as well as \HULLAC\ data from two versions: V1 (purple) \cite{2020MNRAS.496.1369T} and V2 (red) \cite{2024MNRAS.535.2670K}. Results from \GRASP\ calculations (green) are also shown, based on datasets from previous studies \cite{2019ApJS..240...29G, 2020ApJS..248...17R, 2021ApJS..257...29R}. For \GRASP, missing data for doubly ionized species were substituted with calibrated \FAC\ results. For ease of comparison across species, the expansion opacity of Nd derived from \FAC\ calculations is included in gray in all panels.}
	\label{fig:expansion_opacity_Tb-Yb}
\end{figure*}

Expansion opacities, computed using Equations~\ref{equ:boltzmann_excitation} through \ref{equ:expansion_opacity}, were evaluated for a gas with $T=5000$\,K and $\rho=10^{-13}$\,g\,cm$^{-3}$ at an epoch of $t=1$\,day. Under these conditions, lanthanides are partially ionized, with the dominant contributions to the opacity arising from singly and doubly ionized species, as shown in figure~\ref{fig:ionization_balance}. Comparative opacity results based on both uncalibrated and calibrated atomic structures are shown in Figs.~\ref{fig:expansion_opacity_La-Gd} and \ref{fig:expansion_opacity_Tb-Yb}, alongside data from previous studies \cite[][]{2019ApJS..240...29G, 2020MNRAS.496.1369T, 2020ApJS..248...17R, 2021ApJS..257...29R, 2024MNRAS.535.2670K, 2025A&A...696A..32D}.

In the majority of cases, the calibrated opacities obtained in this work are higher than the original V1 \HULLAC\ results reported in earlier work \cite[][]{2020MNRAS.496.1369T}. This increase is partly attributable to the inclusion of a more extensive set of configuration state functions (CSFs), which leads to a denser atomic-level structure. At typical ejecta temperatures during the first week following a kilonova (KN) event ($3000$\,--\,$10\,000$,K), only a limited number of low-lying energy levels are thermally populated due to Boltzmann suppression of higher-energy states. Consequently, the inclusion of additional atomic levels near the ionization threshold increases the number of available upper states for transitions originating from thermally populated low-lying levels. This leads to an enhanced absorption probability at short wavelengths (with $hc/\lambda\,\sim\,E_{\mathrm{ion}}$), thereby increasing the overall opacity in the ultraviolet and optical regimes. In contrast, atomic models that omit these near-threshold configurations underestimate the number of such transitions and thus yield systematically lower opacities in this spectral region. 

Although the inclusion of additional atomic states enhances the opacity at short wavelengths, the astrophysical impact of this increase remains limited. In KN ejecta, the emergent radiation at early times (e.g., $t=1$\,day post-merger) is predominantly emitted redward of $\sim$3000\,\AA\ \citep[][]{2022MNRAS.515..631G}. At shorter wavelengths, such as $\sim$1000\,\AA\ --- corresponding to photon energies near the ionization thresholds of singly ionized lanthanides ($\sim$10\,eV) --- the radiation field is intrinsically weak. Despite elevated opacities in this spectral region, the limited number of photons capable of interacting with these transitions means that their contribution to radiative transfer is minimal. Consequently, opacity enhancements blueward of $\sim$3000\,\AA\ are unlikely to exert a significant influence on the overall radiation transport in kilonova models.

The updated \HULLAC\ calculations exhibit substantially improved agreement with both the present \FAC\ results and the \GRASP-based datasets from earlier studies \cite[][]{2024MNRAS.535.2670K, 2019ApJS..240...29G, 2020ApJS..248...17R, 2021ApJS..257...29R}. In contrast, atomic data reported in a recent \HFR-based study yield significantly higher opacities for many lanthanide species \cite[][]{2025A&A...696A..32D}. This discrepancy is primarily attributable to elevated level densities at low and intermediate excitation energies, as illustrated in Figs.~\ref{fig:Level_densities_singly} and \ref{fig:Level_densities_doubly}. In particular, for the latter half of the singly ionized series and for most doubly ionized species, the \HFR-based level densities exceed those from \FAC, \HULLAC\ (V2), and \GRASP\ by a considerable margin, resulting in opacity enhancements of up to a factor of two.

Discrepancies in the red tail of the opacity between the \FAC, V2 \HULLAC, and \GRASP\ calculations shown in Figs.~\ref{fig:expansion_opacity_La-Gd} and \ref{fig:expansion_opacity_Tb-Yb} reflect differences in ground-state configurations, the density of low-lying levels, and the atomic structure methods employed. While previous efforts have focused on maximizing completeness in opacity datasets \cite[][]{2020MNRAS.496.1369T, 2024MNRAS.535.2670K}, the approach adopted in this work places additional emphasis on accurately reproducing low-lying energy levels, particularly those below 30\,000\,cm$^{-1}$. One notable exception is \ion{Gd}{ii}, where our calculated ground state differs from that reported in NIST ASD by approximately 2\,500\,cm$^{-1}$. In general, a higher level density near the ground state correlates with increased opacity due to the larger number of accessible transitions. While the long-wavelength tail of the opacity remains relatively consistent across the \FAC, \HULLAC, and \GRASP\ results, differences of up to an order of magnitude in peak opacities are observed. 

For ions lacking robust experimental constraints, expansion opacities are subject to larger uncertainties. This effect is especially pronounced for \ion{Pm}{iii}, where the \HULLAC-based opacities reported in recent work exceed those obtained from the present \FAC\ calculations by approximately a factor of two \cite[][]{2024MNRAS.535.2670K}. Such discrepancies are exacerbated at $T=5000$\,K, where doubly ionized species dominate for several elements, yet experimental data are sparse or entirely lacking.

\subsubsection{Planck Mean Opacity}

Line-by-line radiative transfer calculations generally provide the highest accuracy for modeling emergent spectra. However, gray opacities remain a widely used simplification in KN light curve models where radiative diffusion is the dominant transport mechanism. The Planck mean opacity is defined by:
\begin{equation}
	\kappa_{\mathrm{mean}} = \frac{\int_0^\infty B_\lambda(T)\kappa_\mathrm{exp}(\lambda)d\lambda}{\int_0^\infty B_\lambda(T)d\lambda},
\end{equation}
where $B_\lambda(T)$ is the Planck function at temperature $T$, and $\kappa_\mathrm{exp}(\lambda)$ is the expansion opacity as introduced in Equation~\eqref{equ:expansion_opacity}. The use of Planck mean opacities assumes the radiation field closely resembles a blackbody spectrum, a condition typically met during the early phases of KN evolution (as inferred from blackbody fits in \cite[][]{2021FrASS...7..108P, 2017Natur.551...75S, 2019Natur.574..497W, 2022MNRAS.515..631G}). However, this assumption has been challenged in recent work, e.g.~\cite{2023ApJ...954L..41S, 2023MNRAS.521.1858C}, which shows that in multi-dimensional, time-dependent simulations, such a blackbody radiation field may not be realized.

Assuming that the line-forming region can be approximated by a blackbody in the early epochs (up to approximately 5 days post-merger), Planck mean opacities remain a useful diagnostic for quantifying the bulk ejecta opacity. We compute Planck mean opacities at $T=5\,000$\,K for each lanthanide ion, based on the atomic structure datasets described in Section~\ref{sec:expansion_opacity}, assuming an ejecta density of $\rho=10^{-13}$\,g\,cm$^{-3}$ and $t=1$\,day. These results are calculated for individual species, and extending them to mixed compositions would require solving the Saha equation with the appropriate abundances. The Planck mean opacity peaks for most lanthanides between $4\,000$ and $5\,000$\,K.

\begin{figure}
	\includegraphics[width=\linewidth]{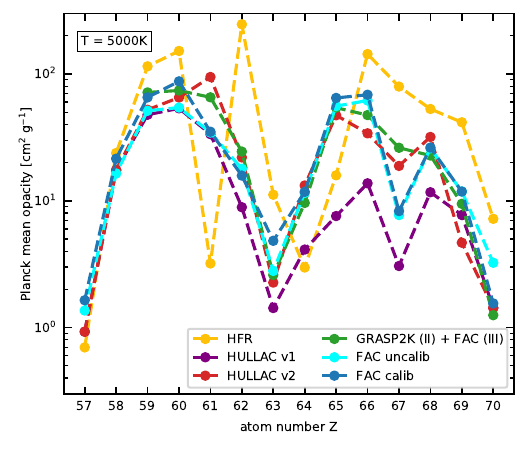}
	\caption{\justifying Planck mean opacities are shown for mono-atomic lanthanide plasmas at $\rho=10^{-13}$\,g\,cm$^{-3}$, evaluated one day after merger at a temperature of 5000\,K. The \FAC\ results from this work are presented in cyan (uncalibrated) and blue (calibrated). For comparison, Planck mean opacities based on previously published datasets are also included, comprising \HULLAC\ results from V1 (purple) and V2 (red) \cite{2020MNRAS.496.1369T, 2024MNRAS.535.2670K}, \HFR\ data (yellow) \cite{2025A&A...696A..32D}, and \GRASP\ calculations (green) \cite{2019ApJS..240...29G, 2020ApJS..248...17R, 2021ApJS..257...29R}.}
	\label{fig:planck_mean_opacity}
\end{figure}

Figure~\ref{fig:planck_mean_opacity} summarizes the Planck mean opacities at $T=5\,000$\,K across the lanthanide series. Among all species, neodymium ($Z=60$) exhibits the highest opacity. A double-peaked structure is evident, with minima corresponding to ions having either an empty, half-filled, or fully filled f-shell. These configurations reduce the number of low-lying levels and consequently the number of accessible transitions. While the low opacity of closed-shell ions is expected, the reduced opacity for half-filled shells is somewhat counterintuitive. As the $f$-subshell is filled, the number of low-energy levels (below 20\,000\,cm$^{-1}$) tends to decrease \citep{2017Natur.551...80K, 2020MNRAS.493.4143F, 2020MNRAS.496.1369T}, resulting in diminished opacity despite the high overall transition count.

The spread in Planck mean opacities among the lanthanides spans nearly two orders of magnitude. For instance, \ion{La}{ii}, \ion{Eu}{ii}, and \ion{Ho}{ii} show opacities around unity, whereas \ion{Pr}{ii}, \ion{Nd}{ii}, and \ion{Dy}{ii} range from 50 to 100\,cm$^2$\,g$^{-1}$. These calculations are based on single-species plasmas and do not reflect the exact contribution of each lanthanide to the total r-process opacity. Nevertheless, figure~\ref{fig:planck_mean_opacity} provides a useful baseline for identifying opacity-dominant species under simplified conditions.

\begin{figure}
	\includegraphics[width=\linewidth]{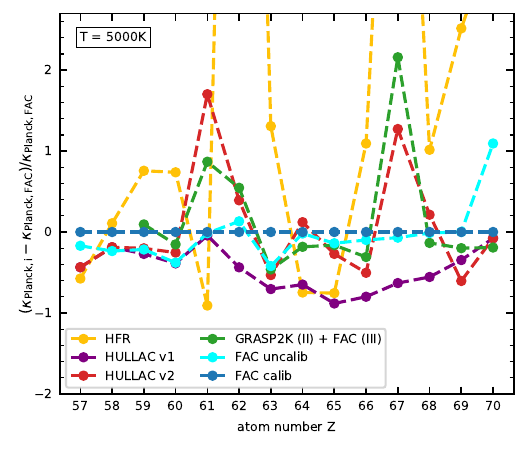}
	\caption{\justifying Planck mean opacities are shown for mono-atomic lanthanide plasmas at $\rho=10^{-13}$\,g\,cm$^{-3}$, evaluated one day after merger at a temperature of 5000\,K relative to the results of the calibrated \FAC\ opacities. The \FAC\ results from this work are presented in cyan (uncalibrated) and blue (calibrated). For comparison, Planck mean opacities based on previously published datasets are also included, comprising \HULLAC\ results from V1 (purple) and V2 (red) \cite{2020MNRAS.496.1369T, 2024MNRAS.535.2670K}, \HFR\ data (yellow) \cite{2025A&A...696A..32D}, and \GRASP\ calculations (green) \cite{2019ApJS..240...29G, 2020ApJS..248...17R, 2021ApJS..257...29R}.}
	\label{fig:planck_mean_opacity_relative}
\end{figure}

We observe good overall consistency between the various atomic structure codes, except for the \HULLAC\ V1 and \HFR\ datasets. Figure~\ref{fig:planck_mean_opacity_relative} quantifies the deviation of each dataset from the calibrated \FAC\ results. Both \HULLAC\ V2 and \GRASP\ agree with \FAC\ to within $20\,\%$. In contrast, \HULLAC\ V1 underestimates opacity in the central lanthanides, while the \HFR-based opacities for \ion{Pm}{ii}, \ion{Sm}{ii}, \ion{Gd}{ii}, \ion{Tb}{ii}, and \ion{Yb}{ii} exceed \FAC\ values by factors of $\sim7$\,--\,10. These differences originate from the significantly enhanced density of low- and intermediate-lying levels in the doubly ionized \HFR\ models.

\begin{figure}
	\includegraphics[width=\linewidth]{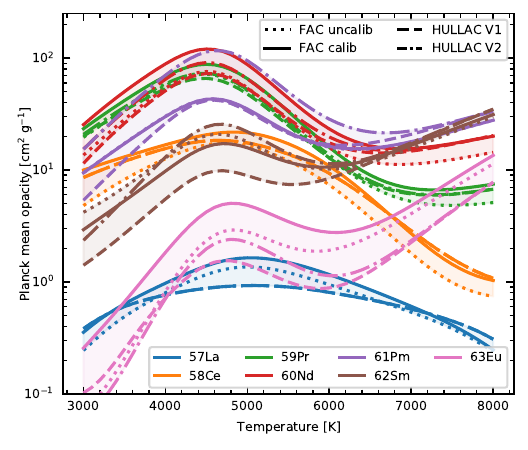}
	\caption{\justifying Planck mean opacities are presented for mono-atomic lanthanide plasmas (La to Eu) at a density of $\rho=10^{-13}$ g\,cm$^{-3}$ one day post-merger, as a function of plasma temperature. The results include both calibrated (solid lines) and uncalibrated (dotted lines) \FAC\ calculations from this work. For comparison, published \HULLAC\ datasets are also shown, with V1 represented by dashed lines \cite{2020MNRAS.496.1369T} and V2 by dash-dotted lines \cite{2024MNRAS.535.2670K}.}
	\label{fig:planck_mean_opacity_with_T_up_to_Eu}
\end{figure}

\begin{figure}
	\includegraphics[width=\linewidth]{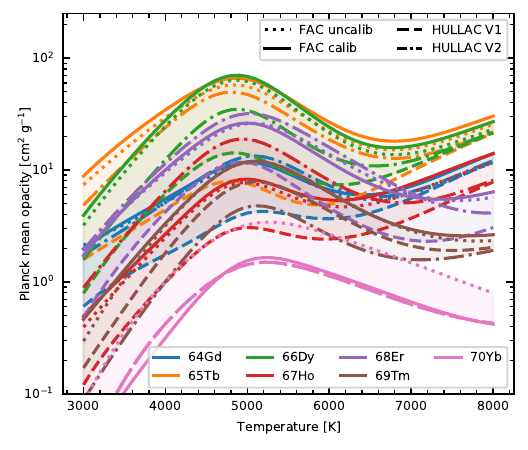}
	\caption{\justifying Planck mean opacities are presented for mono-atomic lanthanide plasmas (Gd to Yb) at a density of $\rho=10^{-13}$ g\,cm$^{-3}$ one day post-merger, as a function of plasma temperature. The results include both calibrated (solid lines) and uncalibrated (dotted lines) \FAC\ calculations from this work. For comparison, published \HULLAC\ datasets are also shown, with V1 represented by dashed lines \cite{2020MNRAS.496.1369T} and V2 by dash-dotted lines \cite{2024MNRAS.535.2670K}.}
	\label{fig:planck_mean_opacity_with_T_up_to_Yb}
\end{figure}

Figures~\ref{fig:planck_mean_opacity_with_T_up_to_Eu} and \ref{fig:planck_mean_opacity_with_T_up_to_Yb} display the temperature dependence of Planck mean opacities, calculated using both calibrated and uncalibrated data as well as previously published datasets. The decline of the Planck mean opacity at high temperatures results from the ionization shift from singly to doubly ionized species near $4500\,$K, which generally yield lower opacities. At lower temperatures ($\lesssim 4\,000$\,K), the reduction in opacity is a result of the Planck spectrum overlapping with the falling tail of the expansion opacity curve, which scales as $\lambda^{-1}$ \citep[][]{2022Atoms..10...18S}. Similarly, at high temperatures above $\sim\,6500\,$K, the convolution of the Planck spectrum and the expansion opacity yields an increased overlap, leading to a rise of the Planck mean opacity. As our calculations are restricted to up to doubly ionized species, we do not evaluate the Planck mean opacity above $T=8\,000$\,K, where contributions from triply ionized species become important (see Section~\ref{sec:opacities}).


\section{Summary and Conclusions}
\label{sec:summary}
In this study, we present comprehensive atomic structure calculations for the energy levels and E1 permitted transitions of all 28 singly and doubly ionized lanthanide ions ($_{57}$La to $_{70}$Yb), employing the \FAC\ code. We use an innovative optimization technique for the local central potential in \FAC~\cite{2025arXiv250213250F}, which significantly reduces discrepancies when compared to experimental atomic data. Furthermore, all experimentally confirmed energy levels available in the literature are utilized to calibrate our computed atomic structures and the corresponding transition energies. This calibration ensures improved accuracy in both level energies and radiative transition parameters. On average, we implemented corrections of 2950\,cm$^{-1}$ and 2571\,cm$^{-1}$ for singly and doubly ionized lanthanides, respectively. The corrections for singly ionized species are generally greater due to the wider availability of highly excited experimental levels, for which the agreement between theory and experiment is lower. For the ions considered in this study, we identified a total of 146\,856 energy levels below the ionization threshold, associated with 28\,690\,443 electric dipole transitions. Of these transitions, 66\,591 have both upper and lower levels precisely calibrated against experimental data, permitting accurate determination of transition wavelengths.

We present an extensive comparison of the atomic energy level structure and oscillator strengths between our calculations and extant experimental and theoretical data. We find very good agreement with the majority of previously published lanthanide atomic structure calculations on the energy level densities. For the oscillator strengths of electric dipole allowed transitions, a strong correlation is observed between our results and both experimental and theoretical values for strong transitions ($\log (gf)\,>\,-1$). Conversely, for weak transitions ($\log (gf)\,<\,-1$), the oscillator strengths calculated by us, as well as those from the semi-empirical calculations in the DREAM database using the \HFR\ code, are markedly weaker than the experimental data. 

We have also computed LTE opacities for our lanthanide ions by solving the Saha ionization and Boltzmann excitation equations. Our results show variations of $\sim$\,2 orders of magnitude in the expansion opacity across the ions in our sample. Differences in expansion opacity among atomic structure calculations from varying codes for the same ion are generally within a factor of 2. In numerous instances, we detect a significant increase in opacity at short wavelengths, attributed to the substantially greater number of electronic configurations compared to previous studies. The calibration procedure implemented to enhance transition accuracy exerts only a marginal influence on the opacity. Gray opacities align well with previous findings, effectively reproducing the previously identified double-peaked structure of the lanthanide group. Our results agree well with published data using the \GRASP\ and \HULLAC\ (V2) codes, but are in moderate disagreement with opacity results from \HFR. We show that this disagreement is mainly caused by a difference in energy level density at intermediate energies ($\sim$ halfway to the ionization threshold). 

While the atomic energy levels and permitted E1 transitions computed in this study suffice for LTE radiative transfer modeling, additional data are required to extend the modeling efforts to the nebular phase. We are currently computing forbidden M1 and E2 transitions between our calibrated energy levels, which are highly relevant for non-LTE radiative transfer, as well as electron-ion collision strengths required to obtain non-LTE occupation numbers. We will publish our non-LTE atomic data in a second paper. 

\begin{acknowledgments}
	AF and GMP acknowledge support by the European Research Council (ERC) under the European Union's Horizon 2020 research and innovation programme (ERC Advanced Grant KILONOVA No.~885281), the Deutsche Forschungsgemeinschaft (DFG, German Research Foundation) - Project-ID 279384907 - SFB 1245, and MA 4248/3-1. RFS acknowledges the support from National funding by FCT (Portugal), through the individual research grant 2022.10009.BD. RFS, JMS and JPM acknowledge support through project funding 2023.14470.PEX "Spectral Analysis and Radiative Data for Elemental Kilonovae Identification (SPARKLE)''\cite{SPARKLE2025}.
		
	We thank G.~Gaigalas, M.~Tanaka, D.~Kato J.~Deprince and C.~J.~Fontes for making some or all of their calculated atomic data public, facilitating benchmarks with other codes.
		
	We want to especially thank Yuri Ralchenko, Alexander Kramida, Joseph N. Tan, Karen Olsen and the whole NIST atomic spectroscopy group for their long-standing community work, without which this study would not have been possible.
		
	This research made use of Astropy, a community-developed core Python package for Astronomy \citep{2013A&A...558A..33A, 2018AJ....156..123A}, as well as numpy \citep{2011CSE....13b..22V}, scipy \citep{scipy}, pandas \citep{mckinney-proc-scipy-2010} and matplotlib \citep{2007CSE.....9...90H}.
\end{acknowledgments}

\begin{table}[hbt]
\centering
\scriptsize
	\caption{\justifying Energy levels (in cm$^{-1}$) relative to the ground state for all ions in this study. }
	\label{tab:level_energies_all}
    \begin{ruledtabular}
	\begin{tabular}{@{}rccrccccc}
        Index & Z\footnote{atomic number}   & Charge\footnote{ion charge} & Energy\footnote{energy in cm$^{-1}$}  & $J$\footnote{total angular momentum} & Parity & Conf. & term & method  \\
        \midrule
         0   &   57   &    1   &     0.00   &    2   &    0    &        5d2  &   3F  &  xmatch \\
         1   &   57   &    1   &  1016.10   &    3   &    0    &        5d2  &   3F  &  xmatch \\
         2   &   57   &    1   &  1394.46   &    2   &    0    &        5d2  &   1D  &  xmatch \\
         3   &   57   &    1   &  1895.15   &    1   &    0    &      5d.6s  &   3D  &  xmatch \\
         4   &   57   &    1   &  1970.70   &    4   &    0    &        5d2  &   3F  &  xmatch \\
         5   &   57   &    1   &  2591.60   &    2   &    0    &      5d.6s  &   3D  &  xmatch \\
         6   &   57   &    1   &  3250.35   &    3   &    0    &      5d.6s  &   3D  &  xmatch \\
         7   &   57   &    1   &  5249.70   &    0   &    0    &        5d2  &   3P  &  xmatch \\
         8   &   57   &    1   &  5718.12   &    1   &    0    &        5d2  &   3P  &  xmatch \\
         9   &   57   &    1   &  6227.42   &    2   &    0    &        5d2  &   3P  &  xmatch \\
        10   &   57   &    1   &  7394.57   &    0   &    0    &        6s2  &   1S  &  xmatch \\
    \end{tabular}
    \end{ruledtabular}
    \newline\\
    \noindent
    \justifying
    This table is available in its entirety in machine-readable form.\\
    Part of the data are shown here for guidance regarding its form and content. The last column shows the type of calibration of the energy level: 'xmatch' indicates that the levels were identified in experimental data, while 'shifted' refers to a calibration based on other identified levels in the P-J group.
\end{table}
\begin{table*}
\centering
\scriptsize

	\caption{\justifying Transitions between levels for all singly and doubly ionised lanthanides in this study.}
	\label{tab:transitions_all}
    \begin{ruledtabular}
	\begin{tabular}{@{}rccrcccccrrccccccrrrr}
          &        &\multicolumn{7}{c}{Lower Level} & \multicolumn{7}{c}{Upper Level} & \multicolumn{5}{c}{Transition information}\\ \cline{3-9} \cline{10-16}\cline{17-21}
        Z\footnote{atomic number} & C\footnote{ion charge} & \# & $E$\footnote{\label{fn:E}energy in cm$^{-1}$} & $J$\footnote{\label{fn:J}total angular momentum} & $P$\footnote{\label{fn:parity}parity} & Conf. & $LS$ & Method & \# & $E$\footnotemark[3] & $J$\footnotemark[4]  & $P$\footnotemark[5]  & Conf. & $LS$ & Method & Type & Energy\footnotemark[3] & $\lambda$\footnote{wavelength in \AA} & Log(gf) & A-value\footnote{values given as $A(B) = A\times10^B$}\\ \hline
   57   &    1   &     0    &    0.00   &   2   &   0     &    5d2   &    3F   &   xmatch  &  14 &  14147.98   &   2   &   1   &    4f.6s   &    1S   &   xmatch  &  E1  &   14147.98  &    7068.15   &    $-$1.0213  & 2.54(6) \\
   57   &    1   &     0    &    0.00   &   2   &   0     &    5d2   &    3F   &   xmatch  &  15 &  14375.17   &   3   &   1   &    4f.6s   &    3G   &   xmatch  &  E1  &   14375.17  &    6956.44   &    $-$1.9309  & 2.31(5) \\
   57   &    1   &     0    &    0.00   &   2   &   0     &    5d2   &    3F   &   xmatch  &  17 &  15773.77   &   3   &   1   &    4f.6s   &    3F   &   xmatch  &  E1  &   15773.77  &    6339.64   &    $-$3.6378  & 5.46(3) \\
   57   &    1   &     0    &    0.00   &   2   &   0     &    5d2   &    3F   &   xmatch  &  19 &  17211.93   &   2   &   1   &    4f.5d   &    3F   &   xmatch  &  E1  &   17211.93  &.   5809.92   &    $-$1.4129  & 1.53(6) \\
   57   &    1   &     0    &    0.00   &   2   &   0     &    5d2   &    3F   &   xmatch  &  21 &  18235.56   &   3   &   1   &    4f.5d   &    3D   &   xmatch  &  E1  &   18235.56  &    5483.79   &    $-$2.0328  & 2.94(5) \\
   57   &    1   &     0    &    0.00   &   2   &   0     &    5d2   &    3F   &   xmatch  &  23 &  18895.41   &   2   &   1   &    4f.5d   &    1D   &   xmatch  &  E1  &   18895.41  &    5292.29   &    $-$1.3232  & 2.26(6) \\
   57   &    1   &     0    &    0.00   &   2   &   0     &    5d2   &    3F   &   xmatch  &  26 &  20402.82   &   3   &   1   &    4f.5d   &    1F   &   xmatch  &  E1  &   20402.82  &    4901.28   &    $-$0.2445  & 2.26(7) \\
   57   &    1   &     0    &    0.00   &   2   &   0     &    5d2   &    3F   &   xmatch  &  28 &  21441.73   &   1   &   1   &    4f.5d   &    3D   &   xmatch  &  E1  &   21441.73  &    4663.80   &    $-$0.6398  & 2.34(7) \\
   57   &    1   &     0    &    0.00   &   2   &   0     &    5d2   &    3F   &   xmatch  &  29 &  22106.02   &   2   &   1   &    4f.5d   &    3D   &   xmatch  &  E1  &   22106.02  &    4523.65   &    $-$1.3769  & 2.74(6) \\
   57   &    1   &     0    &    0.00   &   2   &   0     &    5d2   &    3F   &   xmatch  &  31 &  22537.30   &   3   &   1   &    4f.5d   &    3D   &   xmatch  &  E1  &   22537.30  &    4437.09.  &    $-$1.4071  & 1.90(6) \\
   57   &    1   &     0    &    0.00   &   2   &   0     &    5d2   &    3F   &   xmatch  &  33 &  22705.15   &   1   &   1   &    4f.5d   &    1D   &   xmatch  &  E1  &   22705.15  &    4404.29   &    $-$1.6800  & 2.40(6) \\
   57   &    1   &     0    &    0.00   &   2   &   0     &    5d2   &    3F   &   xmatch  &  34 &  23246.93   &   2   &   1   &    4f.5d   &    3D   &   xmatch  &  E1  &   23246.93  &    4301.64.  &    $-$1.7556  & 1.27(6) \\
   57   &    1   &     0    &    0.00   &   2   &   0     &    5d2   &    3F   &   xmatch  &  35 &  24462.66   &   2   &   1   &    5d.6p   &    3F   &   xmatch  &  E1  &   24462.66  &    4087.86   &    $-$0.1804  & 5.27(7) \\
   57   &    1   &     0    &    0.00   &   2   &   0     &    5d2   &    3F   &   xmatch  &  36 &  24522.70   &   3   &   1   &    4f.5d   &    3P   &   xmatch  &  E1  &   24522.70  &    4077.85   &    $-$1.7765  & 9.59(5) \\
   57   &    1   &     0    &    0.00   &   2   &   0     &    5d2   &    3F   &   xmatch  &  37 &  25973.37   &   1   &   1   &    5d.6p   &    3S   &   xmatch  &  E1  &   25973.37  &    3850.10   &    $-$0.2697  & 8.06(7) \\
   57   &    1   &     0    &    0.00   &   2   &   0     &    5d2   &    3F   &   xmatch  &  38 &  26414.01   &   2   &   1   &    5d.6p   &    3D   &   xmatch  &  E1  &   26414.01  &    3785.87   &    $-$0.8328  & 1.37(7) \\
   57   &    1   &     0    &    0.00   &   2   &   0     &    5d2   &    3F   &   xmatch  &  39 &  26837.66   &   3   &   1   &    5d.6p   &    3F   &   xmatch  &  E1  &   26837.66  &    3726.11   &    $-$1.3915  & 2.79(6) \\
   57   &    1   &     0    &    0.00   &   2   &   0     &    5d2   &    3F   &   xmatch  &  40 &  27388.11   &   2   &   1   &    5d.6p   &    3G   &   xmatch  &  E1  &   27388.11  &    3651.22   &    $-$1.0682  & 8.55(6) \\
   57   &    1   &     0    &    0.00   &   2   &   0     &    5d2   &    3F   &   xmatch  &  42 &  28154.55   &   1   &   1   &    6s.6p   &    3P   &   xmatch  &  E1  &   28154.55  &    3551.82   &    $-$2.7414  & 3.20(5) \\
   57   &    1   &     0    &    0.00   &   2   &   0     &    5d2   &    3F   &   xmatch  &  43 &  28315.25   &   3   &   1   &    5d.6p   &    1G   &   xmatch  &  E1  &   28315.25  &    3531.67   &    $-$2.5111  & 2.36(5) \\
	\end{tabular}
    \end{ruledtabular}
    \newline\\
    \noindent
    \justifying
    This table is available in its entirety in machine-readable form. Part of the data are shown here for guidance regarding its form and content.
\end{table*}
\appendix

\section{Data Availability}

The complete dataset underlying this study is publicly available in machine-readable format on Zenodo (\href{https://zenodo.org/records/15835361}{https://zenodo.org/records/15835361}). It includes all calculated and calibrated energy levels, radiative transition data, and configuration information for singly and doubly ionized lanthanide ions as described in the text. This dataset is intended to support reproducibility and facilitate future applications in radiative transfer modeling, spectroscopic analysis, and atomic physics research. Tables~\ref{tab:level_energies_all} and \ref{tab:transitions_all} provide representative excerpts of the published data, illustrating the structure and format included in the full release.

\section{Energy Levels of Calculated Ions}

\begin{figure*}
	\includegraphics[width=0.99\linewidth]{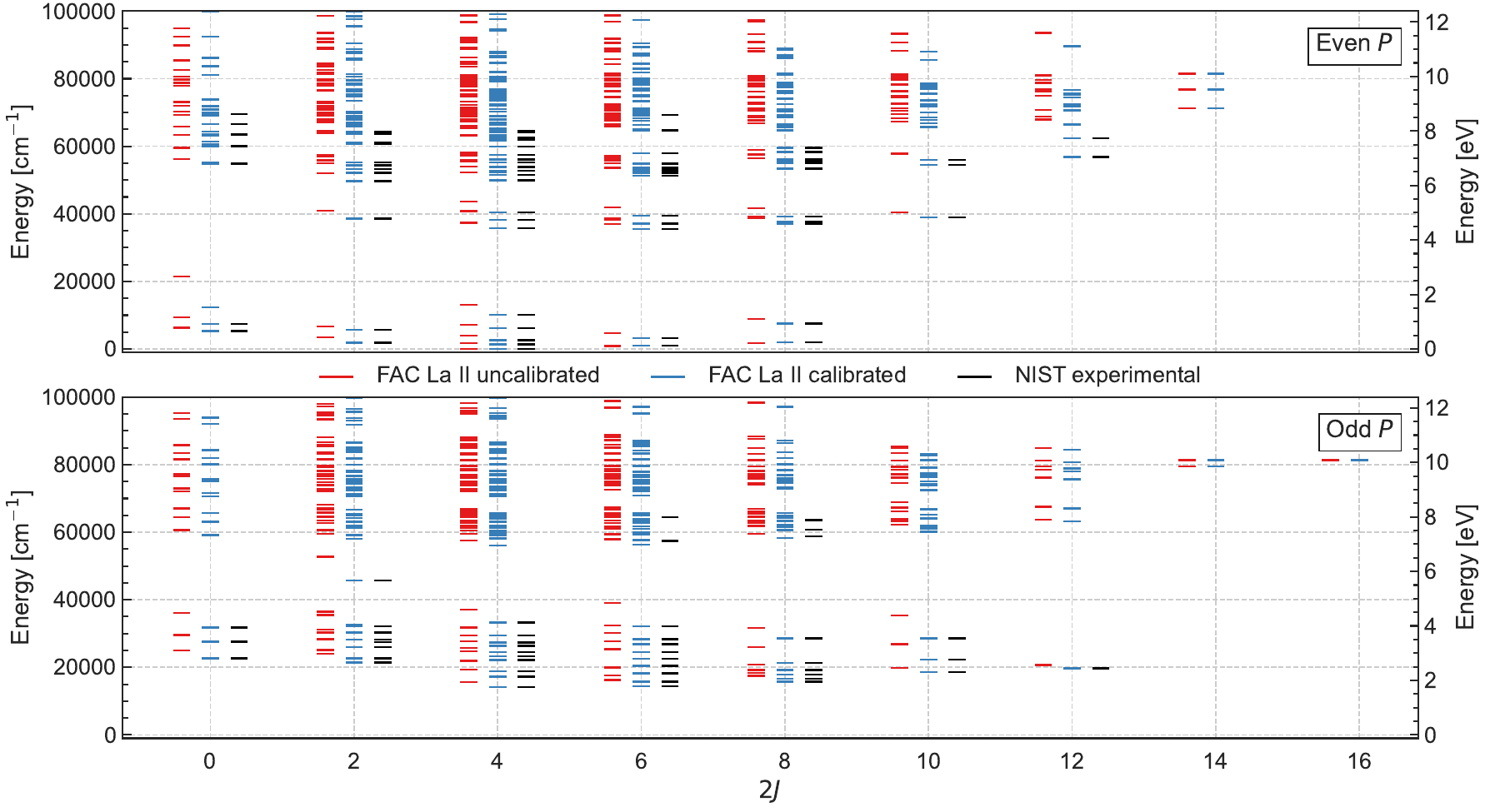}
	\caption{\justifying Energy levels for \ion{La}{ii} for even (top) and odd (bottom) parity for the model given in table~\ref{tab:FAC_configs1}    
		using the \FAC\ code. Red horizontal lines indicate the initial model, while blue lines indicate the level energies of the calibrated model. Black horizontal lines show experimental data from the NIST database \citep{NIST_ASD} used to calibrate the calculated levels.}  
	\label{fig:LaII_levels}
\end{figure*}

\begin{figure*}
	\includegraphics[width=0.99\linewidth]{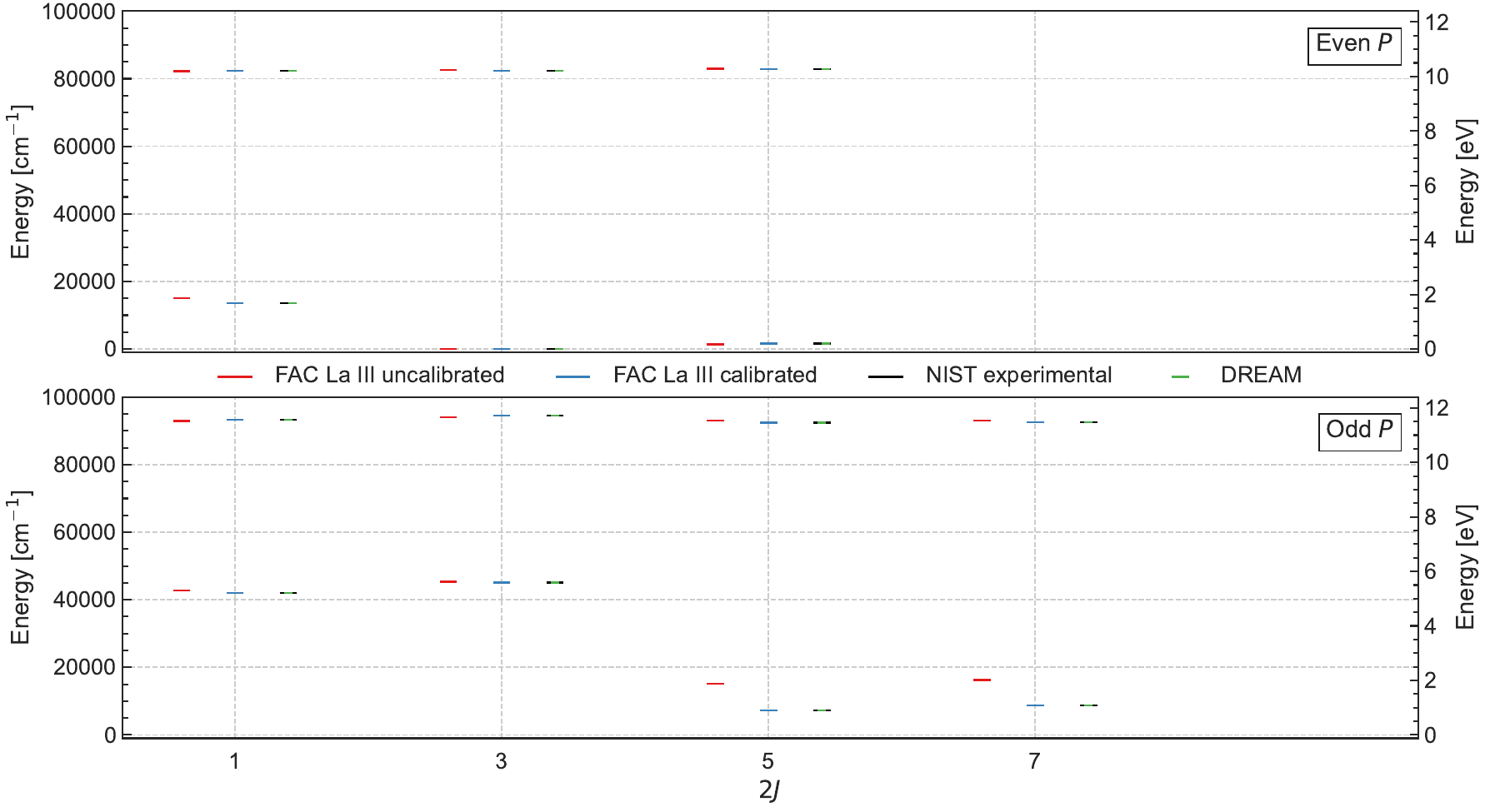}
	\caption{\justifying Energy levels for \ion{La}{iii} for even (top) and odd (bottom) parity for the model given in table~\ref{tab:FAC_configs1}    
		using the \FAC\ code. Red horizontal lines indicate the initial model, while blue lines indicate the level energies of the calibrated model. Black horizontal lines show experimental data from the NIST database \citep{NIST_ASD} used to calibrate the calculated levels. Levels present in the DREAM database are shown in green.}  
	\label{fig:LaIII_levels} 
\end{figure*}

\begin{figure*}
	\includegraphics[width=0.99\linewidth]{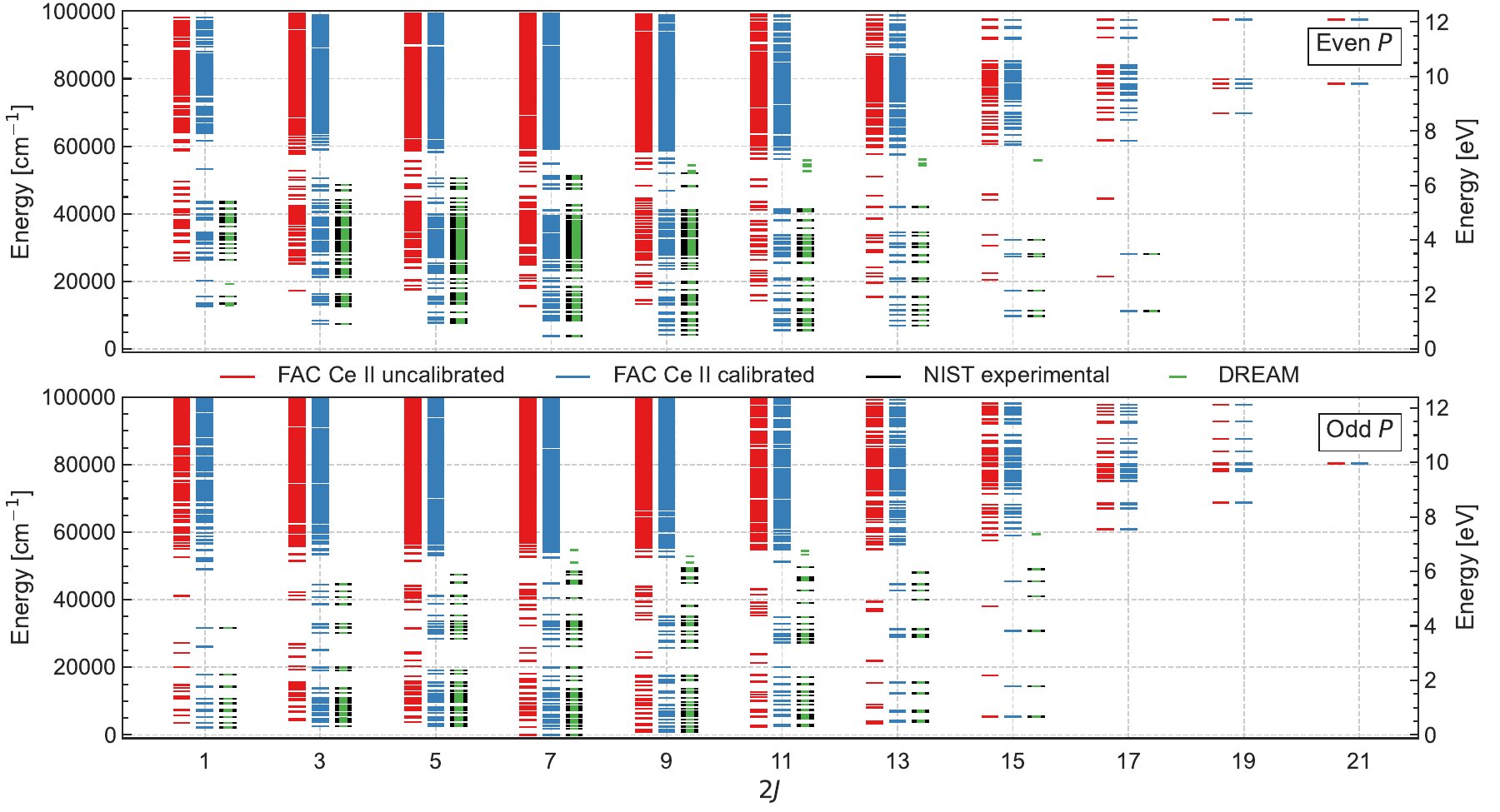}
	\caption{\justifying Energy levels for \ion{Ce}{ii} for even (top) and odd (bottom) parity for the model given in table~\ref{tab:FAC_configs1}    
		using the \FAC\ code. Red horizontal lines indicate the initial model, while blue lines indicate the level energies of the calibrated model. Black horizontal lines show experimental data from the NIST database \citep{NIST_ASD} used to calibrate the calculated levels. Levels present in the DREAM database are shown in green.}  
	\label{fig:CeII_levels} 
\end{figure*}

\begin{figure*}
	\includegraphics[width=0.99\linewidth]{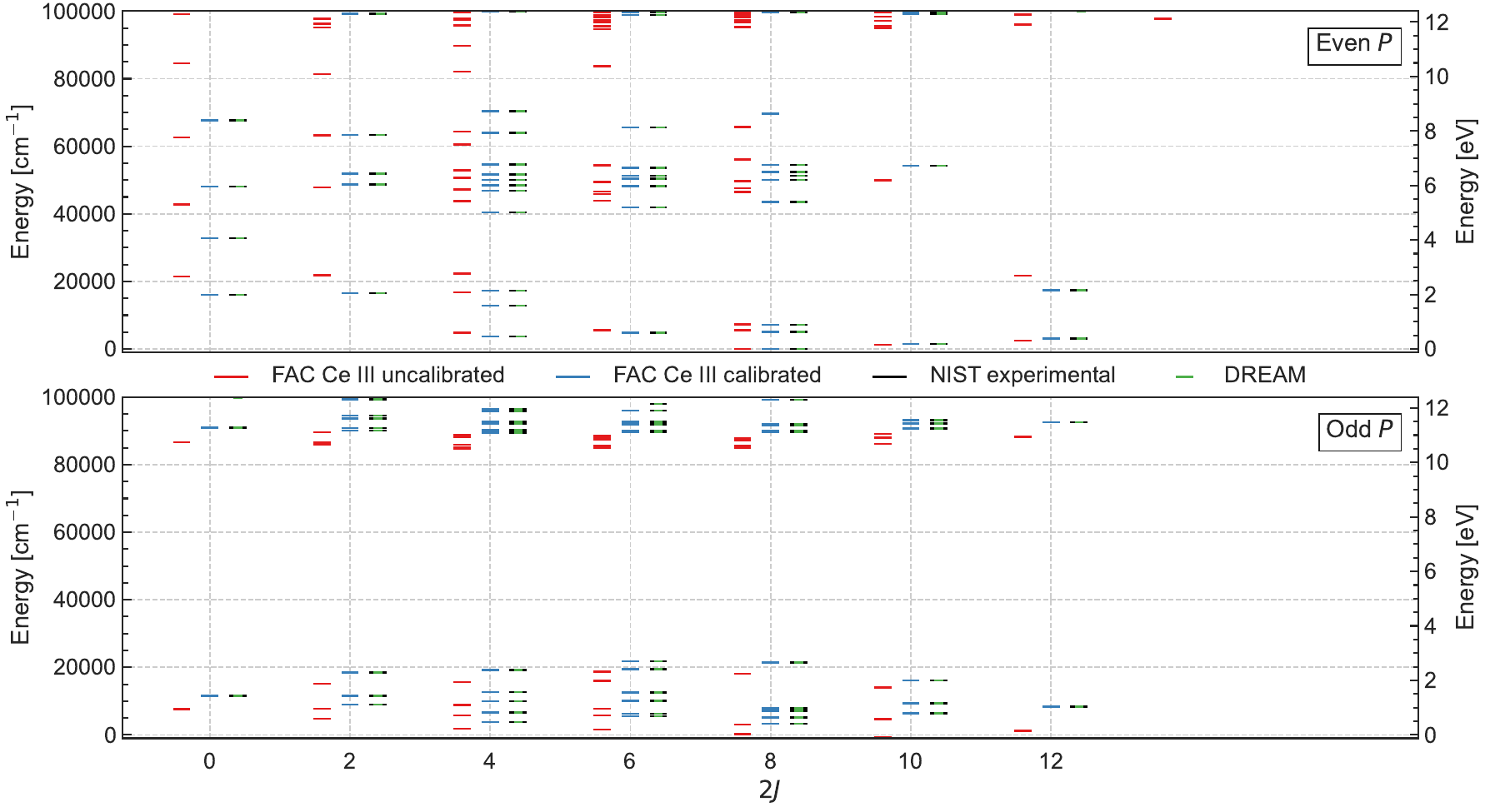}
	\caption{\justifying Energy levels for \ion{Ce}{iii} for even (top) and odd (bottom) parity for the model given in table~\ref{tab:FAC_configs1}    
		using the \FAC\ code. Red horizontal lines indicate the initial model, while blue lines indicate the level energies of the calibrated model. Black horizontal lines show experimental data from the NIST database \citep{NIST_ASD} used to calibrate the calculated levels. Levels present in the DREAM database are shown in green.}  
	\label{fig:CeIII_levels} 
\end{figure*}

\begin{figure*}
	\includegraphics[width=0.99\linewidth]{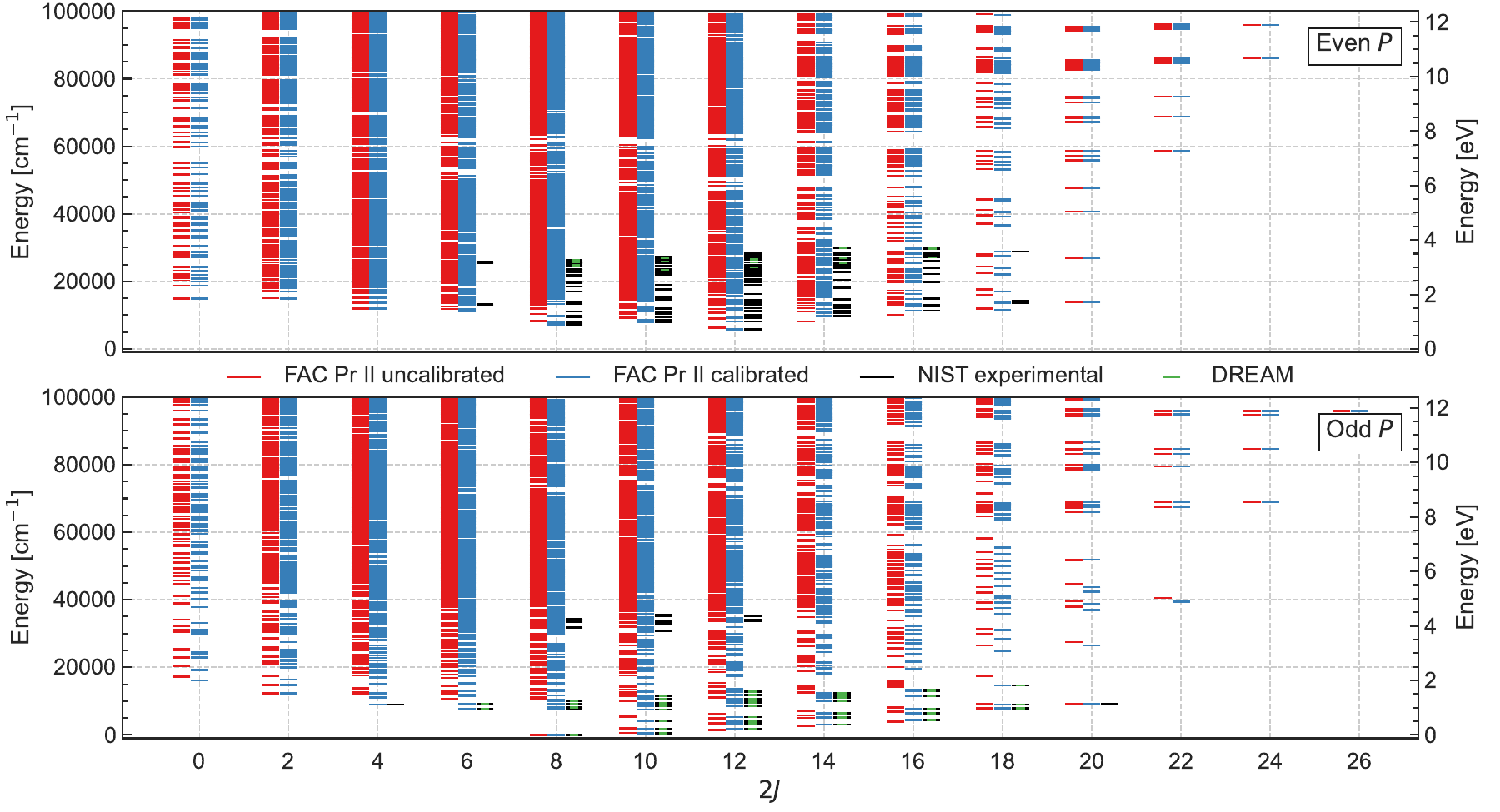}
	\caption{\justifying Energy levels for \ion{Pr}{ii} for even (top) and odd (bottom) parity for the model given in table~\ref{tab:FAC_configs1}    
		using the \FAC\ code. Red horizontal lines indicate the initial model, while blue lines indicate the level energies of the calibrated model. Black horizontal lines show experimental data from the NIST database \citep{NIST_ASD} used to calibrate the calculated levels. Levels present in the DREAM database are shown in green.}  
	\label{fig:PrII_levels} 
\end{figure*}

\begin{figure*}
	\includegraphics[width=0.99\linewidth]{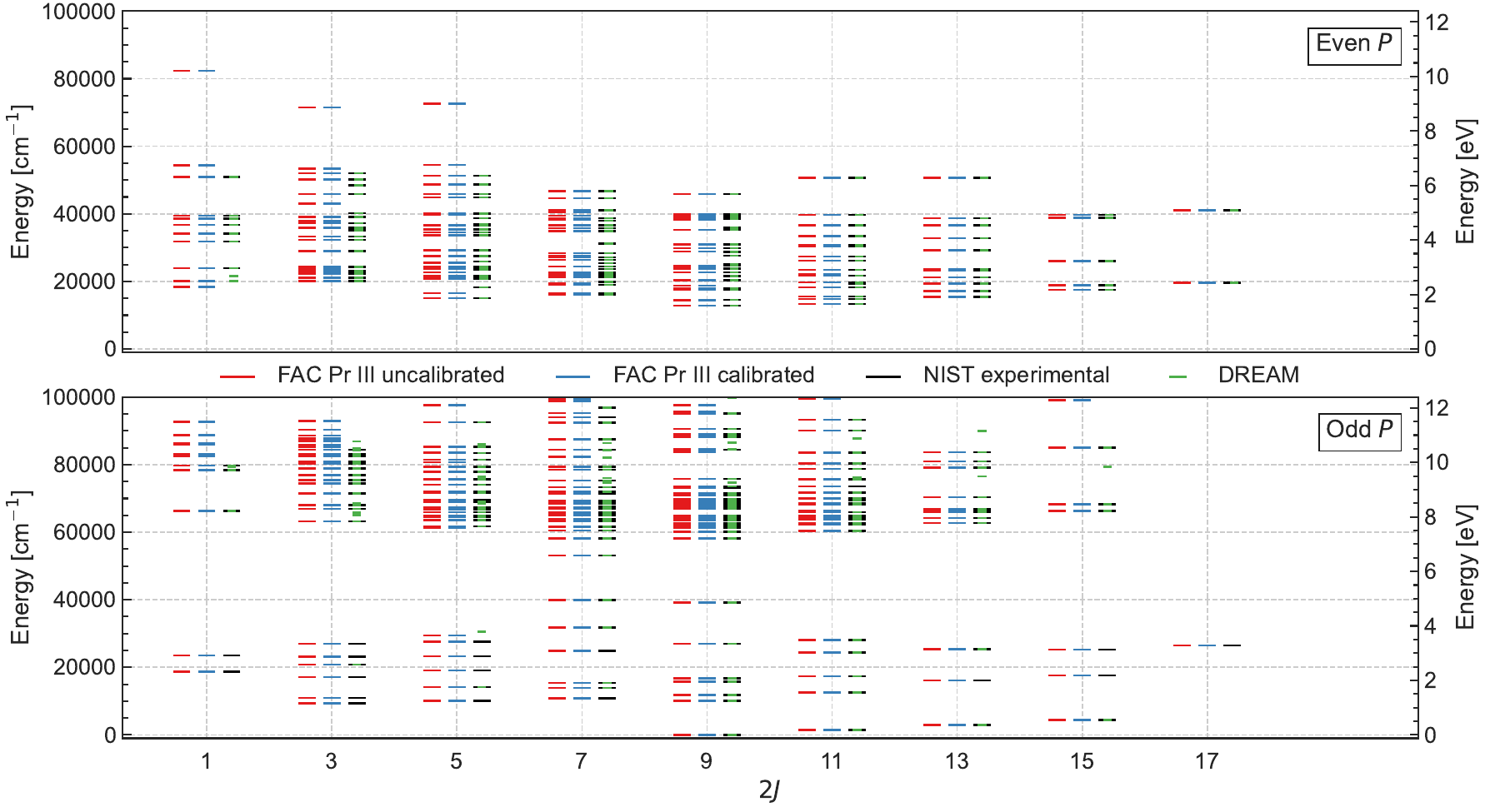}
	\caption{\justifying Energy levels for \ion{Pr}{iii} for even (top) and odd (bottom) parity for the model given in table~\ref{tab:FAC_configs1}    
		using the \FAC\ code. Red horizontal lines indicate the initial model, while blue lines indicate the level energies of the calibrated model. Black horizontal lines show experimental data from the NIST database \citep{NIST_ASD} used to calibrate the calculated levels. Levels present in the DREAM database are shown in green.}  
	\label{fig:PrIII_levels} 
\end{figure*}

\begin{figure*}
	\includegraphics[width=0.99\linewidth]{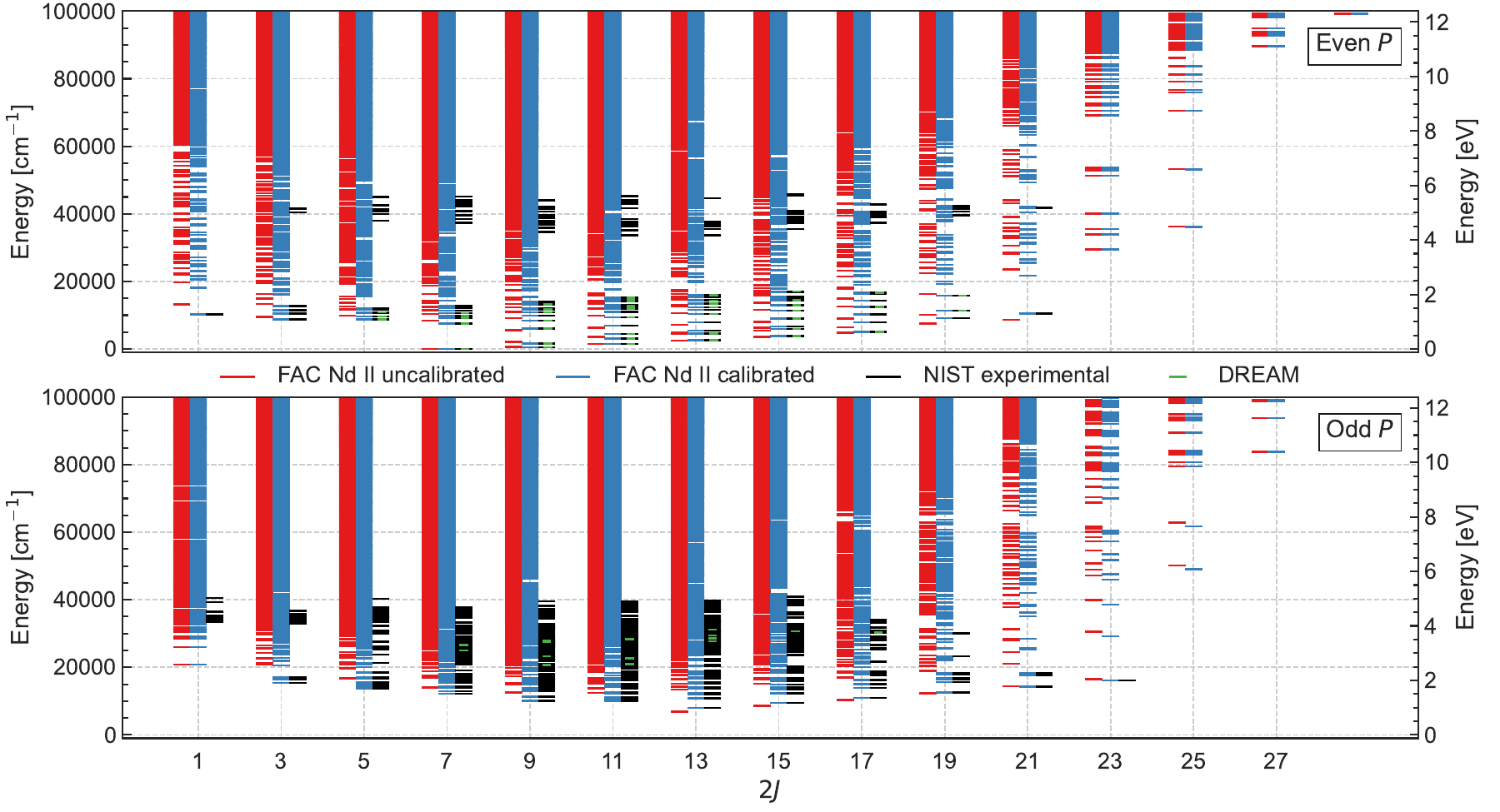}
	\caption{\justifying Energy levels for \ion{Nd}{ii} for even (top) and odd (bottom) parity for the model given in table~\ref{tab:FAC_configs1}    
		using the \FAC\ code. Red horizontal lines indicate the initial model, while blue lines indicate the level energies of the calibrated model. Black horizontal lines show experimental data from the NIST database \citep{NIST_ASD} used to calibrate the calculated levels. Levels present in the DREAM database are shown in green.}  
	\label{fig:NdII_levels} 
\end{figure*}

\begin{figure*}
	\includegraphics[width=0.99\linewidth]{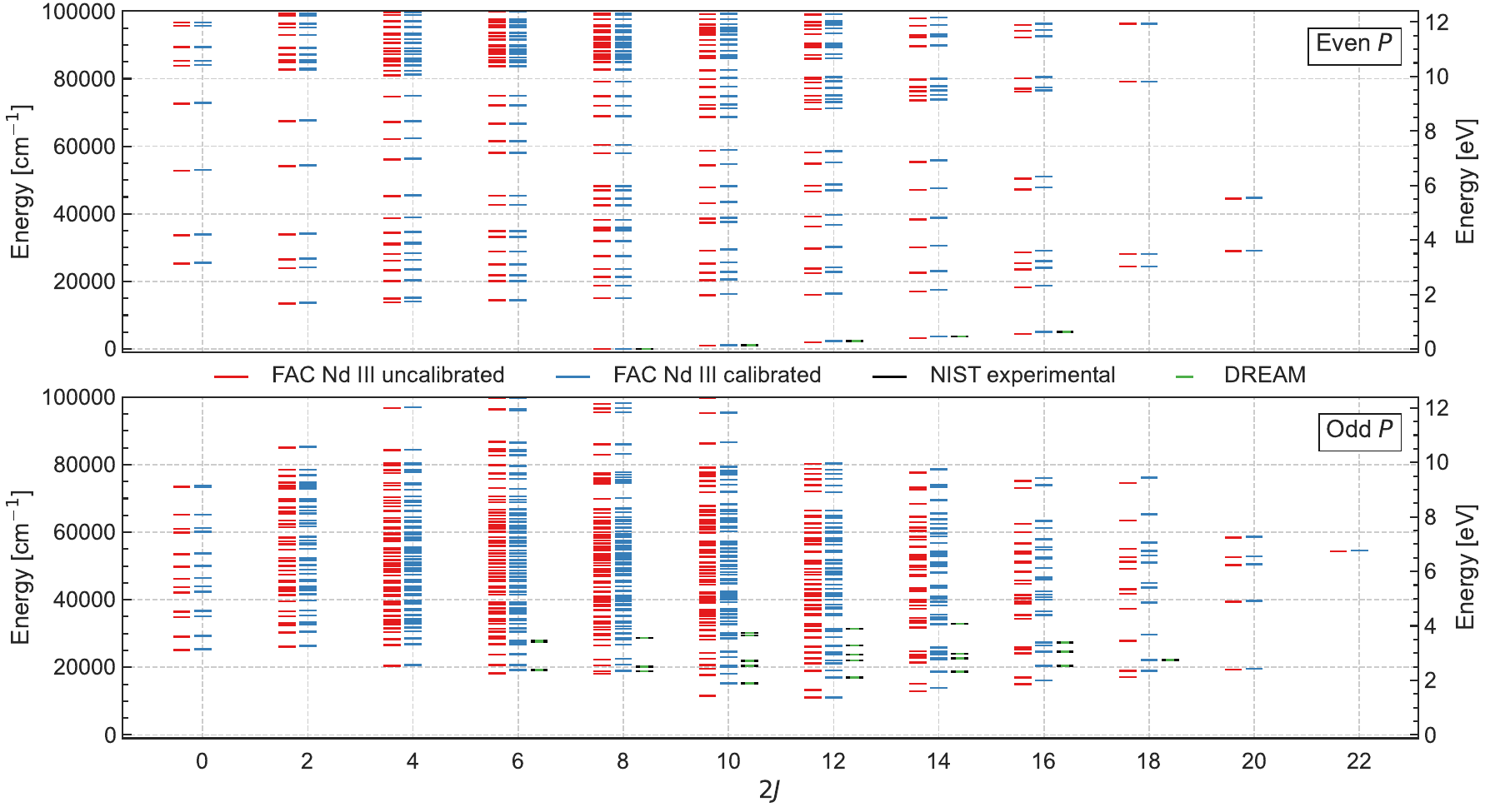}
	\caption{\justifying Energy levels for \ion{Nd}{iii} for even (top) and odd (bottom) parity for the model given in table~\ref{tab:FAC_configs1}    
		using the \FAC\ code. Red horizontal lines indicate the initial model, while blue lines indicate the level energies of the calibrated model. Black horizontal lines show experimental data from the NIST database \citep{NIST_ASD} used to calibrate the calculated levels. Levels present in the DREAM database are shown in green. In contrast to Fig.~\ref{fig:NdIII_levels_comparison}, this shows the calibration to NIST data only.}
	\label{fig:NdIII_levels} 
\end{figure*}

\begin{figure*}
	\includegraphics[width=0.99\linewidth]{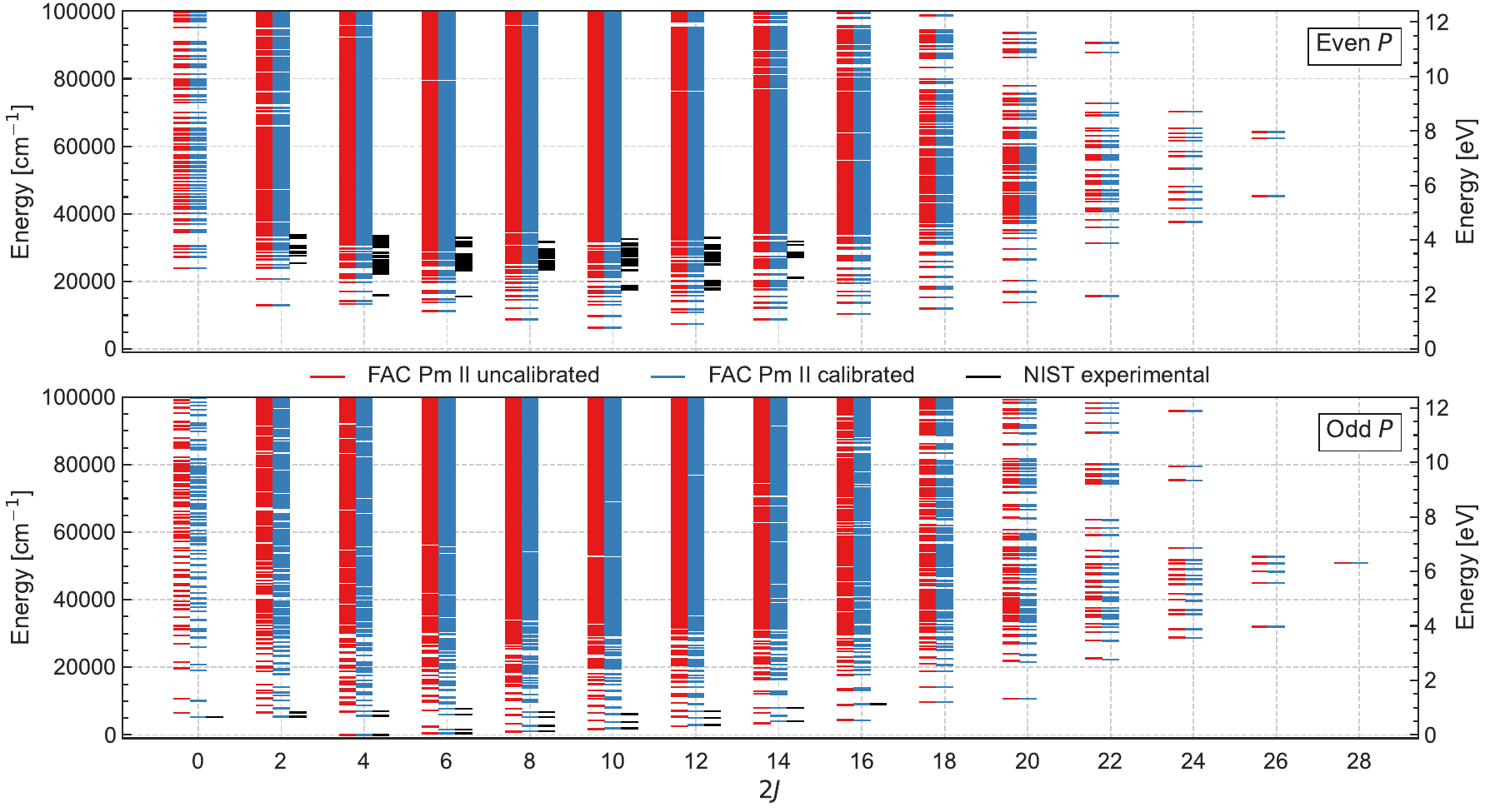}
	\caption{\justifying Energy levels for \ion{Pm}{ii} for even (top) and odd (bottom) parity for the model given in table~\ref{tab:FAC_configs1}    
		using the \FAC\ code. Red horizontal lines indicate the initial model, while blue lines indicate the level energies of the calibrated model. Black horizontal lines show experimental data from the NIST database \citep{NIST_ASD} used to calibrate the calculated levels.}  
	\label{fig:PmII_levels} 
\end{figure*}

\begin{figure*}
	\includegraphics[width=0.99\linewidth]{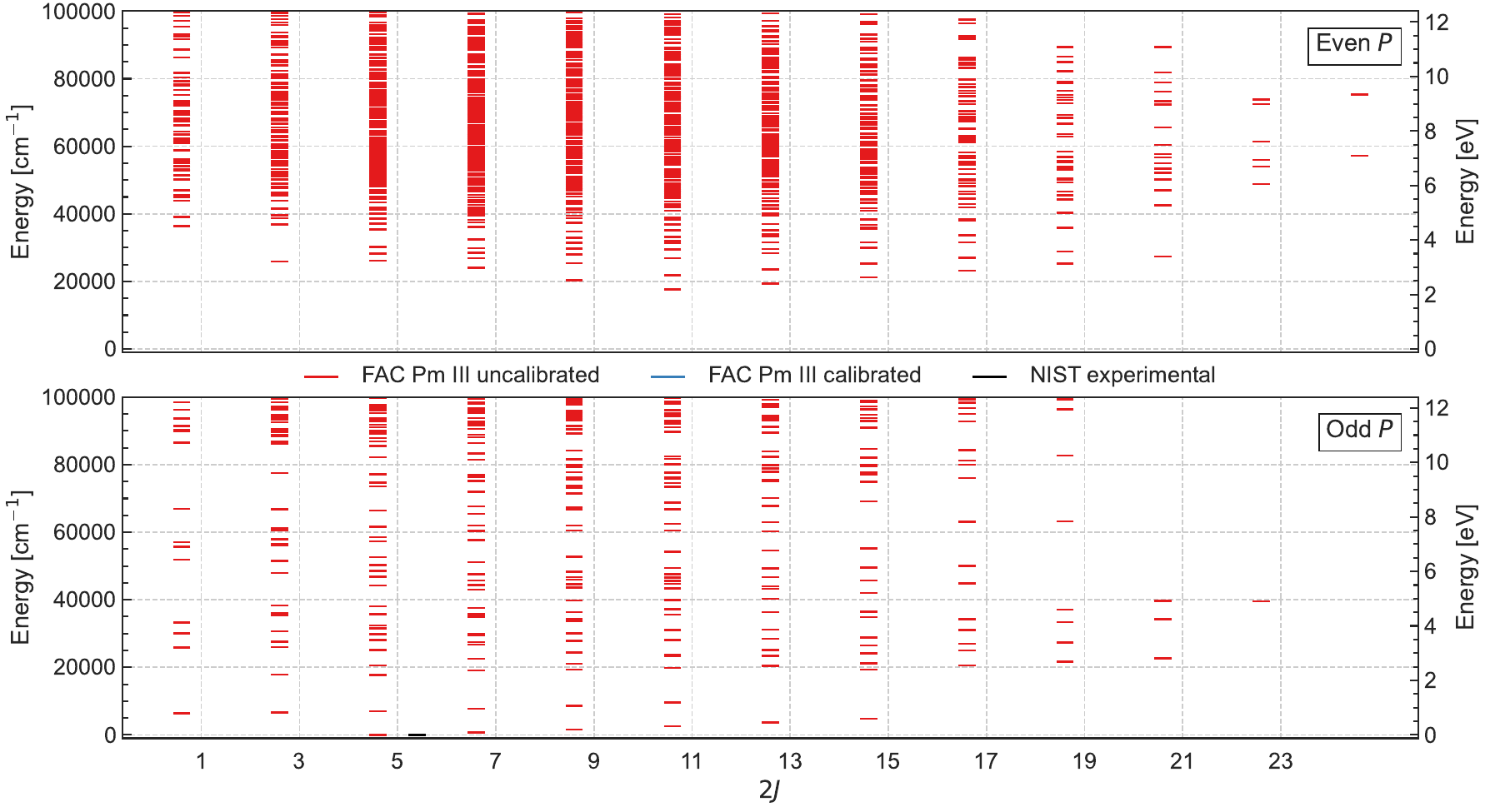}
	\caption{\justifying Energy levels for \ion{Pm}{iii} for even (top) and odd (bottom) parity for the model given in table~\ref{tab:FAC_configs1}    
		using the \FAC\ code. Red horizontal lines indicate the initial model, while blue lines indicate the level energies of the calibrated model. Black horizontal lines show experimental data from the NIST database \citep{NIST_ASD} used to calibrate the calculated levels.}  
	\label{fig:PmIII_levels} 
\end{figure*}

\begin{figure*}
	\includegraphics[width=0.99\linewidth]{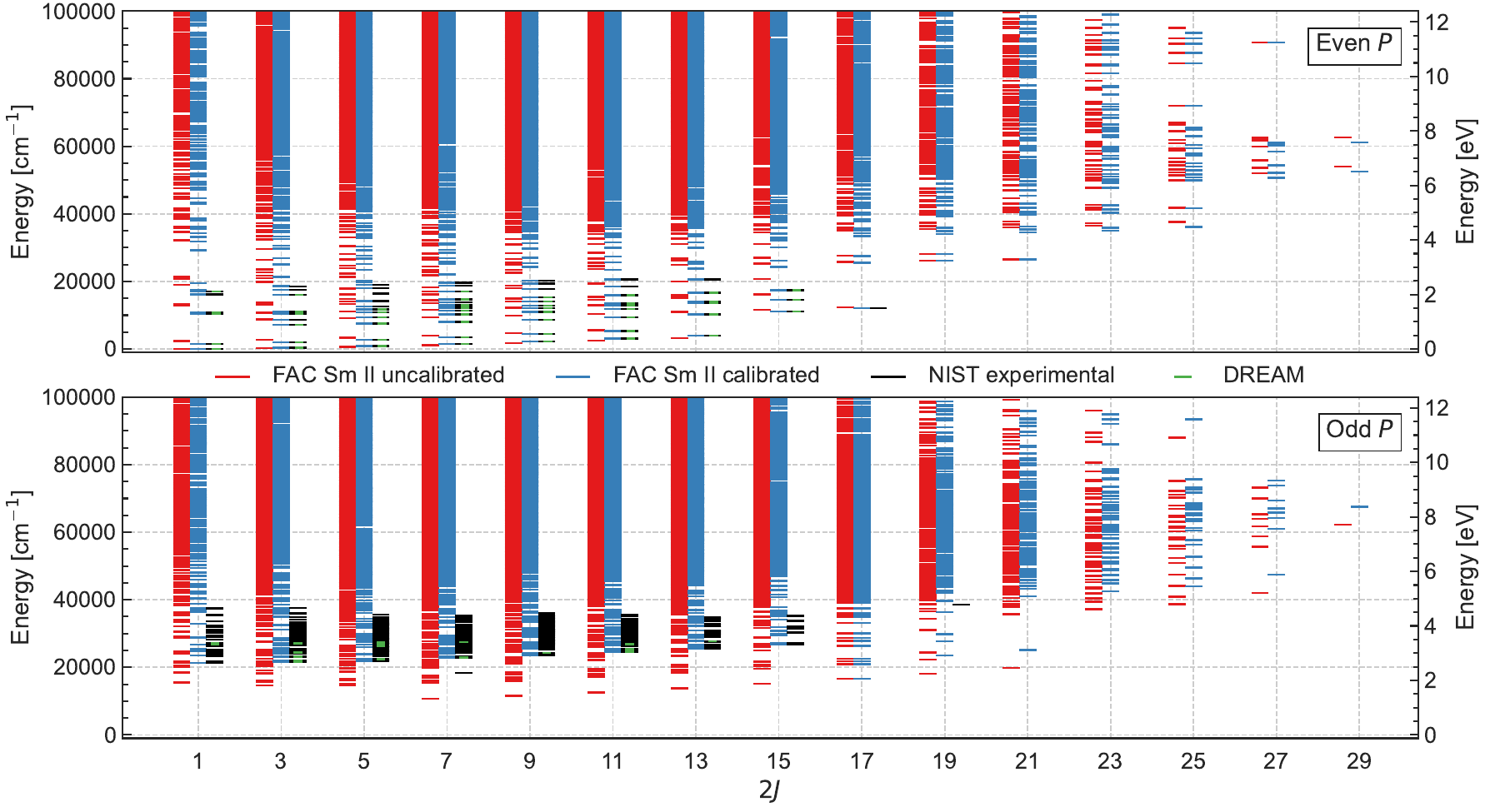}
	\caption{\justifying Energy levels for \ion{Sm}{ii} for even (top) and odd (bottom) parity for the model given in table~\ref{tab:FAC_configs1}    
		using the \FAC\ code. Red horizontal lines indicate the initial model, while blue lines indicate the level energies of the calibrated model. Black horizontal lines show experimental data from the NIST database \citep{NIST_ASD} used to calibrate the calculated levels. Levels present in the DREAM database are shown in green.}  
	\label{fig:SmII_levels} 
\end{figure*}

\begin{figure*}
	\includegraphics[width=0.99\linewidth]{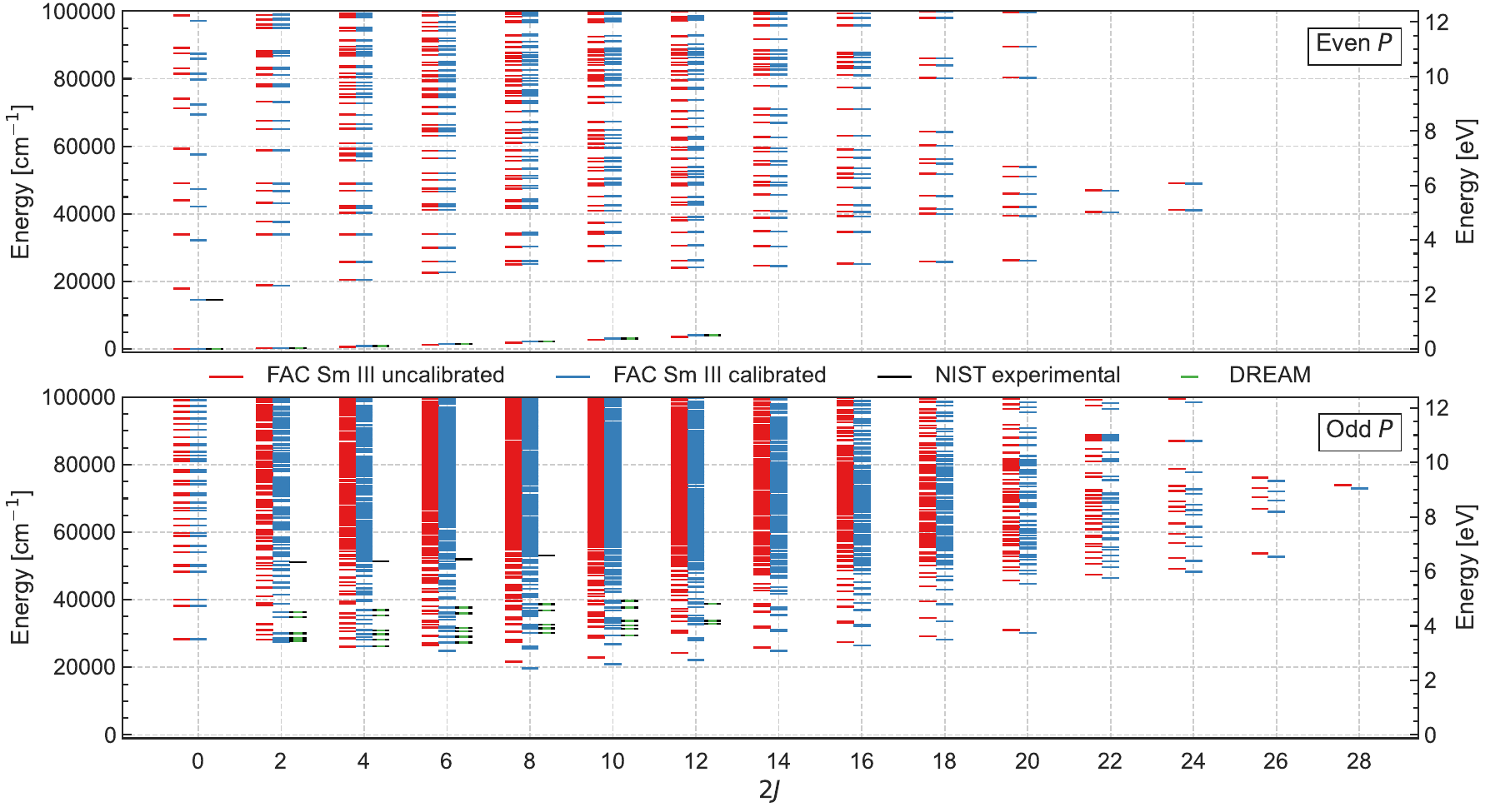}
	\caption{\justifying Energy levels for \ion{Sm}{iii} for even (top) and odd (bottom) parity for the model given in table~\ref{tab:FAC_configs1}    
		using the \FAC\ code. Red horizontal lines indicate the initial model, while blue lines indicate the level energies of the calibrated model. Black horizontal lines show experimental data from the NIST database \citep{NIST_ASD} used to calibrate the calculated levels. Levels present in the DREAM database are shown in green.}  
	\label{fig:SmIII_levels} 
\end{figure*}

\begin{figure*}
	\includegraphics[width=0.99\linewidth]{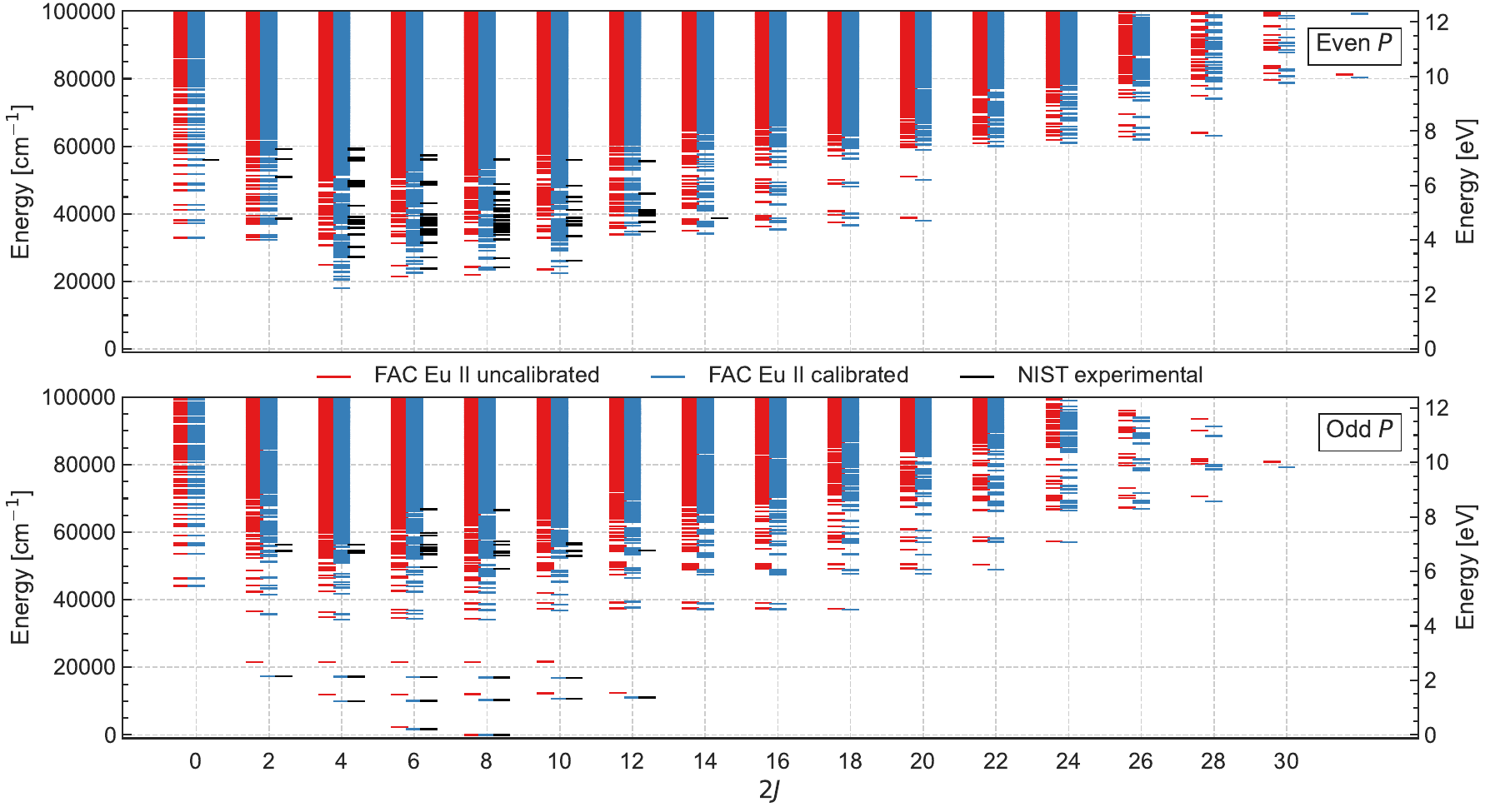}
	\caption{\justifying Energy levels for \ion{Eu}{ii} for even (top) and odd (bottom) parity for the model given in table~\ref{tab:FAC_configs1}    
		using the \FAC\ code. Red horizontal lines indicate the initial model, while blue lines indicate the level energies of the calibrated model. Black horizontal lines show experimental data from the NIST database \citep{NIST_ASD} used to calibrate the calculated levels.}  
	\label{fig:EuII_levels} 
\end{figure*}

\begin{figure*}
	\includegraphics[width=0.99\linewidth]{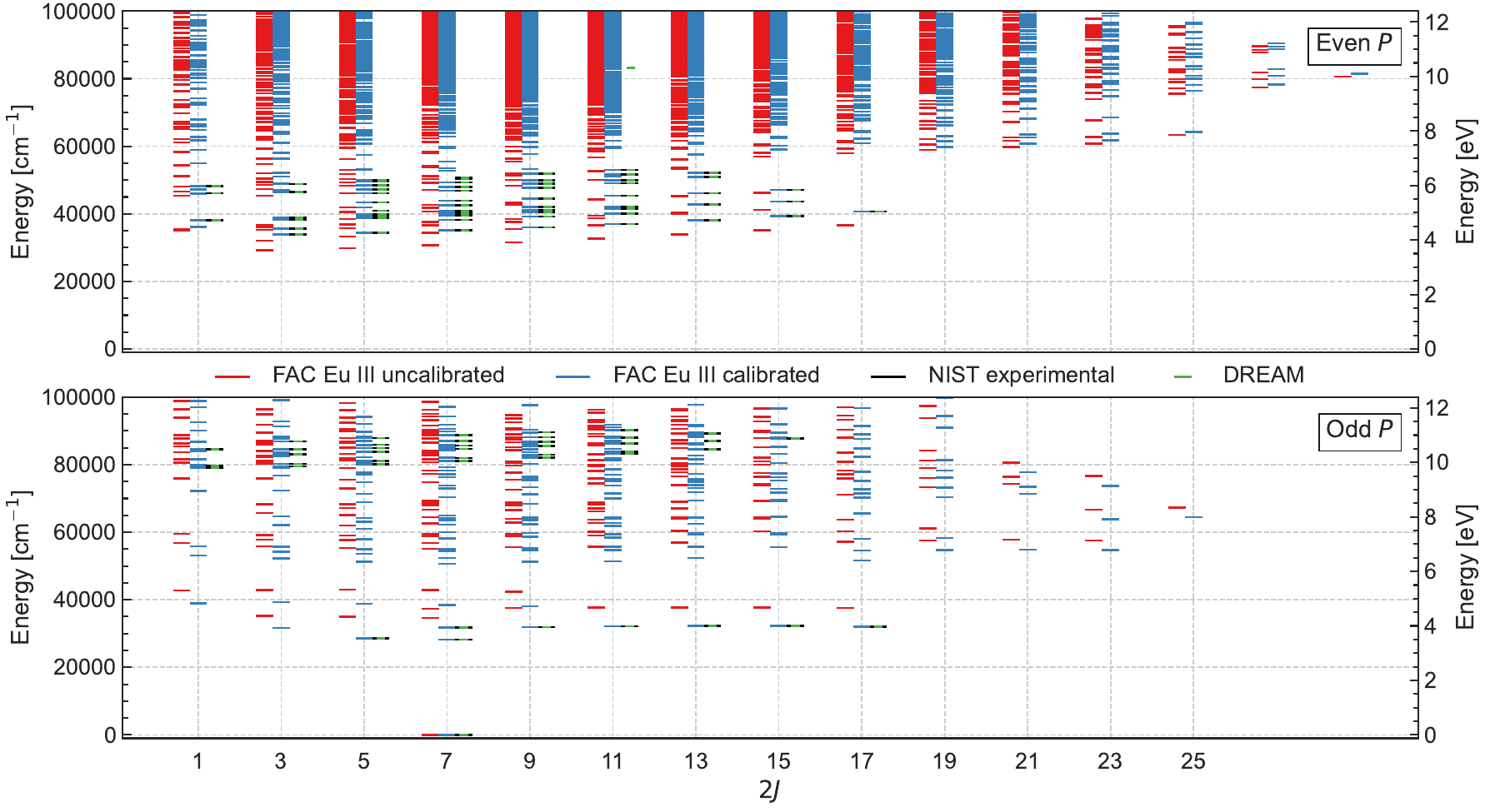}
	\caption{\justifying Energy levels for \ion{Eu}{iii} for even (top) and odd (bottom) parity for the model given in table~\ref{tab:FAC_configs1}    
		using the \FAC\ code. Red horizontal lines indicate the initial model, while blue lines indicate the level energies of the calibrated model. Black horizontal lines show experimental data from the NIST database \citep{NIST_ASD} used to calibrate the calculated levels. Levels present in the DREAM database are shown in green.}  
	\label{fig:EuIII_levels} 
\end{figure*}

\begin{figure*}
	\includegraphics[width=0.99\linewidth]{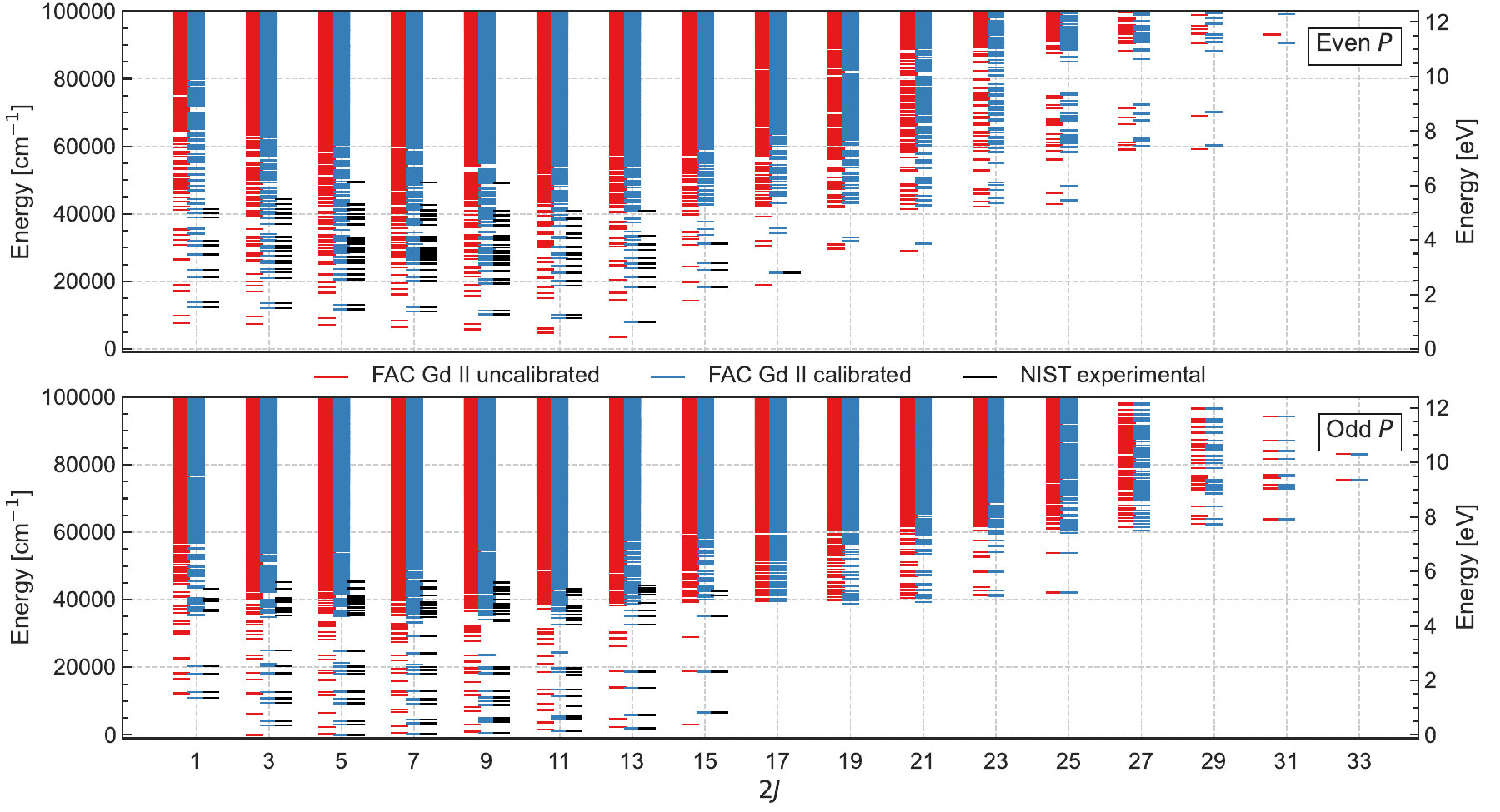}
	\caption{\justifying Energy levels for \ion{Gd}{ii} for even (top) and odd (bottom) parity for the model given in table~\ref{tab:FAC_configs1}    
		using the \FAC\ code. Red horizontal lines indicate the initial model, while blue lines indicate the level energies of the calibrated model. Black horizontal lines show experimental data from the NIST database \citep{NIST_ASD} used to calibrate the calculated levels.}  
	\label{fig:GdII_levels} 
\end{figure*}

\begin{figure*}
	\includegraphics[width=0.99\linewidth]{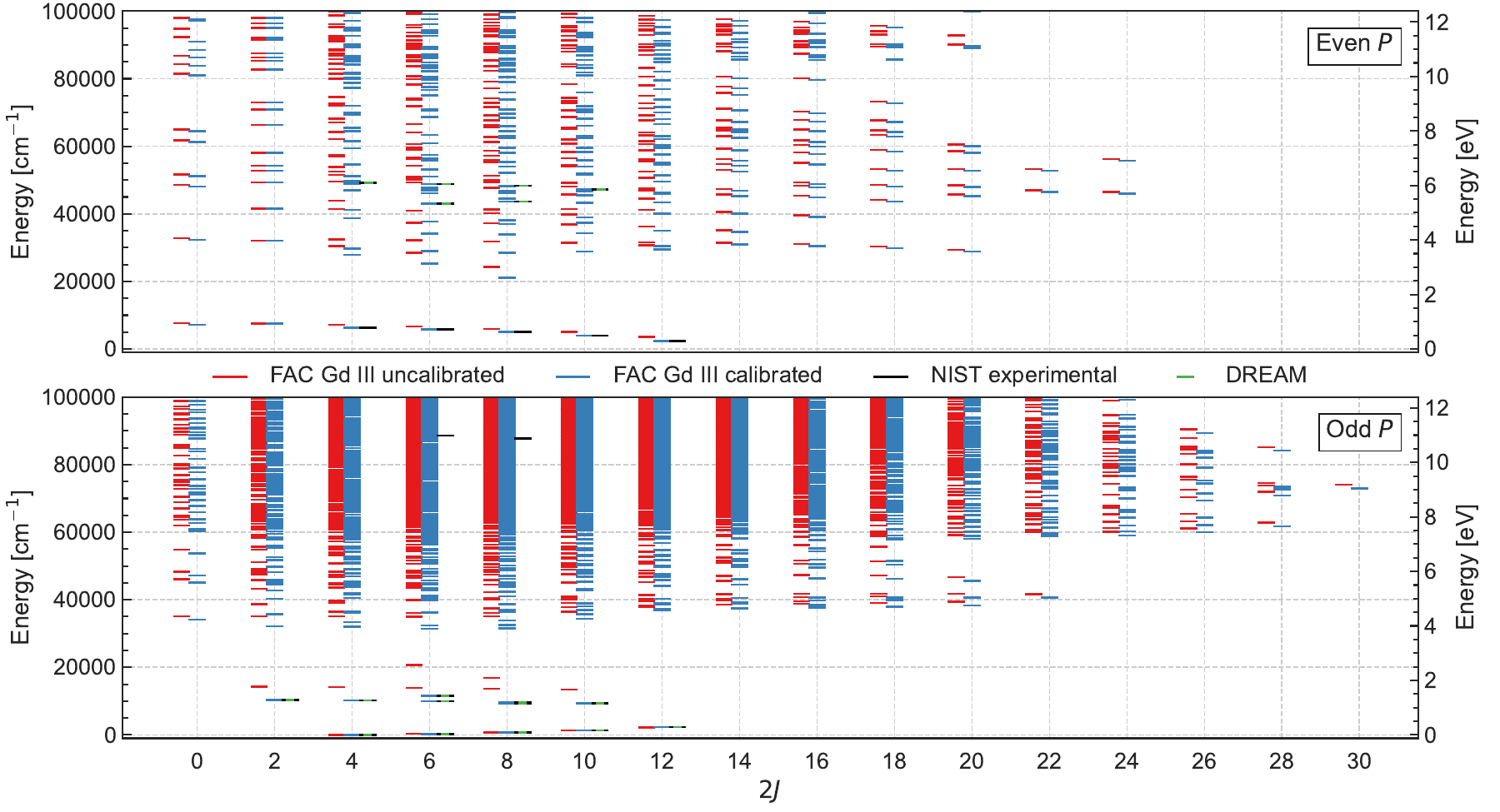}
	\caption{\justifying Energy levels for \ion{Gd}{iii} for even (top) and odd (bottom) parity for the model given in table~\ref{tab:FAC_configs1}    
		using the \FAC\ code. Red horizontal lines indicate the initial model, while blue lines indicate the level energies of the calibrated model. Black horizontal lines show experimental data from the NIST database \citep{NIST_ASD} used to calibrate the calculated levels. Levels present in the DREAM database are shown in green.}  
	\label{fig:GdIII_levels} 
\end{figure*}

\begin{figure*}
	\includegraphics[width=0.99\linewidth]{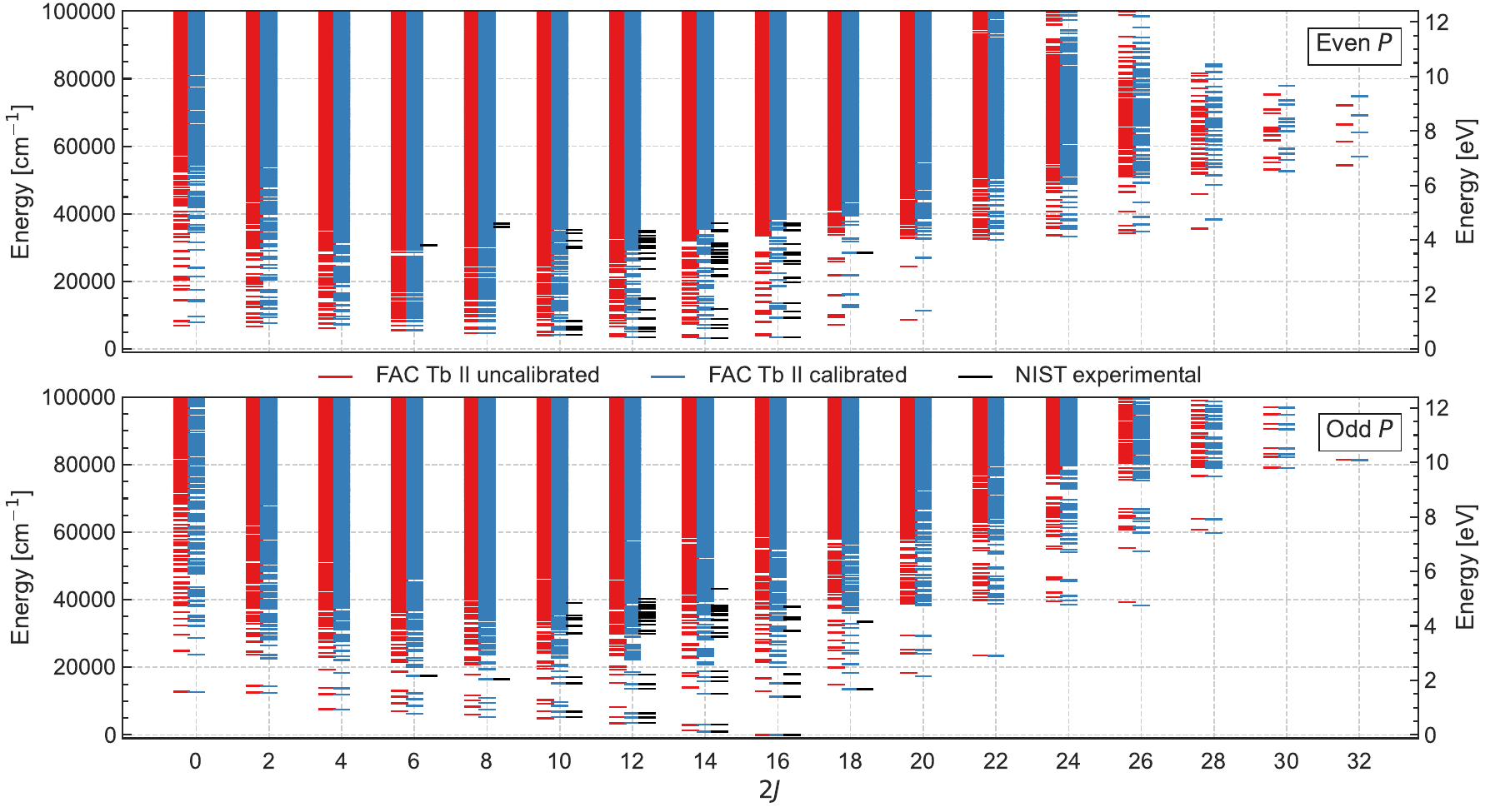}
	\caption{\justifying Energy levels for \ion{Tb}{ii} for even (top) and odd (bottom) parity for the model given in table~\ref{tab:FAC_configs1}    
		using the \FAC\ code. Red horizontal lines indicate the initial model, while blue lines indicate the level energies of the calibrated model. Black horizontal lines show experimental data from the NIST database \citep{NIST_ASD} used to calibrate the calculated levels.}  
	\label{fig:TbII_levels} 
\end{figure*}

\begin{figure*}
	\includegraphics[width=0.99\linewidth]{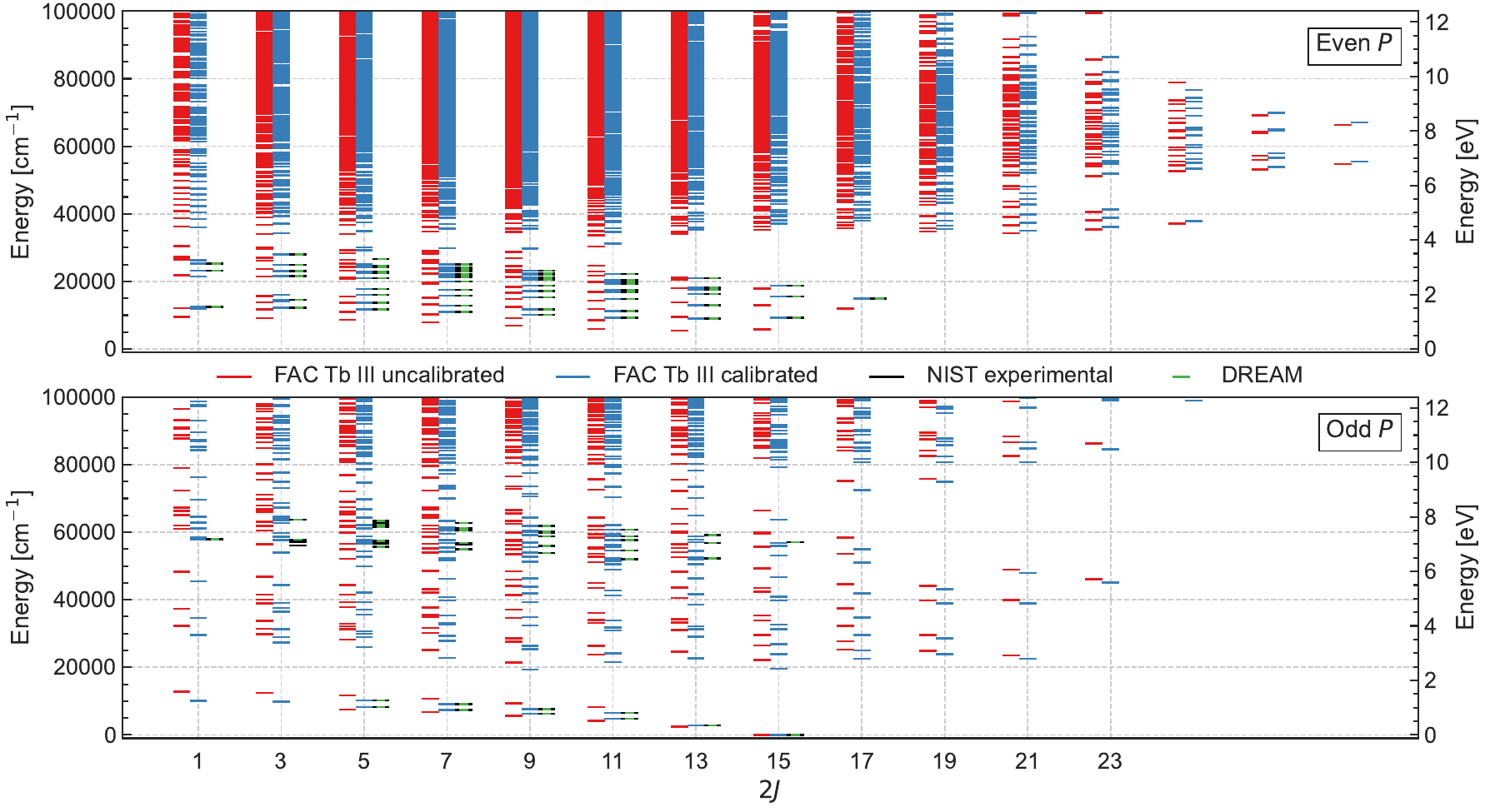}
	\caption{\justifying Energy levels for \ion{Tb}{iii} for even (top) and odd (bottom) parity for the model given in table~\ref{tab:FAC_configs1}    
		using the \FAC\ code. Red horizontal lines indicate the initial model, while blue lines indicate the level energies of the calibrated model. Black horizontal lines show experimental data from the NIST database \citep{NIST_ASD} used to calibrate the calculated levels. Levels present in the DREAM database are shown in green.}  
	\label{fig:TbIII_levels} 
\end{figure*}

\begin{figure*}
	\includegraphics[width=0.99\linewidth]{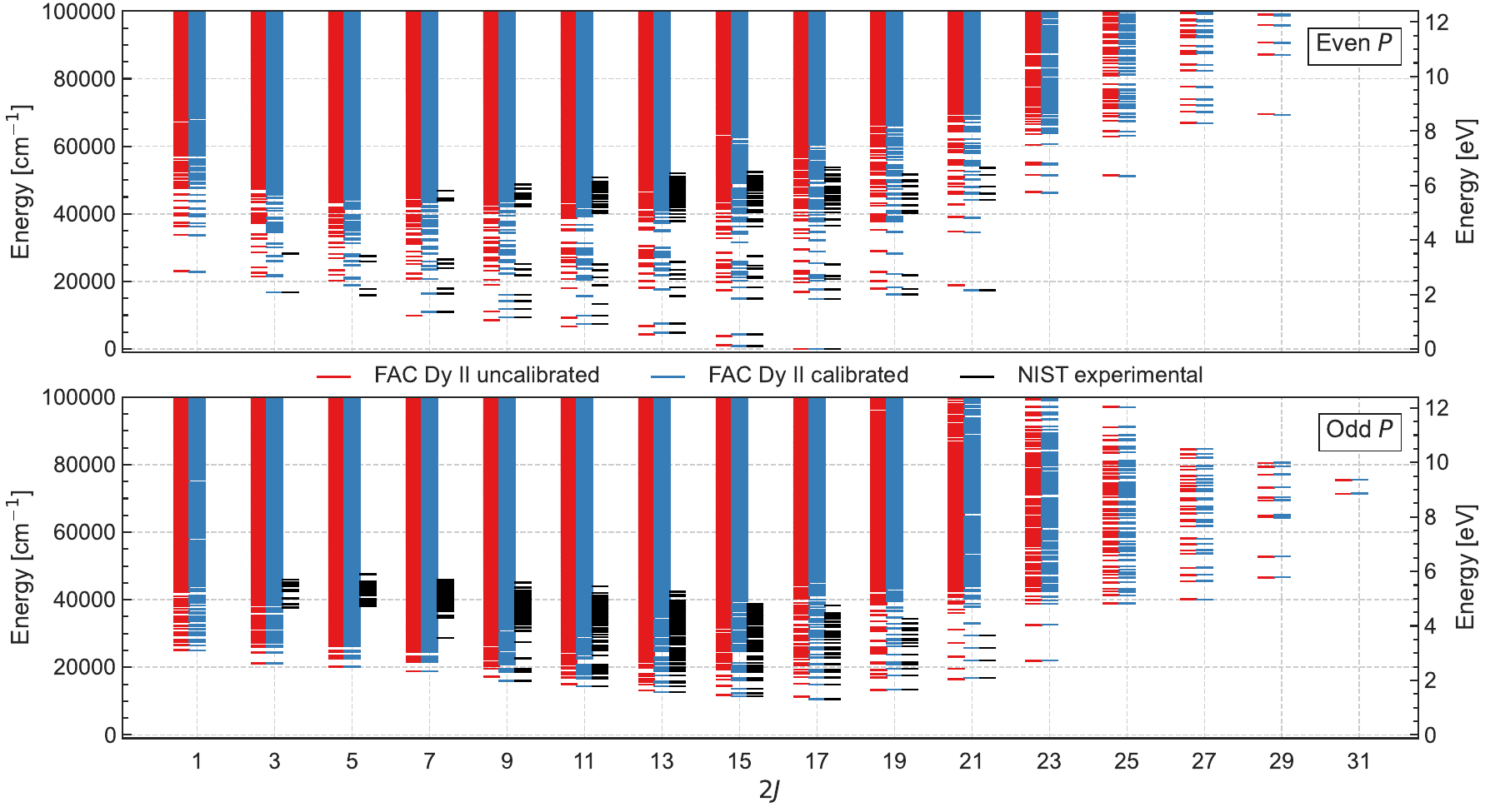}
	\caption{\justifying Energy levels for \ion{Dy}{ii} for even (top) and odd (bottom) parity for the model given in table~\ref{tab:FAC_configs1}    
		using the \FAC\ code. Red horizontal lines indicate the initial model, while blue lines indicate the level energies of the calibrated model. Black horizontal lines show experimental data from the NIST database \citep{NIST_ASD} used to calibrate the calculated levels.}  
	\label{fig:DyII_levels} 
\end{figure*}

\begin{figure*}
	\includegraphics[width=0.99\linewidth]{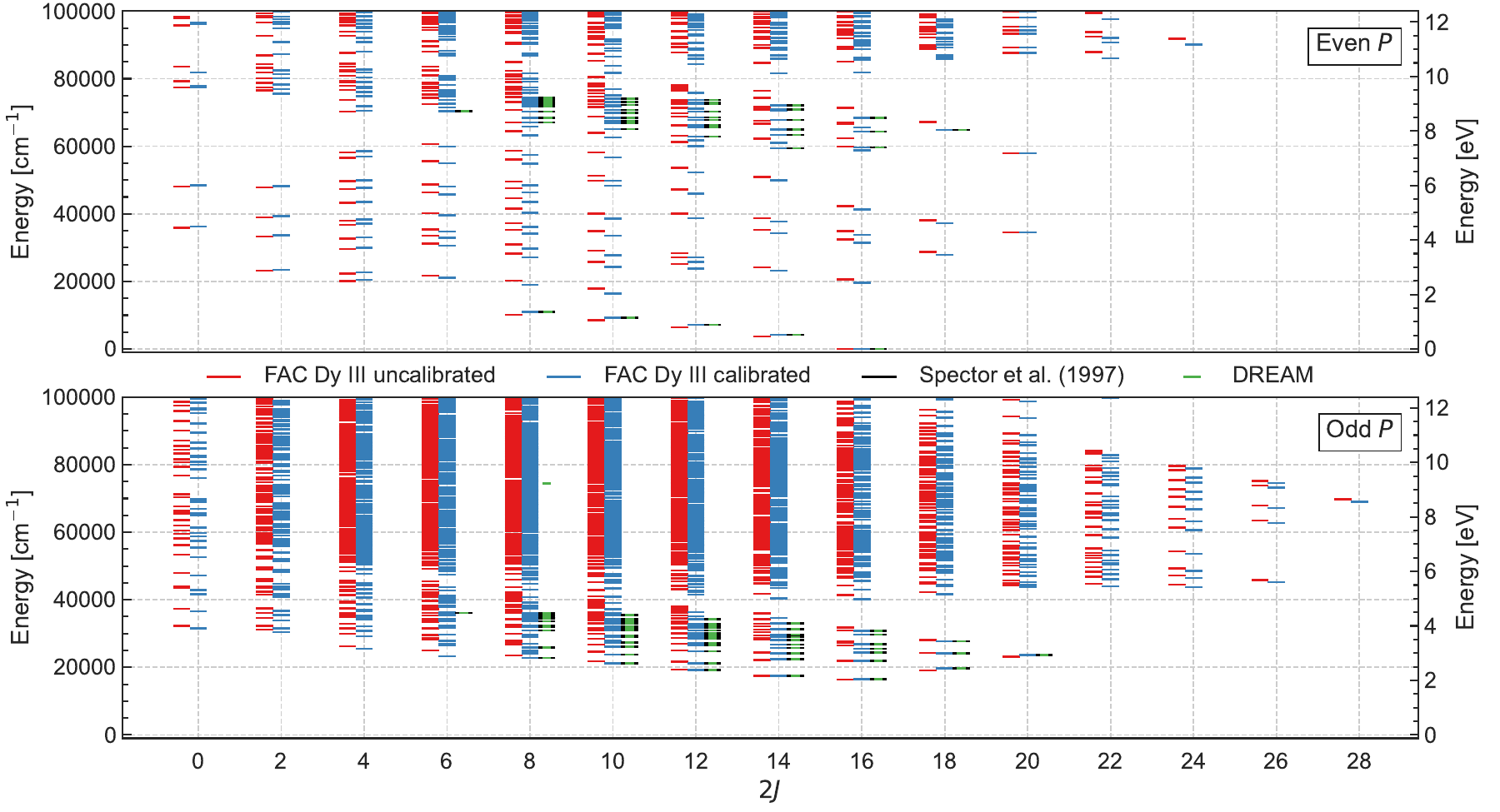}
	\caption{\justifying Energy levels for \ion{Dy}{iii} for even (top) and odd (bottom) parity for the model given in table~\ref{tab:FAC_configs1}    
		using the \FAC\ code. Red horizontal lines indicate the initial model, while blue lines indicate the level energies of the calibrated model. Black horizontal lines show experimental data from the NIST database \citep{NIST_ASD} used to calibrate the calculated levels. Levels present in the DREAM database are shown in green.}  
	\label{fig:DyIII_levels} 
\end{figure*}

\begin{figure*}
	\includegraphics[width=0.99\linewidth]{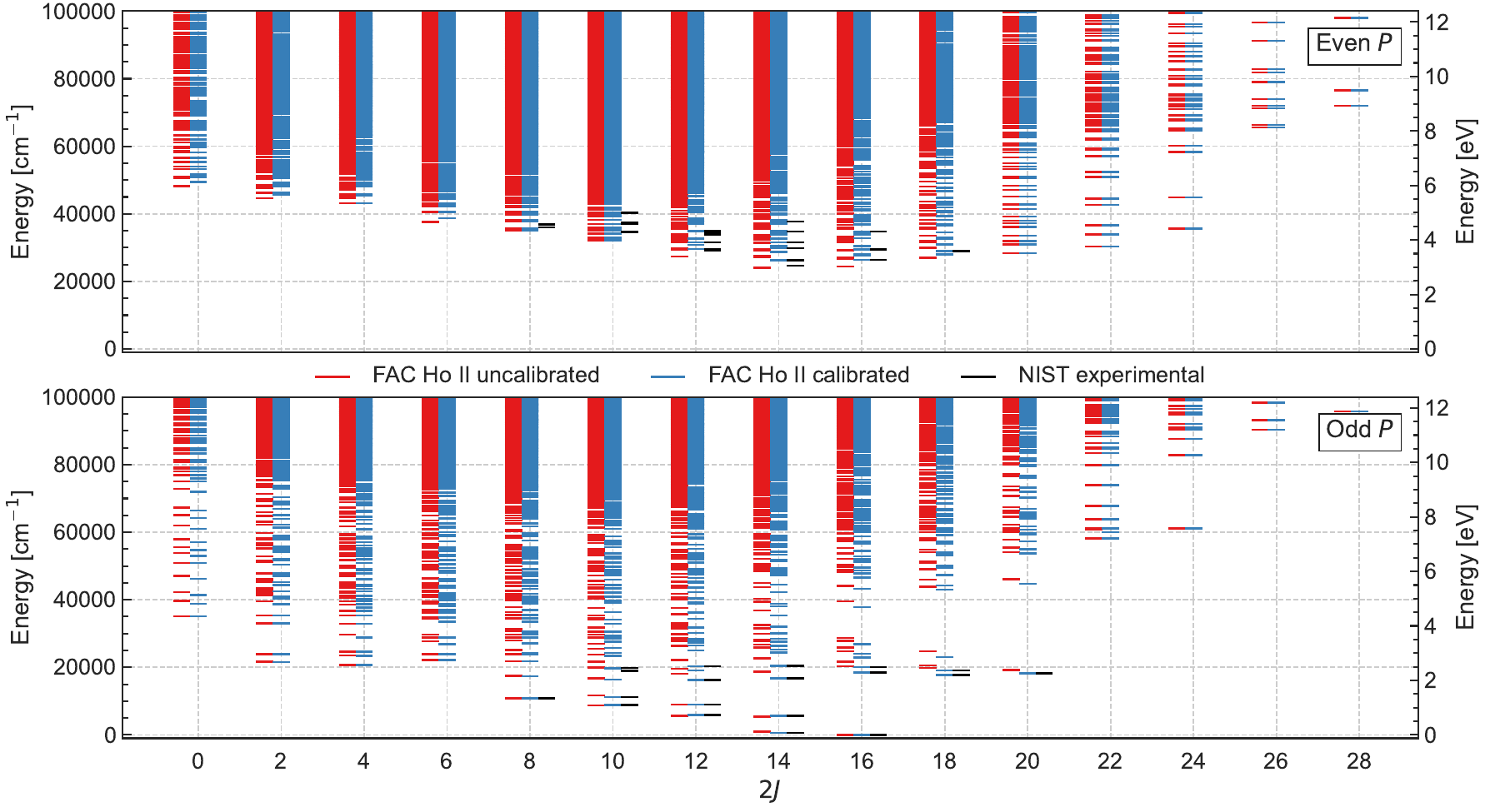}
	\caption{\justifying Energy levels for \ion{Ho}{ii} for even (top) and odd (bottom) parity for the model given in table~\ref{tab:FAC_configs1}    
		using the \FAC\ code. Red horizontal lines indicate the initial model, while blue lines indicate the level energies of the calibrated model. Black horizontal lines show experimental data from the NIST database \citep{NIST_ASD} used to calibrate the calculated levels.}  
	\label{fig:HoII_levels} 
\end{figure*}

\begin{figure*}
	\includegraphics[width=0.99\linewidth]{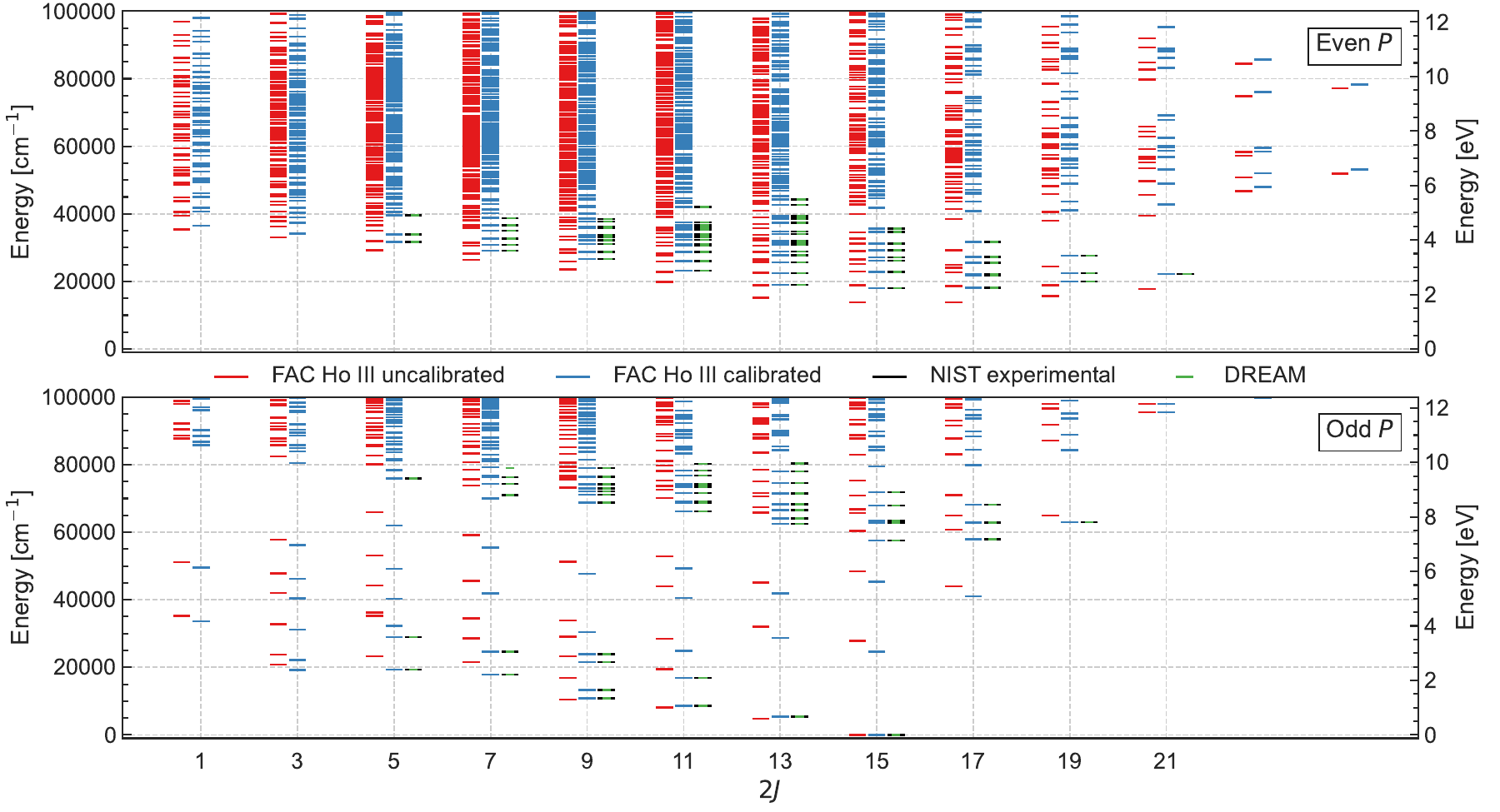}
	\caption{\justifying Energy levels for \ion{Ho}{iii} for even (top) and odd (bottom) parity for the model given in table~\ref{tab:FAC_configs1}    
		using the \FAC\ code. Red horizontal lines indicate the initial model, while blue lines indicate the level energies of the calibrated model. Black horizontal lines show experimental data from the NIST database \citep{NIST_ASD} used to calibrate the calculated levels. Levels present in the DREAM database are shown in green.}  
	\label{fig:HoIII_levels} 
\end{figure*}

\begin{figure*}
	\includegraphics[width=0.99\linewidth]{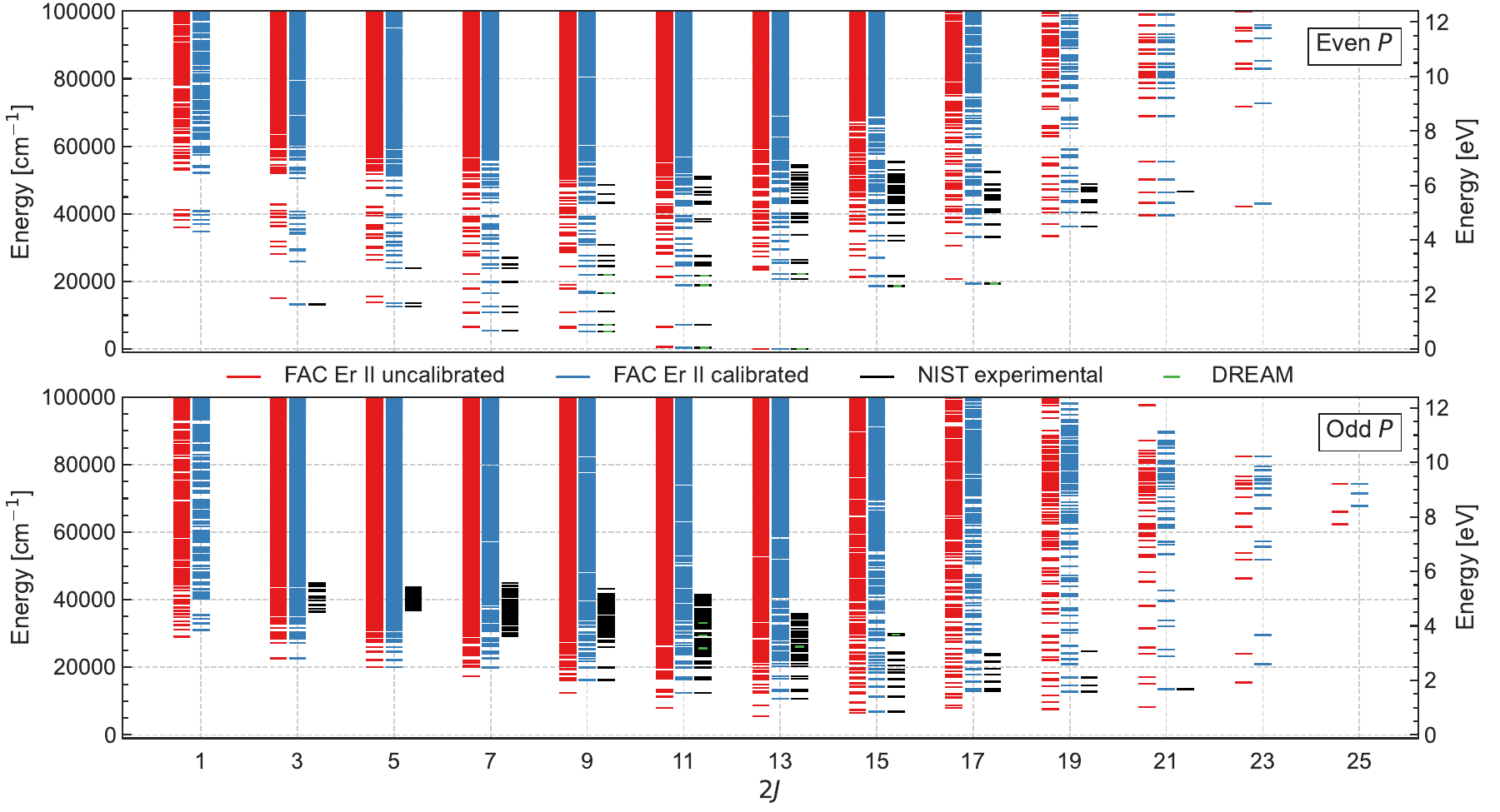}
	\caption{\justifying Energy levels for \ion{Er}{ii} for even (top) and odd (bottom) parity for the model given in table~\ref{tab:FAC_configs1}    
		using the \FAC\ code. Red horizontal lines indicate the initial model, while blue lines indicate the level energies of the calibrated model. Black horizontal lines show experimental data from the NIST database \citep{NIST_ASD} used to calibrate the calculated levels. Levels present in the DREAM database are shown in green.}  
	\label{fig:ErII_levels} 
\end{figure*}

\begin{figure*}
	\includegraphics[width=0.99\linewidth]{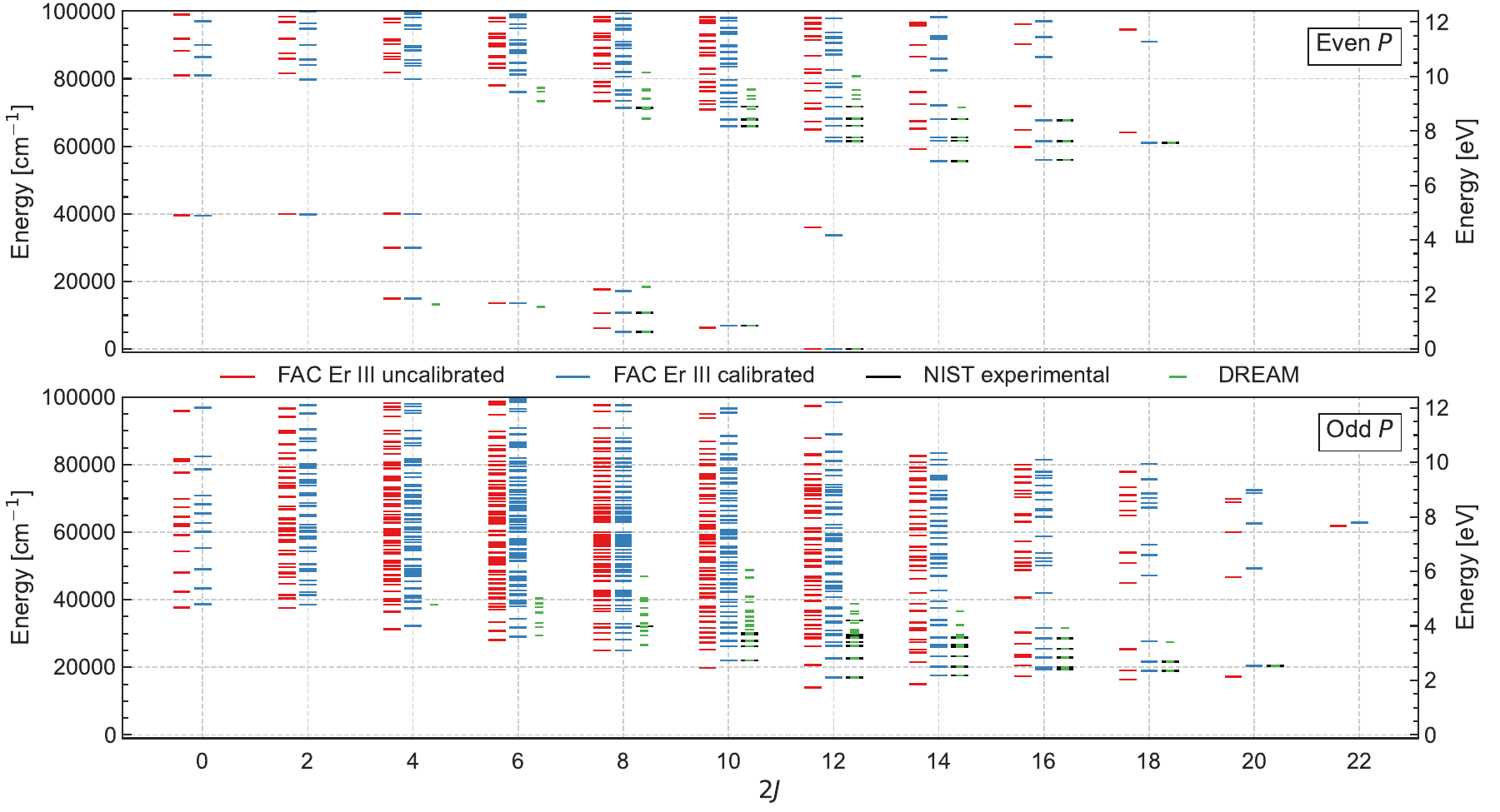}
	\caption{\justifying Energy levels for \ion{Er}{iii} for even (top) and odd (bottom) parity for the model given in table~\ref{tab:FAC_configs1}    
		using the \FAC\ code. Red horizontal lines indicate the initial model, while blue lines indicate the level energies of the calibrated model. Black horizontal lines show experimental data from the NIST database \citep{NIST_ASD} used to calibrate the calculated levels. Levels present in the DREAM database are shown in green.}  
	\label{fig:ErIII_levels} 
\end{figure*}

\begin{figure*}
	\includegraphics[width=0.99\linewidth]{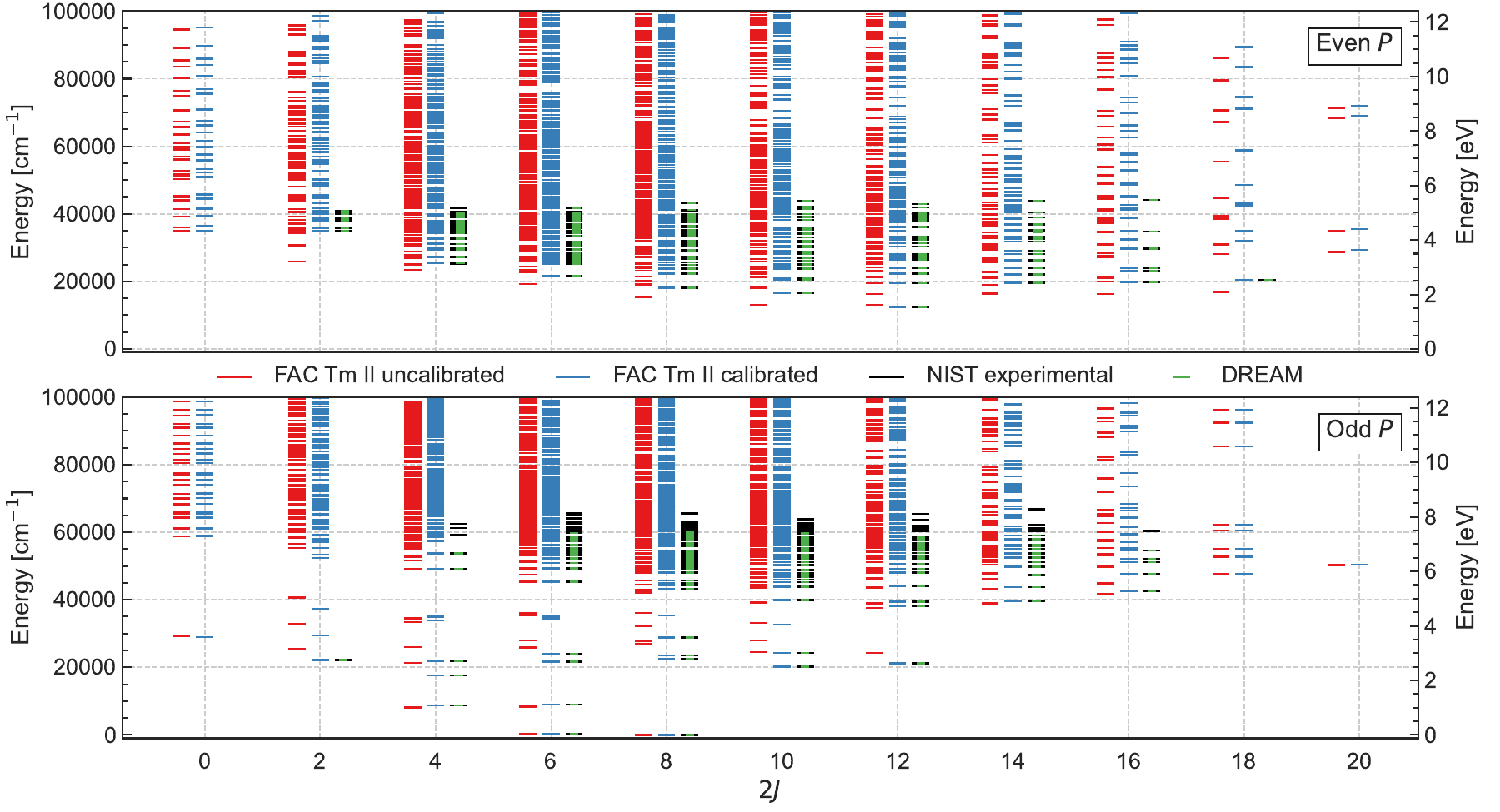}
	\caption{\justifying Energy levels for \ion{Tm}{ii} for even (top) and odd (bottom) parity for the model given in table~\ref{tab:FAC_configs1}    
		using the \FAC\ code. Red horizontal lines indicate the initial model, while blue lines indicate the level energies of the calibrated model. Black horizontal lines show experimental data from the NIST database \citep{NIST_ASD} used to calibrate the calculated levels. Levels present in the DREAM database are shown in green.}  
	\label{fig:TmII_levels} 
\end{figure*}

\begin{figure*}
	\includegraphics[width=0.99\linewidth]{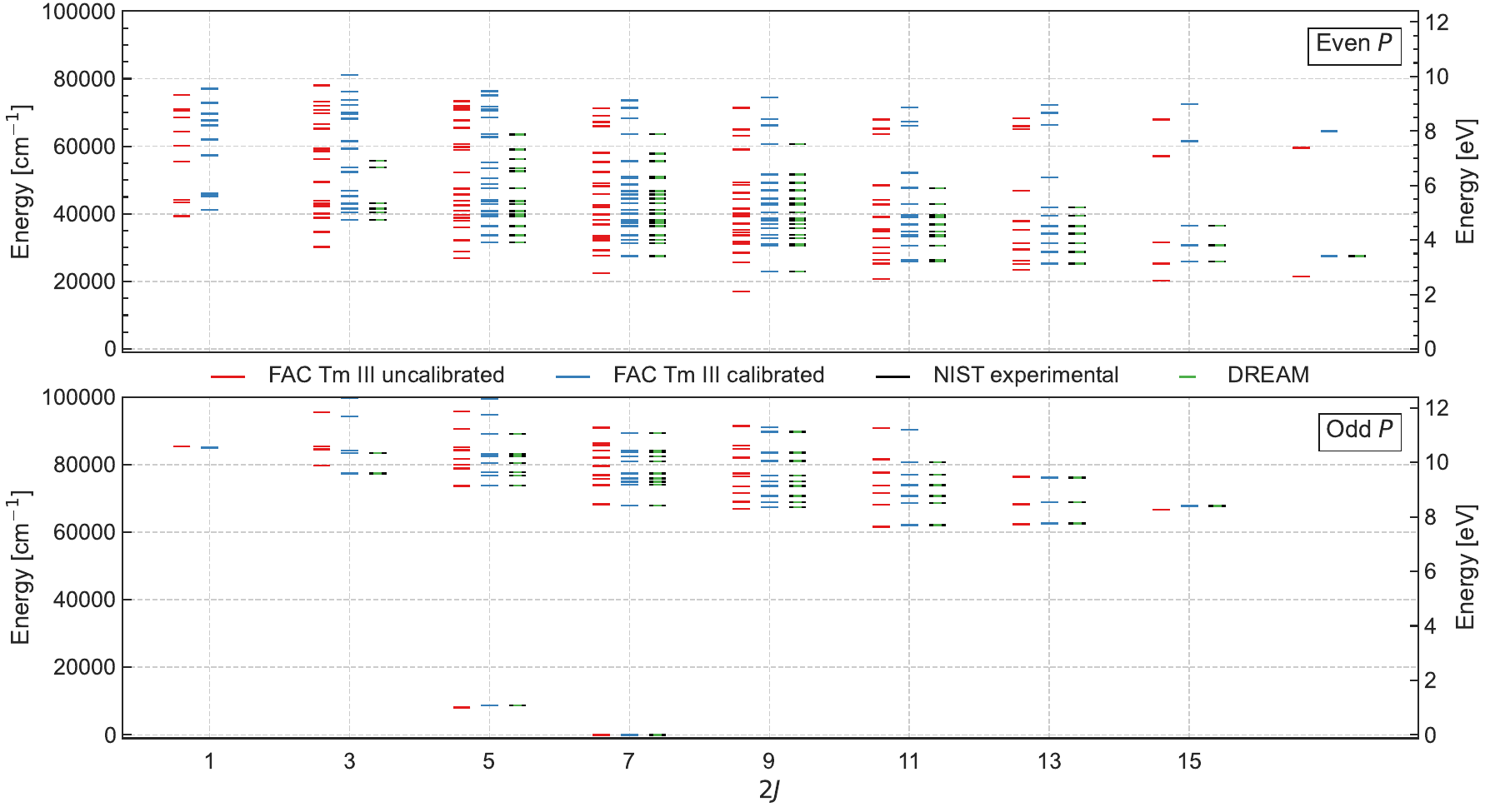}
	\caption{\justifying Energy levels for \ion{Tm}{iii} for even (top) and odd (bottom) parity for the model given in table~\ref{tab:FAC_configs1}    
		using the \FAC\ code. Red horizontal lines indicate the initial model, while blue lines indicate the level energies of the calibrated model. Black horizontal lines show experimental data from the NIST database \citep{NIST_ASD} used to calibrate the calculated levels. Levels present in the DREAM database are shown in green.}  
	\label{fig:TmIII_levels} 
\end{figure*}

\begin{figure*}
	\includegraphics[width=0.99\linewidth]{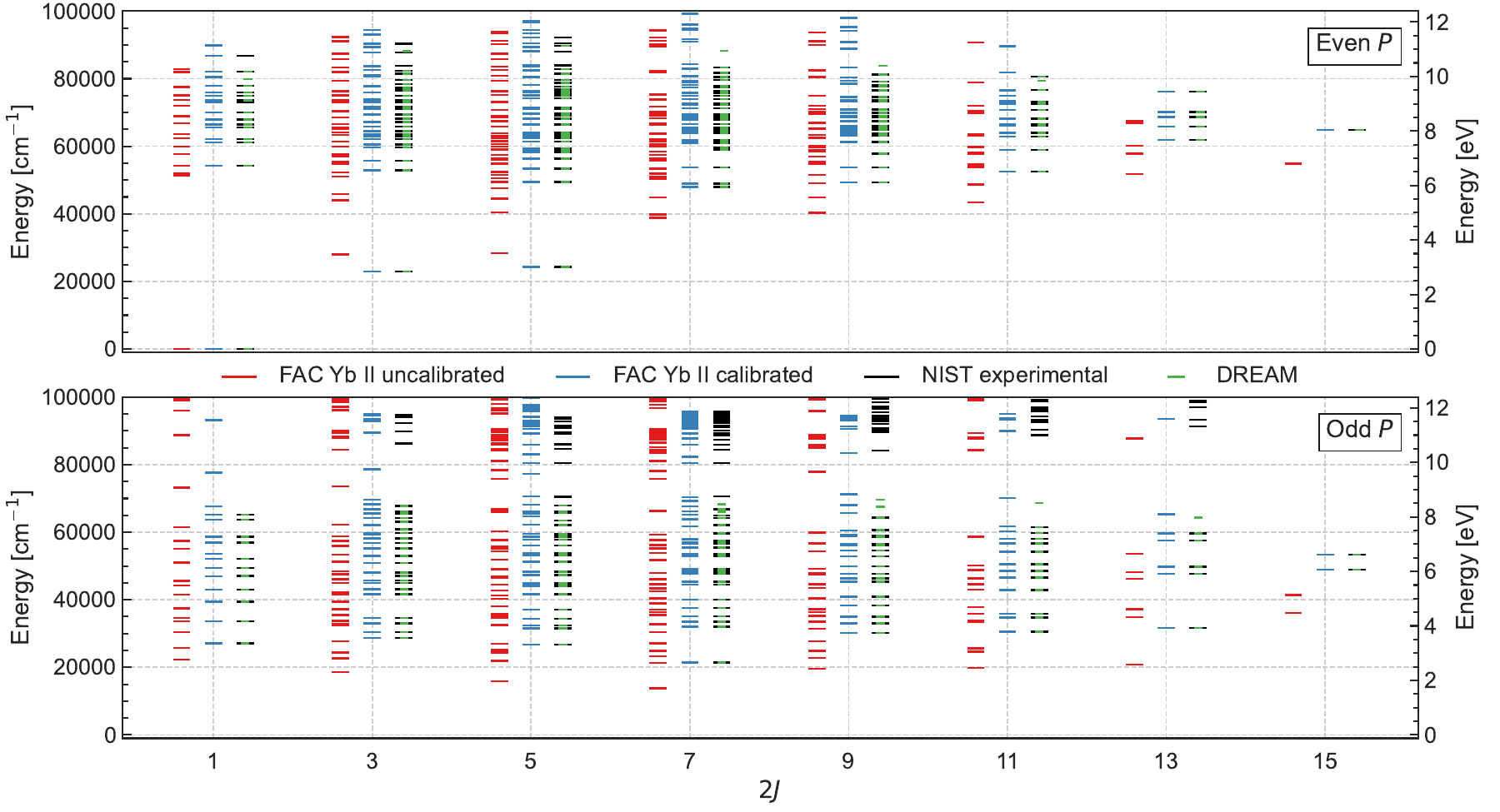}
	\caption{\justifying Energy levels for \ion{Yb}{ii} for even (top) and odd (bottom) parity for the model given in table~\ref{tab:FAC_configs1}    
		using the \FAC\ code. Red horizontal lines indicate the initial model, while blue lines indicate the level energies of the calibrated model. Black horizontal lines show experimental data from the NIST database \citep{NIST_ASD} used to calibrate the calculated levels. Levels present in the DREAM database are shown in green.}  
	\label{fig:YbII_levels} 
\end{figure*}

\begin{figure*}
	\includegraphics[width=0.99\linewidth]{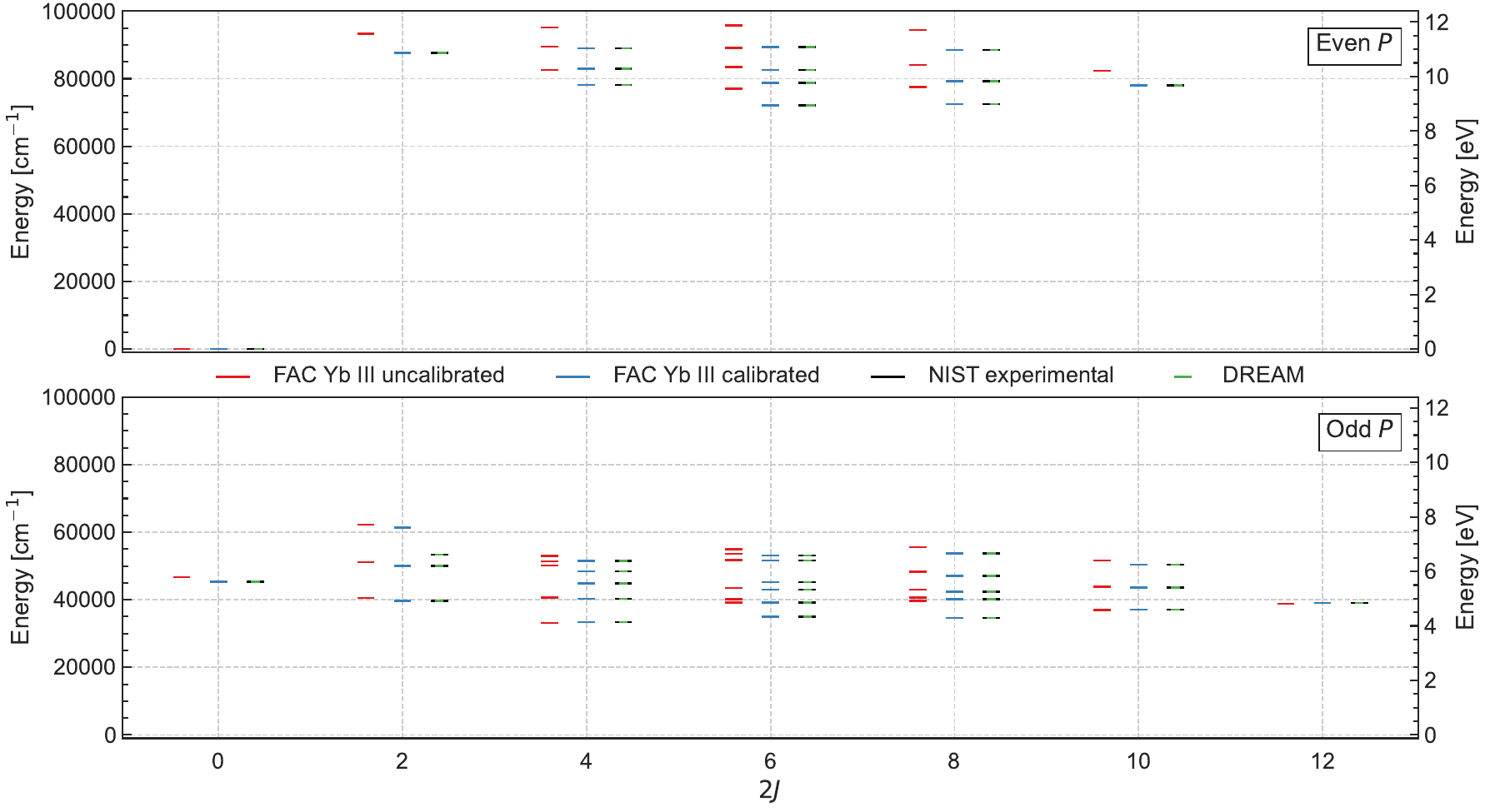}
	\caption{\justifying Energy levels for \ion{Yb}{iii} for even (top) and odd (bottom) parity for the model given in table~\ref{tab:FAC_configs1}    
		using the \FAC\ code. Red horizontal lines indicate the initial model, while blue lines indicate the level energies of the calibrated model. Black horizontal lines show experimental data from the NIST database \citep{NIST_ASD} used to calibrate the calculated levels. Levels present in the DREAM database are shown in green.}  
	\label{fig:YbIII_levels} 
\end{figure*}

In figures~\ref{fig:LaII_levels} -- \ref{fig:YbIII_levels} we give an overview of the energy levels for all 28 ions studied in this work. Instead of showing the electronic configuration bands, we display energy levels split into parity and angular momentum $J$, as these are the true quantum numbers and they do not depend on mixing between configurations. Particularly, at high-level energies, multiple electronic configurations contribute on a similar level to the eigenvector compositions, making it challenging to assign the energy level to one specific configuration. 

\bibliography{references}











\end{document}